\documentclass[3p,preprint,number,sort&compress]{elsarticle}

\usepackage{lineno}
\usepackage{graphicx}  
\usepackage{dcolumn}   
\usepackage{bm}        
\usepackage{amssymb}   
\usepackage{amsmath}
\usepackage{booktabs}
\DeclareMathOperator{\Tr}{Tr}
\usepackage{blkarray, multirow, graphicx, diagbox, color, xcolor, colortbl}
\usepackage{bbm, bbold}
\usepackage{hyphenat}
\usepackage{ifthen}
\usepackage{xkeyval}
\usepackage{moreverb}
\usepackage{rotating}
\usepackage{slashbox}
\usepackage{xspace}
\usepackage{nicefrac}
\usepackage[]{units}

\usepackage{braket}
\usepackage[inline]{enumitem}
\usepackage{tabto}
\usepackage{listings}
\usepackage[super]{nth}
\usepackage{xstring}
\def\ReplaceStr#1{%
	\IfSubStr{#1}{p}{%
		\StrSubstitute{#1}{p}{.}}{#1}}

\usepackage{capt-of}
\usepackage[caption=false]{subfig}

\usepackage{algpseudocode}
\usepackage{algorithm}

\usepackage{fp}
\usepackage[customcolors]{hf-tikz} 
\usepackage{tikz}
\usepackage{calc}
\usetikzlibrary{external}
\graphicspath{{figures/}}
\usepackage{pgffor}
\usepackage{pgfplots}
\pgfplotsset{compat=newest}
\usepackage{pgfplotstable}
\usepgfplotslibrary{groupplots}
\usepgfplotslibrary{fillbetween}

\tikzstyle{n} = [draw,shape=ellipse,minimum size=1.5em,inner sep=0pt,fill=white!20, minimum width=2.5em]
\tikzstyle{Init} = [n,color=green,fill=green!20,text=black]
\tikzstyle{Fin} = [n,color=red,fill=red!20,text=black]
\tikzstyle{Ghost} = [minimum size=1.5em,inner sep=0pt,color=white,text=black]
\tikzstyle{Multiple} = [draw,shape=rect,minimum size=2em,inner sep=0pt]

\tikzstyle{site} = [regular polygon,regular polygon sides=4,rounded corners=2pt, thick, inner sep=0.5pt, minimum size=1.25em, font=\footnotesize, draw=black!70, text opacity=1]
\tikzstyle{ghost} = []
\tikzstyle{op} = [regular polygon, regular polygon sides=4, rounded corners=2pt, draw=blue!50, thick, inner sep=0.2pt, minimum width=1.25em, minimum height=1.5em,font=\footnotesize]
\tikzstyle{ld} = [inner sep=1pt, font=\footnotesize]
\tikzstyle{intersite} = [regular polygon, regular polygon sides=4, shape border rotate= 45, draw=black!50,fill=black!20,thick,inner sep=0pt,minimum width=1.5em]

\tikzstyle{ghostA} = [text=red!70,thick, minimum size=2*(5pt-\pgflinewidth), inner sep=0pt, outer sep=0pt]
\tikzstyle{ghostB} = [text=blue!70,thick, minimum size=2*(5pt-\pgflinewidth), inner sep=0pt, outer sep=0pt]
\tikzstyle{siteA} = [regular polygon, regular polygon sides=3, shape border rotate= 30, draw=red!50,thick,inner sep=0pt,minimum width=1.5em,rounded corners=2pt]
\tikzstyle{ghostsiteA} = [regular polygon, regular polygon sides=3, shape border rotate= 30, draw=white,fill=white,thick,inner sep=0pt,minimum width=1.5em]
\tikzstyle{siteB} = [regular polygon, regular polygon sides=3, shape border rotate= -30, draw=green!50,thick,inner sep=0pt,minimum width=1.5em]
\tikzstyle{siteC} = [regular polygon, regular polygon sides=3, shape border rotate= 180, draw=red!50,thick,inner sep=0pt,minimum width=1.5em,rounded corners=2pt]
\tikzstyle{siteD} = [regular polygon, regular polygon sides=3, shape border rotate= 0, draw=green!50,thick,inner sep=0pt,minimum width=1.5em,rounded corners=2pt]
\tikzstyle{operatorA} = [cross out, draw=red!70, thick, minimum size=2*(5pt-\pgflinewidth), inner sep=0pt, outer sep=0pt]
\tikzstyle{operatorB} = [cross out, draw=blue!70, thick, minimum size=2*(5pt-\pgflinewidth), inner sep=0pt, outer sep=0pt]
\tikzstyle{nosite} = [circle,draw=white,fill=white,thick,inner sep=0.1pt,minimum width=1.5em]
\tikzstyle{unsite} = [circle, outer sep=0pt,inner sep=0.5pt,minimum width=1.45em]
\tikzstyle{hiddensite} = [circle,draw=white!50,fill=white!20,thick,inner sep=0pt,minimum width=1.25em]
\tikzstyle{ghost} = []
\tikzstyle{intersite} = [regular polygon, regular polygon sides=4, shape border rotate= 45, draw=black!50,thick,inner sep=0pt,minimum width=1.5em,rounded corners=2pt]

\definecolor{colorA}{rgb} {0.58,0,0.8275}
\definecolor{colorB}{rgb} {0.11,0.663,0.51}
\definecolor{colorC}{rgb} {0.3373,0.7059,0.9137}
\definecolor{colorD}{rgb} {0.902,0.6235,0}
\definecolor{colorE}{rgb} {0.9451,0.902,0.3255}
\definecolor{colorF}{rgb} {0.3373,0.3255,0.902}
\definecolor{colorG}{rgb} {0.9451,0.3255,0.3373}

\usetikzlibrary
{
	calc,
	decorations,
	plotmarks,
	patterns,
	positioning,
	petri,
	arrows,
	intersections,
	decorations.markings,
	backgrounds,
	fit,
	matrix,
	graphs,
	shapes.geometric,
	decorations.pathmorphing,
	shapes.misc,
	shapes,
	tikzmark,
	arrows.meta,
	pgfplots.colorbrewer,
	fpu,
}
\usepackage{ocgx}

\usetikzlibrary{ocgx}
\pgfplotsset{
        cycle from colormap manual style/.style={
            x=3cm,y=10pt,ytick=\empty,
            stack plots=y,
            every axis plot/.style={line width=2pt},
        },
}

\tikzset{>=stealth}
\tikzset{->-/.style={decoration={
			markings,
			mark=at position .5 with {\arrow{>}}},postaction={decorate}}}
\tikzset{->-rc/.style={decoration={
			markings,
			mark=at position .5 with {\arrow{>}}},postaction={decorate},rounded corners}}
\tikzset{-<-/.style={decoration={
			markings,
			mark=at position .5 with {\arrow{<}}},postaction={decorate}}}

\tikzstyle{orientedsnake} = [
decorate, 
decoration={snake},
->
]  
\tikzstyle{orientedshortarrow} = [
decoration={markings,
	mark=at position .33 with {\arrow{>}}},
postaction={decorate}
]  
\tikzstyle{orientedlongarrow} = [
decoration={markings,
	mark=at position .67 with {\arrow{>}}},
postaction={decorate}
]
\tikzset{dbl/.style={double,
		double equal sign distance,
		-implies,
		shorten >=10pt,
		shorten <=10pt}}
\tikzset{
	between/.style args={#1 and #2}{
		at = ($(#1)!0.5!(#2)$)
	}
}
\pgfmathdeclarefunction{linearFct}{2}%
{%
	\pgfmathparse{#1*x+#2}%
}
\pgfmathdeclarefunction{logFct}{2}%
{%
	\pgfmathparse{#1*log10(x)+#2}%
}
\pgfmathdeclarefunction{quadFct}{3}%
{%
	\pgfmathparse{#1*x*x+#2*x+#3}%
}

\newcommand{\hopping}[0]{t_0}
\newcommand{\nodagger}[0]{{\phantom{\dagger}}}
\newcommand{\noprime}[0]{{\phantom{\prime}}}

\newcommand{\overbar}[1]{\mkern 1.5mu\overline{\mkern-1.5mu#1\mkern-1.5mu}\mkern 1.5mu}

\pgfmathdeclarefunction{peierlspotential}{6}%
{
	\pgfmathparse
	{
		#2 / (#3 * #5) 
		* sin(deg(#1 * #3 - #3 * #5 * (x)))
		* exp(-( ( #1 - #5 * ((x) - #4) ) )^2.0 / ( #6^2.0 ) )
	}%
}
\newboolean{buildCurrentBlockInline}
\setboolean{buildCurrentBlockInline}{false}
\newboolean{rebuildCurrentBlock}
\setboolean{rebuildCurrentBlock}{false}

\newboolean{rebuildData}
\setboolean{rebuildData}{false}

\newboolean{removeData}
\setboolean{removeData}{false}

\newboolean{onlyReadFitData}
\setboolean{onlyReadFitData}{false}

\newboolean{buildtikzpics}
\setboolean{buildtikzpics}{true}

\newboolean{includeFits}
\setboolean{includeFits}{false}

\newif\ifrebuildtikz
\newif\ifChangeMode
\ChangeModetrue
\ChangeModefalse
\ifthenelse{\boolean{buildtikzpics}}
{
	\rebuildtikztrue
	\usetikzlibrary{external}
	\tikzexternalize[optimize=false,prefix=figures/autogen/]
}
{
	\rebuildtikzfalse
}

\usepackage[acronym,nohypertypes={acronym,notation}]{glossaries}
\usepackage{cleveref}
\Crefname{appendix}{Appendix}{Appendices}
\Crefname{equation}{Equation}{Equations}
\Crefname{figure}{Figure}{Figures}
\Crefname{section}{Section}{Sections}
\Crefname{tabular}{Tabular}{Tabulars}
\crefname{appendix}{appendix}{appendizes}
\crefname{equation}{}{}
\crefname{figure}{figure}{figures}
\crefname{section}{section}{sections}
\crefname{tabular}{table}{tables}

\newcommand{\symmps}{\textsc{SymMPS}}

\newcommand{\ingoing}[1]{{#1}}
\newcommand{\outgoing}[1]{{#1}}
\newcommand{\going}[1]{#1}

\def\bi{m}

\def\biwqn{\alpha}

\ExplSyntaxOn
\DeclareExpandableDocumentCommand \eval { m } { \fp_eval:n { #1 } }
\ExplSyntaxOff

\makeatletter
\def\pgfplotsutil@decstringcounter#1{%
 \begingroup
  \c@pgf@counta=#1\relax
  \advance\c@pgf@counta by -1
  \edef#1{\the\c@pgf@counta}%
  \pgfmath@smuggleone#1%
 \endgroup
}%

\pgfplotsset{
/pgfplots/each nth point**/.style 2 args={%
/pgfplots/x filter/.append code={%
 \ifnum\coordindex=0
  \def\c@pgfplots@eachnthpoint@xfilter{#2}%
  \def\c@pgfplots@eachnthpoint@xfilter@zero{0}%
 \fi
 \ifx\c@pgfplots@eachnthpoint@xfilter@zero\c@pgfplots@eachnthpoint@xfilter
  \def\c@pgfplots@eachnthpoint@xfilter{#1}%
 \else
  \let\pgfmathresult\pgfutil@empty
 \fi
 \pgfplotsutil@decstringcounter\c@pgfplots@eachnthpoint@xfilter
}%
},
}

\ExplSyntaxOn
\DeclareExpandableDocumentCommand \eval { m } { \fp_eval:n { #1 } }
\ExplSyntaxOff

\newcommand{\printpgfnumberwitherror}[2]%
{%
	\pgfmathfloatparsenumber{#1}%
	\pgfmathfloattomacro{\pgfmathresult}{\Fn}{\Mn}{\En}%
	\pgfmathparse{\Fn==2 ? "-" : ""}%
	\edef\Sn{\pgfmathresult}%
	\pgfmathfloatparsenumber{#2}%
	\pgfmathfloattomacro{\pgfmathresult}{\Fe}{\Me}{\Ee}%
	\pgfmathparse{int(sqrt((\Ee-\En)^2))}%
	\edef\precisionAbsEe{\pgfmathresult}%
	\pgfmathparse{int(\Ee-\En)}%
	\edef\precisionE{\pgfmathresult}%
	\pgfmathparse{\eval{\Me*10^(\precisionE)}}%
	\ifthenelse{\En=0}%
	{%
		$\Sn\pgfmathprintnumber[fixed, precision=\precisionAbsEe, zerofill]{\Mn} \pm (\pgfmathprintnumber[std, precision=0, zerofill]{#2})$%
	}%
	{%
		\ifthenelse{\En=-1}%
		{%
			\pgfmathparse{\eval{\Mn/10}}%
			\edef\Mn{\pgfmathresult}%
			\pgfmathparse{int(\precisionAbsEe+1)}%
			\edef\precisionAbsEe{\pgfmathresult}%
			$\Sn\pgfmathprintnumber[fixed, precision=\precisionAbsEe, zerofill]{\Mn} \pm (\pgfmathprintnumber[std, precision=0, zerofill]{#2})$%
		}%
		{%
			$\left(\Sn\pgfmathprintnumber[std, precision=\precisionAbsEe, zerofill]{\Mn} \pm \pgfmathprintnumber[fixed, precision=\precisionAbsEe, zerofill]{\pgfmathresult}\right)\cdot10^{\En}$%
		}%
	}%
}

\newacronym{1D}{1D}{one\hyp dimensional}
\newacronym{MPS}{MPS}{matrix\hyp product state}
\newacronym{MPO}{MPO}{matrix\hyp product operator}
\newacronym{SVD}{SVD}{singular\hyp value decomposition}
\newacronym{QCS}{QCS}{quantum\hyp computer simulator}
\newacronym{QC}{QC}{quantum computer}
\newacronym{FSM}{FSM}{finite\hyp state machine}
\newacronym{ACA}{ACA}{adaptive cross\hyp approximation}
\newacronym{CDW}{CDW}{charge\hyp density wave}
\newacronym{LL}{LL}{Luttinger-liquid}
\newacronym{SDW}{SDW}{spin\hyp density wave}
\newacronym{ARPES}{ARPES}{angle-resolved photoemission spectroscopy}
\newacronym{OBC}{OBC}{open-boundary conditions}
\newacronym{PBC}{PBC}{periodic-boundary conditions}
\newacronym{TEBD}{TEBD}{time-evolution block-decimation}
\newacronym{TDVP}{TDVP}{time-dependent variational principle}
\newacronym{iff}{iff}{if and only if}
\newacronym{DFT}{DFT}{density\hyp functional theory}
\newacronym{DMFT}{DMFT}{dynamical mean\hyp field theory}
\newacronym{DMRG}{DMRG}{density\hyp matrix renormalization group}
\newacronym{LBO}{LBO}{local basis optimization}
\newacronym{PS-DMRG}{PS-DMRG}{pseudosite \gls{DMRG}}
\newacronym{PP-2DMRG}{PP-2DMRG}{projected purified two-site \gls{DMRG}}
\newacronym{DMRG3S+LBO}{DMRG3S+LBO}{strictly single-site \gls{DMRG} with \gls{LBO}}
\newacronym{PP}{PP}{projected purified}
\newacronym{QMC}{QMC}{quantum Monte Carlo}
\newacronym{AIM}{AIM}{Anderson impurity model}
\newacronym{SIAM}{SIAM}{single impurity Anderson model}
\newacronym{LDA}{LDA}{local\hyp density approximation}
\newacronym{LBNL}{LBNL}{Lawrence Berkeley National Laboratory}
\newacronym{VQE}{VQE}{variational\hyp quantum eigensolver}
\newacronym{ED}{ED}{exact diagonalization}
\newacronym{QPT}{QPT}{quantum phase transition}
\newacronym{QCP}{QCP}{quantum critical point}
\newacronym{ETH}{ETH}{eigenstate thermalization hypothesis}
\newacronym{AKLT}{AKLT}{Affleck\hyp Lieb\hyp Kennedy\hyp Tasaki}
\newglossaryentry{QR}{name={QR},description={QR decomposition}}
\newacronym{TNS}{TNS}{tensor\hyp network state}
\newacronym{VMPS}{VMPS}{variational matrix-product state}
\newacronym{DMRG-LBO}{DMRG-LBO}{\gls{DMRG} with local basis optimization}
\newacronym{1RDM}{1RDM}{single\hyp site reduced density\hyp matrix}

\begin{document}
	
\begin{frontmatter}
\title{Comparative Study of State-of-the-Art Matrix-Product-State Methods for Lattice Models with Large Local Hilbert Spaces without $U(1)$ symmetry}

\author[goe]{Jan Stolpp}
\author[upp]{Thomas K\"ohler}
\author[goe]{Salvatore R. Manmana\fnref{fn1}}
\author[han]{Eric Jeckelmann}
\author[goe]{Fabian Heidrich-Meisner}
\author[lmu]{Sebastian Paeckel\corref{mycorrespondingauthor}}
\cortext[mycorrespondingauthor]{Corresponding author}

\address[goe]{Institut f\"ur Theoretische Physik, Georg-August-Universit\"at G\"ottingen, 37077 G\"ottingen, Germany}

\fntext[fn1]{Present Address: Fachbereich Physik, Philipps-Universit\"at Marburg, 35032 Marburg, Germany}

\address[upp]{Department of Physics and Astronomy, Uppsala University, Box 516, S-751 20 Uppsala, Sweden}

\address[han]{Leibniz Universit\"at Hannover, Institut f\"ur Theoretische Physik, 30167 Hannover, Germany}

\address[lmu]{Department of Physics, Arnold Sommerfeld Center for Theoretical Physics (ASC), Munich Center for Quantum Science and Technology (MCQST), Ludwig-Maximilians-Universit\"{a}t M\"{u}nchen, 80333 M\"{u}nchen, Germany}

\ead{sebastian.paeckel@physik.uni-muenchen.de}

\begin{abstract}
	Lattice models consisting of high\hyp dimensional local degrees of freedom without global particle\hyp number conservation constitute an important problem class in the field of strongly correlated quantum many\hyp body systems.
	For instance, they are realized in electron-phonon models, cavities, atom-molecule resonance models, or superconductors.
	In general, these systems elude a complete analytical treatment and need to be studied using numerical methods where matrix-product states (MPS) provide a flexible and generic ansatz class.
	Typically, MPS algorithms scale at least quadratic in the dimension of the local Hilbert spaces.
	Hence, tailored methods, which truncate this dimension, are required to allow for efficient simulations.
	Here, we describe and compare three state\hyp of\hyp the\hyp art MPS methods each of which exploits a different approach to tackle the computational complexity.
	We analyze the properties of these methods for the example of the Holstein model, performing high\hyp precision calculations as well as a finite\hyp size\hyp scaling analysis of relevant ground\hyp state obervables.
	The calculations are performed at different points in the phase diagram yielding a comprehensive picture of the different approaches.
\end{abstract}
\end{frontmatter}

\section{\label{sec:introduction}Introduction}
In the past two decades, \gls{TNS} based methods have become very successful tools to study strongly correlated, low\hyp dimensional quantum systems in and out of equilibrium \cite{PhysRevLett.69.2863,PhysRevB.48.10345,PhysRevLett.75.3537,PhysRevLett.93.040502,Verstraete2006,Schollwoeck201196,Paeckel2019}.
In principle, these methods can reach a very high accuracy that is controlled by a successive scaling in the bond dimensions.
The prerequisites are a sufficiently small amount of entanglement encoded in the wave function and an efficient means of computing energies and expectation values \cite{Schollwoeck201196}.
Re-expressing the early algorithms of the \gls{DMRG} \cite{PhysRevLett.69.2863,PhysRevB.48.10345} using \gls{MPS} \cite{PhysRevLett.75.3537,verstraete-cirac_periodic} representations paved the way for the formulation of new efficient and flexible variational algorithms to study correlated quantum systems with excellent precision in one dimension~\cite{Verstraete2006,MPSreviewVerstraete,PhysRevLett.100.040501,Schollwoeck201196}.
Exploring the tensor-product structure of the many\hyp body Hilbert space to decompose the coefficients of a quantum state into \textit{local} tensors yields a powerful formulation of \gls{DMRG} and allows the exploitation of ideas from quantum\hyp information theory \cite{Verstraete2006,PhysRevLett.99.220405,Orus2014}.
Thereby, numerical optimization schemes that work on the local tensors can be formulated, which is at the heart of their high efficiency.
Unfortunately, such a decomposition becomes numerically very costly if the local quantum systems have a large number of internal degrees of freedom that do not obey conservation laws.
Indeed, these situations are not too exotic as may be illustrated by the simple example of lattice fermions locally coupled to Einstein phonons, the Holstein model \cite{Holstein1959}.
Describing such a system by means of \gls{MPS} requires, in principle, an infinite number of local degrees of freedom, while in practice, one needs to introduce a cutoff in the local Hilbert space dimension~\cite{Tezuka2005,Matsueda2007}.
As is well known, these systems can feature rich physics.
Modeling the influence of phonons on interacting fermions on a numerically unbiased footing \cite{Marsiglio1995,Wellein1996,Weisse1998,Bonfmmodeheckclsecia1999,Weise2000,Hohenadler2004,Wall2016} is an important problem with many intensively studied questions, such as the formation, stability, and dynamics of (bi-)polarons in the (Hubbard-)Holstein model \cite{Jeckelmann1998,Loos2006,PhysRevB.79.235336,Brockt2015,Hashimoto2017,Weber2018,Kloss2019,Stolpp2020}.
Various numerical approaches have been employed to study these systems ranging from exact diagonalization~\cite{Jeckelmann2007,Weisse2008} over quantum Monte-Carlo methods~\cite{Assaad2008,Hohenadler2008} to \gls{DMFT}~\cite{Werner2007,Aoki2014,Kemper2017}.

Ultracold atoms provide another relevant platform in which large local Hilbert spaces play an important role for the theoretical description.
For instance, models involving Feshbach-resonances feature a molecular and an atomic channel as well as a term that converts two atoms into one molecule. 
As a consequence, the particle numbers of molecules and atoms are not conserved individually. 
These models are called Bose-Fermi or Bose-Bose resonance models \cite{Bloch:2005p988,Heidrich-Meisner2010,Ejima2011,Dorfner2016}.
Models of single atoms coupled to optical cavities \cite{Purdy2010,Thompson2013,Ritsch2013} have to account for the bosonic nature of the light field and are important systems to understand the interplay between light and matter on the atomic scale.
Finally, in heterostructures of proximity\hyp coupled s\hyp wave superconductors, the condensate can be described separately, yielding an effective bosonic bath with pairwise creation and annihilation of particles.
Recently, these systems have been suggested as promising candidates for an experimental realization of Majorana quasi particles, potential building blocks of topological qubits \cite{Brouwer2011,Cook2011,Hui2015,Keselman2019}.
Here, we discuss three state\hyp of\hyp the\hyp art \gls{MPS} methods that can be utilized to efficiently deal with such a large number of required local degrees of freedom: The \gls{PS-DMRG} method~\cite{Jeckelmann1998}, \gls{DMRG-LBO}~\cite{Zhang1998,Brockt2015,PhysRevB.60.1643,Friedman2000,Wong2008}, and the \gls{PP-2DMRG}~\cite{10.21468/SciPostPhys.10.3.058}.
The first method was introduced by Jeckelmann and White to extend the applicability of \gls{DMRG} to describe electrons coupled to phonons \cite{Jeckelmann1998}.
\Gls{LBO} proposed by Zhang \textit{et al.}~\cite{Zhang1998} is based on a rotation into the eigenbasis of the \gls{1RDM}.
It allows for a faithful truncation yielding an optimal approximation of the \gls{1RDM}.
The \gls{PP-2DMRG} developed recently by some of the authors consists of a mapping of the initial problem into a purified Hilbert space with a subsequent projection into an invariant subspace to generate an artificial $U(1)$ symmetry that collapses the local dimensions of the bosonic degrees of freedom and allows to exploit canonical \gls{MPS} truncation schemes \cite{10.21468/SciPostPhys.10.3.058}.

We compare the methods by applying them to the Holstein model at half filling.
This system undergoes a phase transition from a gapless \gls{LL} phase to a gapped \gls{CDW} phase upon changing the electron\hyp phonon\hyp coupling strength\cite{Hirsch1983,Weisse1998,Bursill1998,Creffield2005}.
To this end, we conduct high\hyp precision calculations at an intermediate system size of $L = 51$ (where $L$ is the number of sites) and perform a finite\hyp size extrapolation of the ground\hyp state energy and the \gls{CDW} order parameter from data at $L = 51,101,151,201$ with relaxed precision demands.
According to our analysis, all three methods agree within their anticipated error margins.
We further find that the \gls{PS-DMRG} and the \gls{PP-2DMRG} require a larger bond dimension in order to reach a comparable precision as \gls{DMRG-LBO}.
However, the efficient representation of states using the \gls{DMRG-LBO} method comes at the cost of additional numerical control parameters in which the calculations have to be converged.
Here, the \gls{PS-DMRG} and \gls{PP-2DMRG} benefit from being conceptually simpler approaches.
\gls{PS-DMRG} can be converged with respect to a scaling analysis in the bond dimension and comparative calculations using different local Hilbert-space dimensions.
Finally, using \gls{PP-2DMRG}, it suffices to converge the calculations in the discarded weight, only.
On the one hand, this comes at the cost of the largest growth in bond dimension compared to the other methods, but, on the other hand, this can be compensated by exploiting the restored global $U(1)$ symmetries.
The paper is organized as follows:
In \cref{sec:holstein-model}, we briefly review the Holstein model and its phase diagram at half filling.
In \cref{sec:methods}, we introduce the different numerical methods.
After recapitulating some basic facts about \glspl{MPS} in \cref{sec:methods:mps}, we introduce the \gls{PS-DMRG} in \cref{sec:methods:ps-dmrg}, the \gls{DMRG3S+LBO} in \cref{sec:methods:lbo-dmrg},
 and the \gls{PP-2DMRG} in \cref{sec:methods:pp-dmrg}.
\Cref{sec:comparison} is devoted to the comparison between the methods, where \cref{sec:comparison:gs-scaling} and \cref{sec:comparison:loc-observables} contain the high\hyp precision calculations,
 while \cref{sec:comparison:finite-site-extrapolation} contains the finite\hyp size extrapolations.
Finally, in \cref{sec:conclusion}, we give a conclusion of our work.

\subsection{\label{sec:holstein-model}Holstein Model}
\begin{figure}
	\centering
	\tikzsetnextfilename{holstein_phases_gs}
	\def\miny{0}
	\def\maxy{4}
	\begin{tikzpicture}
		\begin{axis}
		[
			width	= 0.95\textwidth,
			height	= 0.35\textheight,
			xmode	= log,
			ymin	= \miny,
			ymax	= \maxy,
			xmin	= 5e-2,
			xmax	= 20,
			xlabel	= {$\nicefrac{\hopping}{\omega_0}$},
			ylabel	= {$\nicefrac{\gamma}{\omega_0}$},
			axis on top,
			ticklabel style	= {font=\scriptsize},
			ylabel style	= {font=\scriptsize},
			xlabel style	= {font=\scriptsize},
		]
			\addplot
			[
				name path=curve,
				smooth,
				thick,
			] 
				table
				[
				]
				{data/holstein/phase_diagram_data_from_PRL.80.5607};
			\addplot
			[
				name path=min,
				domain=5e-2:20,
			]
				{\miny};
			\addplot
			[
				name path=max,
				domain=5e-2:20,
			]
				{\maxy};
			\addplot 
			[
				fill=colorC!40,
			]
			fill between
			[
				of = min and curve,
			];
			\addplot
			[
				fill=magenta!20!red!30,
			]
			fill between
			[
				of = max and curve,
			];
			
			\node[star, star points=5, draw=colorD!60!black, fill=colorD!60] at (axis cs:1,0.5) (P1) {};
			\node[anchor=west] at (P1.east) {(P.1)};
			
			\node[star, star points=5, draw=colorD!60!black, fill=colorD!60] at (axis cs:1,1.5) (P2) {};
			\node[anchor=west] at (P2.east) {(P.2)};
			
			\node[star, star points=5, draw=colorD!60!black, fill=colorD!60] at (axis cs:1,2) (P3) {};
			\node[anchor=west] at (P3.east) {(P.3)};
			
			\node[] at (0.2,3) (CDW) {\large CDW};
			\node[] at (7,1) (metal) {\large metallic \gls{LL}};
		\end{axis}
	\end{tikzpicture}
	\caption
	{
		\label{fig:holstein-model:phase-diagram}
		Schematic reproduction of the phase diagram of the $1$D Holstein model computed in \cite{Bursill1998}.
		The transition between the \gls{CDW} and the metallic \gls{LL} phase is controlled by the coupling $\gamma$ between the local oscillator displacement and the fermion density as well as the phonon frequency $\omega_0$.
		Marked points indicate the position of the parameter sets studied in this paper and we always chose $\omega_0 = 1$ in our calculations.
	}
\end{figure}
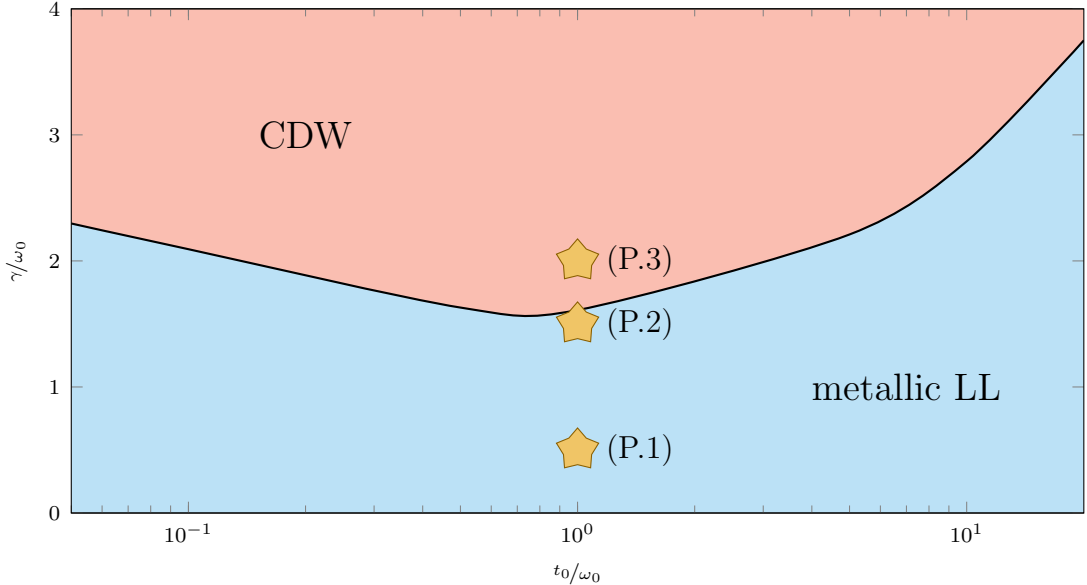

In order to demonstrate the applicability of the different methods, we conduct numerical ground\hyp state calculations of the half\hyp filled $1$D Holstein model with spinless fermions \cite{Holstein1959}. 
The model is defined by a set of spinless fermions with creation/annihilation operators $\hat c^{[\dagger]}_j$ on a lattice, where every lattice site possesses an associated harmonic oscillator with bosonic creation/annihilation operators $\hat b^{[\dagger]}_j$.
The oscillators are decoupled from each other and play the role of a single Einstein-phonon branch.
The density of the fermions $\hat n^f_j = \hat c^\dagger_j \hat c^\nodagger_j$ at a specific site $j$ couples to the displacement $(\hat b^\dagger_j + \hat b^\nodagger_j)$ of the associated harmonic oscillator:
\begin{align}
	\hat H
	&=
	\hat H_{f} +
	\hat H_{ph}+
	\hat H_{f-ph} \\
	\hat H_{f} &= -\hopping \sum_j \left(\hat c^\dagger_j \hat c^{\nodagger}_{j+1} + \mathrm{h.c.} \right) \\
	\hat H_{ph} &= \omega_0 \sum_j \hat b^\dagger_{j} \hat b^{\nodagger}_{j} \\
	\hat H_{f-ph} &= \gamma \sum_j \hat n^f_j\left(\hat b^\dagger_j + \hat b^{\nodagger}_j \right) \; .
\end{align}
An important exact relation between the different particle species and for eigenstates is given by~\cite{Hirsch1983,Jeckelmann1999}
\begin{equation}
	\braket{ \hat b^\dagger_j + \hat b^\nodagger_j } = 2\frac{\gamma}{\omega_0} \braket{ \hat n^f_j } \; .
\end{equation}
At half filling, the ground\hyp state phase diagram of this model exhibits a transition from a metallic \gls{LL} phase into a \gls{CDW} insulating phase when increasing the ratio $\nicefrac{\gamma}{\hopping}$ for fixed $\nicefrac{\hopping}{\omega_0} \neq 0$ \cite{Hirsch1983,Weisse1998,Bursill1998,Creffield2005} (see \cref{fig:holstein-model:phase-diagram}).

We investigate the Holstein model with open boundary conditions at different points in its parameter space.
In particular, we fixed $\hopping \equiv 1.0$ as unit of energy and chose the following sets of parameters
\begin{align}
	\tag{P.1} &\nicefrac{\omega_0}{\hopping}=1.0,& &\nicefrac{\gamma}{\hopping}=0.5 \; , \label{eq:comp-pp-with-lbo:1} \\
	\tag{P.2} &\nicefrac{\omega_0}{\hopping}=1.0,& &\nicefrac{\gamma}{\hopping}=1.5 \; , \label{eq:comp-pp-with-lbo:2} \\
	\tag{P.3} &\nicefrac{\omega_0}{\hopping}=1.0,& &\nicefrac{\gamma}{\hopping}=2.0 \; , \label{eq:comp-pp-with-lbo:3}
\end{align}
so that we expect the ground state to realize a \gls{LL} for the parameter set \cref{eq:comp-pp-with-lbo:1}, \cref{eq:comp-pp-with-lbo:2} is close to the phase boundary, and a \gls{CDW} is realized for \cref{eq:comp-pp-with-lbo:3} (cf., \cref{fig:holstein-model:phase-diagram}).
An important aspect is the absence of particle-number conservation in the phonon system.
As a consequence, techniques to decompose tensors into irreducible representations with respect to global $U(1)$-symmetries \cite{1742-5468-2007-10-P10014,PhysRevA.82.050301,PhysRevB.83.115125} cannot be applied to the phononic degrees of freedom in order to reduce the numerical complexity.
Combined with the large dimension of the local Hilbert spaces, this paradigmatic model is particularly challenging for standard \gls{MPS} methods and specialized algorithms have to be utilized, which are discussed in the next section.
\section{\label{sec:methods}Methods}
The applicability of the \gls{MPS} ansatz is rooted in the fact that in one dimension, an efficient parameterization of so\hyp called area\hyp law states is possible by means of mixed-canonical representations, exploiting the inherent gauge freedom of~\gls{MPS}~\cite{MPSreviewVerstraete,PhysRevLett.100.040501,Schollwoeck201196}.
In area\hyp law states, the entanglement entropy
\begin{align}
	S_{\rm N} = -\operatorname{Tr} \left\{ \hat \rho_{A/B} \log \left( \rho_{A/B} \right) \right\}
\end{align}
scales only with the surface area of the partitioned subsystems~\cite{Vidal2003}.
For instance, for gapped \gls{1D} systems, $S_{\rm N}$ is bounded by a constant.
Here, $\rho_{A/B}$ is the reduced density matrix of a bipartition of the many-particle Hilbert space $\mathcal H = \mathcal H_A \otimes \mathcal H_B$.
There has been great effort to characterize the entanglement properties of generic models and it has been shown that for a wide range of gapped systems with local couplings, the ground states obey the area law and thus can be represented efficiently using \gls{MPS} \cite{Peschel2004,calabrese04,Eisert_RMP_arealaw}.
For pure fermion or small-spin systems, the dominating numerical costs are caused by contracting the site tensors along the auxiliary index between two lattice sites.
However, in typical algorithms such as a ground\hyp state search, the local dimension $d$ enters the numerical complexity as $d^2$ or $d^3$ in case of a single- or two-site algorithm~\cite{Schollwoeck201196,Hubig2015}.
Therefore, a generic representation allowing to numerically treat large $d$ values is essential, particularly in the intermediate coupling regime.
A loophole is to exploit global $U(1)$ symmetries, decomposing the local degrees of freedom into \gls{1D} representations \cite{1742-5468-2007-10-P10014,PhysRevB.83.115125}.
However, in the Holstein model, the number of phonons is not conserved, which is one major reason for the numerically challenging situation.
\subsection{\label{sec:methods:mps}Matrix-Product States}
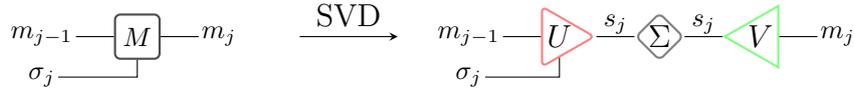
\begin{figure}[!t]
	\centering
	\tikzsetnextfilename{SVD_site_tensor}
	\begin{tikzpicture}
		\begin{scope}[node distance = 0.2 and 0.5]
			\node[ld]			(site0c)								{$m_{j-1}$};
			\node[ld]			(sigma0)		[below=of site0c]		{$\sigma_j$};
			\node[site]			(site1c)		[right=of site0c]		{$M$};
			\node[ld]			(site2c)		[right=of site1c]		{$m_{j}$};	
			\node[ghost]		(site3c)		[right=of site2c]		{};
			
			\draw[] (site0c) -- (site1c);
			\draw[] (site2c) -- (site1c);
			\draw[] (site1c) |- (sigma0);

			\node[ghost]		(interce)		[right=of site3c] 		{};
			\node[ghost]		(interce2)		[right=of interce]		{};
			\node[ghost]		(interce3)		[right=of interce2] 	{};
			\node[ld]			(site0e)		at (interce3|-site0c)	{$m_{j-1}$};
			\node[ld]			(sigma0e)		[below=of site0e]		{$\sigma_j$};
			\node[siteA]		(site1e)		[right=of site0e]		{$U$};
			\node[intersite]	(site2e)		[right=of site1e]		{$\Sigma$};
			\node[siteB]		(site3e)		[right=of site2e]		{$V$};
			\node[ld]			(site4e)		[right=of site3e]		{$m_{j}$};
			
			\draw[]		(site0e) -- (site1e);
			\draw[]		(site1e) -- (site2e)	node [ld,midway,above]	{$s_j$};
			\draw[] 	(site2e) -- (site3e)	node [ld,midway,above]	{$s_j$};
			\draw[] 	(site3e) -- (site4e);
			\draw[] 	(site1e) |- (sigma0e);
			\draw[->]	(site3c) -- (interce2)	node [midway,above,sloped] {SVD};
		\end{scope}
	\end{tikzpicture}
	\caption
	{
		\label{fig:SVD_site_tensor} 
		\Gls{SVD} of a site tensor $M$ into a right orthonormal tensor $U$ (red triangle), a diagonal matrix $\Sigma$ (grey diamond), and a left orthonormal tensor $V$ (green triangle).
		This operation is used to bring a site tensor $M$ into left canonical form via setting $A^{\sigma_j}_{m_{j-1}, m_j}=U_{(\sigma, m_{j-1}),s_j}$ and contracting $\Sigma$ and $V$ with the following site tensor.
	}
\end{figure}

\Glspl{MPS} supply both a flexible and numerically efficient ansatz class for quantum many-body states.
The coefficients $\psi_{\sigma^\noprime_1,\ldots,\sigma^\noprime_L} \in \mathbb C$ of a wave function representing a state $\ket{\psi}\in\mathcal H$ in a Hilbert space $\mathcal H$ composed of the tensor product of $L\in\mathbb N$ $d$-dimensional local degrees of freedom are expanded in terms of $d \cdot L$ matrices $M^{\sigma^\noprime_j} \in \mathbb C^{m_{j-1}\times m_j}$~\cite{Schollwoeck201196}
\begin{align}
	\ket{\psi}
	=
	\sum_{\sigma^\noprime_1,\ldots,\sigma^\noprime_L} \psi_{\sigma^\noprime_1,\ldots,\sigma^\noprime_L} \ket{\sigma^\noprime_1, \ldots, \sigma^\noprime_L} 
	\equiv
	\sum_{\sigma^\noprime_1,\ldots,\sigma^\noprime_L} M^{\sigma^\noprime_1} \cdots M^{\sigma^\noprime_L} \ket{\sigma^\noprime_1, \ldots, \sigma^\noprime_L} \; .
\end{align}
Encoding the information of the wave function locally in rank $3$ site tensors $M_j$ (i.e., the set of matrices $\left\{M^{\sigma^\noprime_j}\right\}$ per site) provides direct access to the Schmidt coefficients if the site tensors are expressed in the canonical gauge~\cite{Schollwoeck201196}.
Site tensors $A_j$ are called left canonical if they fulfill $\sum_{\sigma^{\noprime}_j, m^{\noprime}_{j-1}} \left(A^{\sigma^{\noprime}_j}_{m^{\noprime}_{j-1}, m^\prime_j}\right)^\dagger A^{\sigma^{\noprime}_j}_{m^{\noprime}_{j-1}, m^{\noprime}_j} = 1^{\noprime}_{m^{\prime}_{j}, m^{\noprime}_{j}}$ and site tensors $B_j$ are called right canonical if they fulfill $\sum_{\sigma^{\noprime}_j, m^{\noprime}_{j}} B^{\sigma^{\noprime}_j}_{m^{\noprime}_{j-1}, m^{\noprime}_j} \left(B^{\sigma^{\noprime}_j}_{m^{\prime}_{j-1}, m^{\noprime}_j}\right)^\dagger = 1^{\noprime}_{m^{\noprime}_{j-1}, m^{\prime}_{j-1}}$.
The \gls{SVD} is one way of obtaining these canonical tensors (see \cref{fig:SVD_site_tensor}).
Expanding the site tensors of a canonically gauged state by means of a \gls{SVD}
\begin{align}
	M^{\sigma^\noprime_j} = M_{(\sigma^\noprime_j, m_{j-1}), m_j} = \sum_{s_j} U_{(\sigma^\noprime_j, m_{j-1}),s_j} \Sigma_{s_j} V_{s_j, m_j}
\end{align}
allows for an optimal approximation with respect to the reduced density matrix $\hat \rho = \operatorname{Tr}_{k\leq j} \ket{\psi}\bra{\psi}$ by truncating the series of Schmidt values $\Sigma_{s_j}$ below a certain threshold $\delta_0$, leading to a cutoff bond dimension $m_{\rm max}$.
The squared and summed discarded parts of the Schmidt spectrum are commonly referred to as discarded weight $\delta(m_{\rm max})$
\begin{equation}
	\delta = \sum_{s_j > m_{\rm max}} \Sigma^2_{s_j} \; .
\end{equation}
Rapidly decaying Schmidt values $\Sigma_{s_j}$ are thus a necessary condition for a compact approximation of a state by means of a \gls{MPS} and small truncation error controlled by $\delta$.
Further numerical benefits can be achieved if $\ket{\psi}$ transforms under a global symmetry group $G$.
In this situation, the Wigner\hyp Eckhardt theorem ensures a decomposition of the site tensors in terms of irreducible representations $\Gamma(g)$ of the elements in the symmetry group $g \in G$ \cite{1742-5468-2007-10-P10014,PhysRevA.82.050301,PhysRevB.83.115125}.
The dimensions of the tensor indices decompose under the action of $G$: $m_j = \sum_{g_j} m_{g_j}$ so that by working on each block separately, the computational costs typically are reduced by an order of magnitude in the case of a global $U(1)$ symmetry.
\subsection{\label{sec:methods:ps-dmrg}Pseudosite Method}
\begin{figure}
	\centering
	\includegraphics[width=0.8\textwidth]{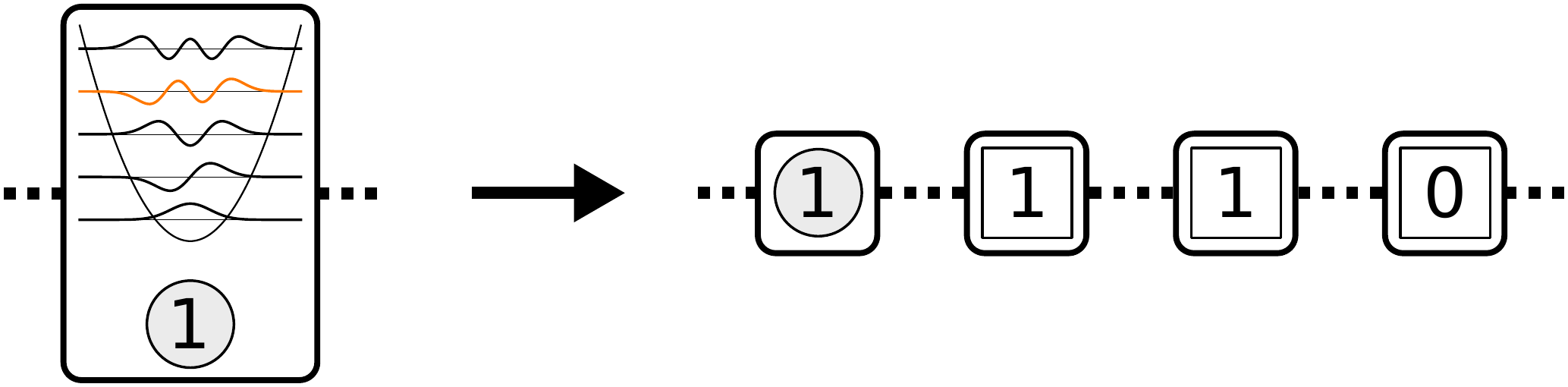}
	\caption
	{
		\label{fig:ps-dmrg:sketch}
		Sketch of local Hilbert space unfolded into pseudosites encoding $n_{\text{pseudo}}$ local degrees of freedom.
		Using a binary number representation, the local dimension is $2^{n_{\text{pseudo}}}-1$.
		The fermionic degree of freeom is indicated by a circle while the squares denote the introduced pseudo sites.
	}
\end{figure}
In this section, we summarize the \gls{PS-DMRG} method for large local dimensions $d$.
Details can be found in \cite{Jeckelmann1998,Jeckelmann2007}.
The pseudosite approach is based on the original two-site \gls{DMRG} method \cite{PhysRevLett.69.2863,PhysRevB.48.10345}.
Without special techniques for large sites, the complexity of the two-site \gls{DMRG} method scales as $S L d^3 m_{\text{max}}^3$, where $L$ is the number of sites in the lattice and $m_{\text{max}}$ is the maximum bond dimension.
In the spinless Holstein model, $d=2(n_{\text{ph}}+1)$, where $n_{\text{ph}}$ is the maximum number of phonons per site.
$S$ represents the number of iterations required to achieve convergence and combines both the iterative diagonalization of the superblock Hamiltonian at each \gls{DMRG} step and the repeated \gls{DMRG} sweeps through the lattice.
The key idea of the \gls{PS-DMRG} method is to substitute $n_{\text{pseudo}} = \log_2(n_{\text{ph}}+1)$ pseudosites of dimension 2 for the phonon site of dimension $n_{\text{ph}}+1$.
As the computational cost increases linearly with the number of sites but with the cube of the site dimension, \gls{PS-DMRG} can handle this representation more efficiently.
The exact mapping between the truncated phonon Hilbert space and the Hilbert space spanned by the $n_{\text{pseudo}}$ pseudosites can be implemented using the binary number representation of the boson number $n \ (0 \leq n \leq n_{\text{ph}})$ on a site.
An occupied pseudosite corresponds to the binary digit 1 while an empty pseudosite corresponds to the binary digit 0 (cf., \cref{fig:ps-dmrg:sketch}).
Boson operators have to be represented in terms of pseudosite operators to perform \gls{PS-DMRG} calculations.
\subsubsection{\label{sec:methods:ps-dmrg:n-pseudo-sites}Number of Pseudosites}
\begin{figure}
	\centering
	\tikzsetnextfilename{ps_dmrg-n_pseudosites-t_1p0_w_1p0_g_0p5}
	\begin{tikzpicture}
		\begin{groupplot}
		[
			group style = 
			{
				group size 		=	2 by 1,
				x descriptions at	=	edge bottom,
				y descriptions at	=	edge left,
				horizontal sep		=	2em,
			},
			height	= 0.25\textheight,
			width 	= 0.475\textwidth,
			ymode	= log,
			ylabel	= {$\Delta E$},
			ticklabel style	= {font=\scriptsize},
			ylabel style	= {font=\scriptsize},
			xlabel style	= {font=\scriptsize},
		]
			\nextgroupplot
			[
				axis on top,
				xlabel style	= {yshift=0.3em},
				xlabel		= {$m_{\mathrm{max}}$},
				extra description/.code={\node[anchor=north,font=\scriptsize] at (0.5,0.99) {$L=51$, $\nicefrac{\omega_0}{\hopping}=1.0$, $\nicefrac{\gamma}{\hopping}=0.5$};},
				title		= {\subfloat[\label{fig:ps-dmrg:n-pseudosites:t-1p0:w-1p0:g-0p5}]{}},
				title style	= {yshift=-0.5em, xshift=-0.25\textwidth},	
				ticklabel style	= {font=\scriptsize},
				xlabel style	= {font=\scriptsize,yshift=0pt},
				ylabel style	= {font=\scriptsize,yshift=0pt},
				ymin		= 1e-12,
				ymax		= 1e-06,
				legend style	= {font=\scriptsize,anchor=south west, at={(0.03,0.03)}},
			]
				\addplot
				[
					color			= colorB,
					mark			= x,
					unbounded coords	= jump,
					thick,
				]
				table
				[
					x expr	= \thisrowno{0},
					y expr	= \thisrowno{1},
				]
				{data/holstein/L_51_N_25_t_1p0_w_1p0_g_0p5/ediff.pseudo_site.nph.7};
				\addlegendentry{$n_{\text{ph}}=\phantom{1}7$}
				
				\addplot
				[
					color			= colorC,
					mark			= triangle,
					unbounded coords	= jump,
					thick,
				]
				table
				[
					x expr	= \thisrowno{0},
					y expr	= \thisrowno{1},
				]
				{data/holstein/L_51_N_25_t_1p0_w_1p0_g_0p5/ediff.pseudo_site.nph.15};
				\addlegendentry{$n_{\text{ph}}=15$}
			\nextgroupplot
			[
				axis on top,
				xlabel style	= {yshift=0.3em},
				xlabel		= {$m_{\mathrm{max}}$},
				extra description/.code={\node[anchor=north,font=\scriptsize] at (0.5,0.99) {$L=51$, $\nicefrac{\omega_0}{\hopping}=1.0$, $\nicefrac{\gamma}{\hopping}=1.5$};},
				title		= {\subfloat[\label{fig:ps-dmrg:n-pseudosites:t-1p0:w-1p0:g-1p5}]{}},
				title style	= {yshift=-0.5em, xshift=-0.225\textwidth},	
				ticklabel style	= {font=\scriptsize},
				xlabel style	= {font=\scriptsize,yshift=0pt},
				ylabel style	= {font=\scriptsize,yshift=0pt},
				ymin		= 1e-12,
				ymax		= 1e-06,
				legend style	= {font=\scriptsize,anchor=south west, at={(0.03,0.03)}},
			]
				\addplot
				[
					color			= colorC,
					mark			= triangle,
					unbounded coords	= jump,
					thick,
				]
				table
				[
					x expr	= \thisrowno{0},
					y expr	= \thisrowno{1},
				]
				{data/holstein/L_51_N_25_t_1p0_w_1p0_g_1p5/ediff.pseudo_site.nph.15};
				\addlegendentry{$n_{\text{ph}}=15$}
				
				\addplot
				[
					color			= colorD,
					mark			= square,
					unbounded coords	= jump,
					thick,
				]
				table
				[
					x expr	= \thisrowno{0},
					y expr	= \thisrowno{1},
				]
				{data/holstein/L_51_N_25_t_1p0_w_1p0_g_1p5/ediff.pseudo_site.nph.31};
				\addlegendentry{$n_{\text{ph}}=31$}
		\end{groupplot}
	\end{tikzpicture}	
	\caption
	{
		\label{fig:ps-dmrg:n-pseudosites}
		Relative deviation of the ground\hyp state energy obtained with PS-DMRG from reference values as a function of the maximum bond dimension $m_{\rm max}$ and varying phonon cutoff $n_{\rm ph}$.
		The reference values are obtained by taking the best approximations to the ground state from high-precision calculations using the \gls{DMRG3S+LBO} and \gls{PP-2DMRG} methods, as discussed in \cref{sec:comparison:gs-scaling}.
		%
		%
		%
	}
\end{figure}
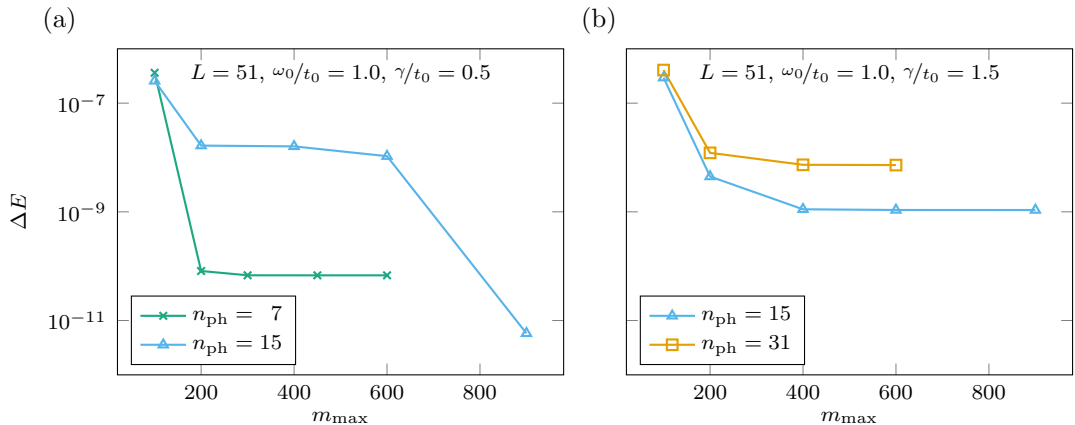
For the Holstein model, the pseudosite Hamiltonian contains $\sim n_{\text{ph}}$ many-particle terms acting over a range of up to $n_{\text{pseudo}}$ pseudosites.
Thus, the computational cost of the \gls{PS-DMRG} method scales as $S L (n_{\text{ph}}+1) \log_2(n_{\text{ph}}+1) m^3_{\text{max}}$, which is significantly more efficient than the usual \gls{DMRG} method applied to large phonon sites.
Due to the longer interaction range in the pseudosite representation, however, the \gls{PS-DMRG} method requires a larger bond dimension $m_{\text{max}}$ and a larger number of iterations $S$ to reach a given accuracy.
The difference becomes more pronounced for larger site dimensions.
Thus, the precision increases first with increasing the number of pseudosites, but may deteriorate for too large $n_{\text{pseudo}}$ if the number of sweeps and the bond dimension are kept constant, as observed in \cref{fig:ps-dmrg:n-pseudosites}\subref{fig:ps-dmrg:n-pseudosites:t-1p0:w-1p0:g-1p5}.
Nonetheless, the \gls{PS-DMRG} method allows much larger numbers of bosons than the usual \gls{DMRG} method \cite{Pai1996,Caron1996,Caron1997}.
It was successfully used to study the metal-insulator transition in various Holstein-type models \cite{Jeckelmann1999,Tezuka2005,Tezuka2007,Fehske2008,Ejima2009}.
\subsubsection{\label{sec:methods:ps-dmrg:benchmark-setup}Benchmark Setup}
For the high-precision calculations presented in \cref{sec:comparison:gs-scaling}, we use $n_{\text{pseudo}}$ from $3$ to $6$ ($7 \leq n_{\text{ph}} \leq 63$).
The \gls{PS-DMRG} calculation always starts with a small bond dimension $m$, which is increased progressively up to $m_{\text{max}}$.
Three to six \gls{DMRG} sweeps (back and forth) through the full lattice are carried out for each bond dimension for a total of up to 30 sweeps.
The bond dimensions are so large that the discarded weight $\delta$ vanishes within double precision ($10^{-16}$).
Thus $\delta$ is not a helpful quantity to decide whether a calculation has converged and extrapolations of observables to vanishing discarded weights are not used~\cite{Jeckelmann2007}.
The accuracy of the calculated energies is limited by the slow convergence with the number of iterations rather than by the \gls{DMRG} truncation error or the phonon Hilbert-space truncation.
For the study of the finite-size scaling of the order parameter in \cref{sec:comparison:finite-site-extrapolation}, we use several bond dimensions $m$ up to $m_{\text{max}}=200$ or $400$ with $3$ to $6$ sweeps for each value of $m$.
From the scaling with varying bond dimension $m$, we estimate the relative errors of the finite-system energies to be smaller than $10^{-6}$, while the absolute errors of the finite-system order parameters are
smaller than $10^{-5}$.
\subsection{\label{sec:methods:lbo-dmrg}DMRG with local basis optimization}
Here, we recapitulate the \gls{DMRG3S+LBO} introduced in \cite{Stolpp2020} (see also \cite{phd:dorfner}).
The method is based on the \gls{LBO} introduced by Zhang {\it et al.} \cite{Zhang1998}.
At the core, the \gls{DMRG3S+LBO} is a combination of the \gls{VMPS} method originally formulated for spin-boson models and introduced by Guo {\it et al.} \cite{Guo2012} (see also \cite{Bruognolo2014,Blunden-Codd2017,Bruognolo2017})
with the single-site \gls{DMRG} method with subspace expansion introduced by Hubig {\it et al.} \cite{Hubig2015}.
\subsubsection{\label{sec:methods:lbo-dmrg:lbo}LBO}
The \gls{LBO} tackles the problem of large local Hilbert spaces by dynamically finding an optimized local basis for the problem at hand.
The optimized basis at site $j$ can by found by computing the \gls{1RDM}:
\begin{align}
	\hat\rho_j = \operatorname{Tr}_{k\neq j} \ket{\psi}\bra{\psi}\;,
\end{align}
where the trace is over all degrees of freedom $\sigma_k$ that are not on site $j$.
Upon diagonalization of $\hat\rho_j$, a local basis transformation matrix $U_l$ as well as the spectrum of $\hat\rho_j$ is found.
Along the lines of the \gls{DMRG} strategy, the local basis can be truncated according to the spectrum of $\hat\rho_j$ and the state can be transformed into the new local basis.
If only a few of the eigenvalues of $\hat\rho_j$ have considerable magnitude, the truncation will retain most of the information in the state.
In case of the Holstein model, this strategy can be motivated by the insight that the ground state of the Holstein model in the atomic limit ($\hopping = 0$) needs just a single state at every site \cite{Hirsch1983}.
In the ground state of the Holstein-polaron model, an exponential decay of the spectrum was found numerically for several sets of parameters \cite{Zhang1998,PhysRevB.91.104302}.
A similar behavior in the spectrum of $\hat\rho_j$ was also found in the Bose-Bose resonance model \cite{Dorfner2016}.
A \gls{MPS} with \gls{LBO} takes the form:
\begin{align}
	\ket{\psi} \approx \underset{\tilde\sigma^\noprime_1, \ldots, \tilde\sigma^\noprime_L}{\sum_{\sigma^\noprime_1,\ldots,\sigma^\noprime_L}}
	 M^{\tilde\sigma^\noprime_1} R^{\tilde\sigma^\noprime_1, \sigma^\noprime_1} \cdots M^{\tilde\sigma^\noprime_L} R^{\tilde\sigma^\noprime_L, \sigma^\noprime_L} \ket{\sigma^\noprime_1, \ldots, \sigma^\noprime_L}\;,
 \end{align}
 where truncated local basis transformations $R_j$ (with entries $R^{\tilde\sigma^\noprime_j \sigma^\noprime_j}$) are attached to every local tensor $M_j$ (i.e., the set of matrices $\left\{M^{\tilde\sigma^\noprime_j}\right\}$) (cf., \cref{fig:lbo-dmrg:sketch} left).
 The matrices $R_j$ have dimension $d_{\rm o}\times d$, where $d_{\rm o}$ is the number of optimal modes, i.e., the number of local basis states that are kept after the truncation.
 %
 
\subsubsection{\label{sec:methods:lbo-dmrg:dmrg3s-lbo}DMRG3S+LBO}

\begin{figure}
	\centering
	\tikzsetnextfilename{dmrg_lbo_sketch}
	\begin{tikzpicture}
		\clip(-2.45,3.6) rectangle (13.2,-5);
		\begin{scope}
		[
			node distance = 1.5em and 1.75em,
		]
			\node (init) 
			{
				\tikzset{external/export next=false}
				\begin{tikzpicture}
					\node[ghost] (left1) {};
					\node[site] (site1) [right = of left1] {$M_j$};
					\node[ghost] (right1) [right = of site1] {};
					\node[siteD, inner sep=-2pt] (rot1) [below = of site1] {$R_j$};
					\node[ghost] (below1) [below = of rot1] {};
					
					\draw (left1) to node[above,ld,xshift=-1em] {$m_{j-1}$} (site1);
					\draw (site1) to node[above,ld,xshift=1em] {$m_{j}$} (right1);
					\draw (site1) to node[right,ld] {$\tilde{\sigma}_j$} (rot1);
					\draw (rot1) to node[right,ld] {$\sigma_j$} (below1);
				\end{tikzpicture}
			};
			
			\node[right = of init] (second) 
			{
				\begin{tikzpicture}
					\node[ghost,minimum height = 2em] (left2) {};
					\node[siteC] (site2) [right = of left2] {$U$};
					\node[ghost,minimum height = 2em] (right2) [right = of site2] {};
					\node[intersite] (Sigma2) [below = of site2] {$\Sigma$};
					\node[siteD] (V2) [below = of Sigma2] {$V$};
					\node[siteD, inner sep=-2pt] (rot2) [below = of V2] {$R_j$};
					\node[ghost] (below2) [below = of rot2] {};
					
					\node
					[
						fit = (Sigma2) (V2) (rot2), 
						draw = blue!50, 
						fill = blue!10, 
						fill opacity = 0.2,
						rounded corners=2pt,
						thick,
					] (comb2) {};
					
					\node[anchor = south west, inner sep = 0] at (comb2.north east) {$Q$};
					
					\draw (left2.north east) -| ($(site2.north)+(-0.1,0)$);
					\draw ($(site2.north)+(0.1,0)$) |- (right2.north west);
					\draw (site2) to (Sigma2);
					\draw (Sigma2) to (V2);
					\draw (V2) to (rot2);
					\draw (rot2) to (below2);
				\end{tikzpicture}
			};			
		
			\draw[->] (init) to node[below,ld,inner sep=2pt] {SVD of $M_j$} node[above,ld,inner sep=2pt] {$(1)$} (second);
			
			\node[right = of second] (third)
			{
				\begin{tikzpicture}
					\node[ghost,minimum height = 2em] (left3) {};
					\node[siteC] (site3) [right = of left3] {$U$};
					\node[ghost,minimum height = 2em] (right3) [right = of site3] {};
					\node[op] (rot3) [below = of site3] {$Q^\prime$};
					\node[ghost] (below3) [below = of rot3] {};
					
					\draw (left3.north east) -| ($(site3.north)+(-0.1,0)$);
					\draw ($(site3.north)+(0.1,0)$) |- (right3.north west);
					\draw (site3) to (rot3);
					\draw (rot3) to (below3);
				\end{tikzpicture}
			};
		
			\draw[->] (second) to node[below,ld,inner sep=2pt] {optimize $Q$} node[above,ld,inner sep=2pt] {$(2)$} (third);
			
			\node[right = of third] (fourth) 
			{
				\begin{tikzpicture}
					\node[ghost,minimum height = 2em] (left4) {};
					\node[siteC] (site4) [right = of left4] {$U$};
					\node[ghost,minimum height = 2em] (right4) [right = of site4] {};
					\node[siteC] (U4) [below = of site4] {$U^\prime$};
					\node[intersite] (Sigma4) [below = of U4] {$\Sigma^\prime$};
					\node[siteD, inner sep=-1pt] (rot4) [below = of Sigma4] {$R^\prime_j$};
					\node[ghost] (below4) [below = of rot4] {};
					
					\node
					[
						fit = (site4) (U4) (Sigma4), 
						draw = black!70, 
						fill = black!10, 
						fill opacity = 0.2,
						rounded corners=2pt,
						inner sep = 8pt,
						thick,
					] (comb4) {};
					\node[anchor = south west, inner sep = 0] at (comb4.north east) {$M^\prime_j$};
					
					\draw (left4.north east) -| ($(site4.north)+(-0.1,0)$);
					\draw ($(site4.north)+(0.1,0)$) |- (right4.north west);
					\draw (site4) to (U4);
					\draw (U4) to (Sigma4);
					\draw (Sigma4) to (rot4);
					\draw (rot4) to (below4);
				\end{tikzpicture}
			};		
		
			\draw[->] (third) to node[below,ld,inner sep=2pt] {SVD of $Q^\prime$} node[above,ld,inner sep=2pt] {$(3)$} (fourth);
			
			\draw[->, in = 230, out = -50, looseness=1.4] (fourth.east) to node[below,ld,inner sep=2pt] {optimize $M^\prime_j$} node[above,ld,inner sep=2pt] {$(4)$} (init.west);
		\end{scope}
	\end{tikzpicture}
	
	\caption
	{
		\label{fig:lbo-dmrg:sketch}
		Sketch of the different steps in the \gls{DMRG3S+LBO} method to optimize the local basis matrix $R_j$ and the local site tensor $M_j$ (see also \cite{Guo2012,Stolpp2020}).
	}
\end{figure}
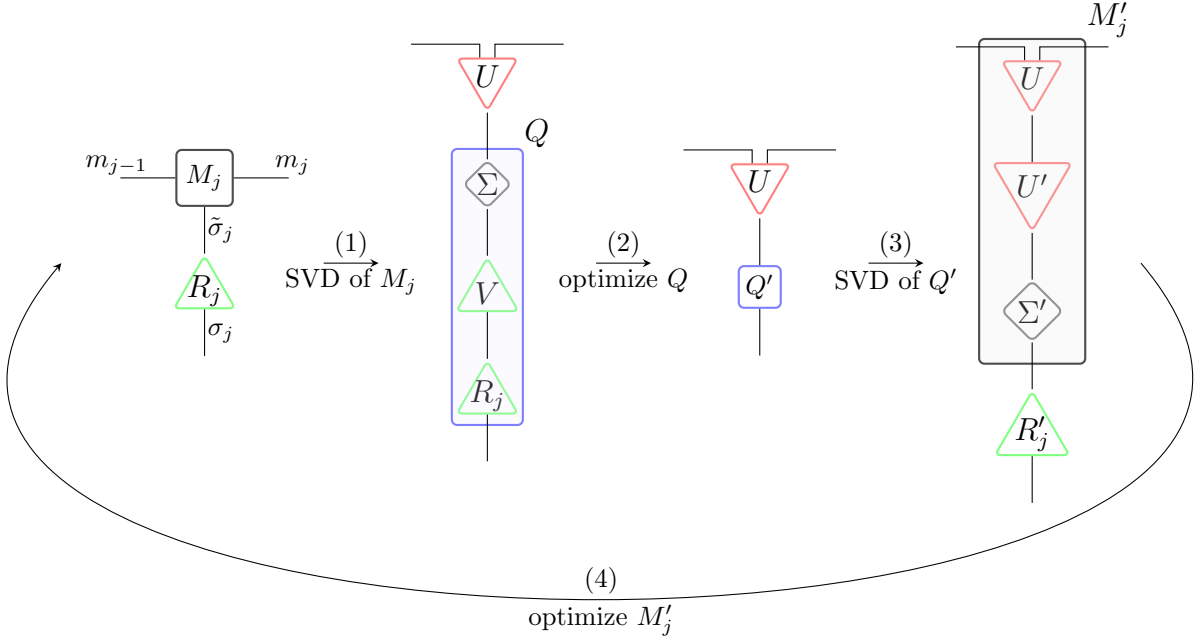
The \gls{DMRG3S+LBO} method consists of two stages, an optimization of the local basis matrix $R_j$ and an optimization of the local site tensor $M_j$.
Both of these are performed by a \gls{DMRG}-like optimization step.
The tensor manipulations necessary are sketched in \cref{fig:lbo-dmrg:sketch}.
Consider a \gls{MPS} with \gls{LBO} in mixed\hyp canonical form where all site tensors left of site $j$ are left canonical and all tensors to the right are right canonical.
In the first step, a \gls{SVD} of the site tensor is performed such that [cf., \cref{fig:lbo-dmrg:sketch}(1)]:
\begin{align}
	M_{(m_{j-1}, m_j), \tilde\sigma^\noprime_j} = \sum_{s_j} U_{(m_{j-1}, m_j),s_j} \Sigma_{s_j} V_{s_j, \tilde\sigma^\noprime_j}\;.
\end{align}
Subsequently, $\Sigma$ and $V$ are contracted with $R_j$ and a \gls{DMRG} optimization of the resulting matrix is performed [cf., \cref{fig:lbo-dmrg:sketch}(2)].
The original form of the \gls{MPS} is then restored by a \gls{SVD} of the optimized matrix [cf., \cref{fig:lbo-dmrg:sketch}(3)].
In this step, a truncation of the optimal basis may be done.
However, note that the optimized basis dimension cannot grow throughout the optimization step.
A truncation is therefore only necessary if the left dimension of the matrix $R_j$ is larger than the desired $d_{\rm o}$ before the optimization.
In practice, it is more common that the dimension has to be enlarged artificially (e.g., in the first sweep if the initial state has a simple structure) so that future optimization steps can explore the full $d_{\rm o}$ local states.
Note further that we utilize particle-number conservation of the fermions in the Holstein model so that the local-basis-transformation matrices consist of two symmetry blocks.
The maximal number of optimal modes $d_{\rm o}$ refers to the maximal number per block.
In the truncation step, the singular values from both blocks are combined in one list and the blocks are filled according to the size of the singular values until one of the blocks reaches the maximal size $d_{\rm o}$.
The smaller block is then filled with zeros so that both blocks have the same size.
The second step is a \gls{DMRG} optimization of the local site tensor $M_j$ [cf., \cref{fig:lbo-dmrg:sketch}(4)].
These two steps of optimizing the local basis and then the site tensor can be done in a loop until no further improvements are detected \cite{Guo2012}.
However, in the implementation of the \gls{DMRG3S+LBO} algorithm used for this work, every optimization is done for a fixed number of two times.

After the optimization of both, the local basis and the site tensor is finished, the center of the mixed-canonical \gls{MPS} is moved to the next site.
To prevent the algorithm from getting stuck in local minima of the energy landscape, an enrichment step on the bonds has to be implemented.
For this purpose, the subspace expansion described in \cite{Hubig2015} is used.
\subsubsection{\label{sec:methods:convergence}Convergence of the DMRG3S+LBO method}
In a single-site \gls{DMRG} method, the bond dimension does not grow on its own during the algorithm.
Therefore, it is crucial to implement a procedure that enhances the bond dimension artificially.
This procedure has the second purpose of preventing the algorithm from getting stuck in local energy minima.
In the \gls{DMRG3S+LBO}, we choose the subspace expansion developed by Hubig {\it et al.} in \cite{Hubig2015} (see also \cite{White2005,1742-5468-2007-10-P10014} for earlier work on single-site \gls{DMRG} algorithms).
The crucial steps of the method are recapitulated here.
In a left-to-right sweep, after the \gls{DMRG} optimization of the local tensor $M_j$, the tensor is enlarged by $P_j$ with entries:
\begin{align}
	P^{\tilde\sigma^\prime_j}_{k_{j-1},(l_j,m_j)} = \alpha \sum_{l_{j-1}, m_{j-1}, \tilde\sigma^\noprime_{j}} L_{k_{j-1}, l_{j-1}, m_{j-1}} M^{\tilde\sigma^\noprime_{j}}_{m_{j-1}, m_j} H^{\tilde\sigma^\noprime_j, \tilde\sigma^\prime_j}_{l_{j-1}, l_j}\;,
\end{align}
where $H_j$ (with entries $H^{\tilde\sigma^\noprime_j, \tilde\sigma^\prime_j}_{l_{j-1}, l_j}$) is the local \gls{MPO} site tensor of the Hamiltonian transformed into the optimal basis:
\begin{align}
	H^{\tilde\sigma^\noprime_j, \tilde\sigma^\prime_j}_{l_{j-1}, l_j} = \sum_{\sigma^\noprime_j, \sigma^\prime_j} R^{\tilde\sigma^\noprime_j, \sigma^\noprime_j} H^{\sigma^\noprime_j, \sigma^\prime_j}_{l_{j-1}, l_j} (R^\dagger)^{\sigma^\prime_j, \tilde\sigma^\prime_j}
\end{align}
and the tensor $L_{j-1}$ is recursively defined as having entries:
\begin{align}
	L_{k_{j-1}, l_{j-1}, m_{j-1}} = \sum_{k_{j-2}, l_{j-2}, m_{j-2}, \tilde\sigma^\noprime_{j-1}, \tilde\sigma^\prime_{j-1}} L_{k_{j-2}, l_{j-2}, m_{j-2}} A^{\tilde\sigma^\noprime_{l-1}}_{m_{j-2}, m_{j-1}} H^{\tilde\sigma^\noprime_{j-1}, \tilde\sigma^\prime_{j-1}}_{l_{j-2}, l_{j-1}} (A^\dagger)^{\tilde\sigma^\prime_{l-1}}_{k_{j-2}, k_{j-1}}\;,
\end{align}
where the $A_j$ are the left-canonical site tensors of the \gls{MPS} and $L_0 = 1$.
The parameter $\alpha$ is a mixing factor that has to be chosen appropriately.
The tensor on site $j$ is enlarged by $P_j$ and the \gls{MPS} tensor on the following site $j+1$ is enlarged by zeros in such a way that the tensors fit together:
\begin{align}
	\tilde M_j = [M_j P_j], \quad \tilde B_{j+1} = \left[ B_{j+1} \atop 0 \right] \;.
\end{align}
This enrichment step can be interpreted as an expansion of the local site tensor into a two-dimensional Krylov space, as $P_j$ provides the left-contraction of the current state multiplied by the system's Hamiltonian.
If the \gls{MPS} is already close to the global ground state, the site tensors are essentially eigenstates of the effective single-site Hamiltonian $H_j$, contracted with the proper left- and right-contractions $L_{j-1}$ and $R_{j+1}$, respectively.
In this case, the mixing factor should be chosen as $\alpha \equiv 0$.
The previous step of enlarging the tensor $M_j$ by $P_j$ does not change the state at all, since the expansion tensor $P_j$ is multiplied by zeros on the next site.
However, the bond dimension between site $j$ and $j+1$ is increased by $w \cdot \chi$, where $w$ is the bond dimension of the \gls{MPO} and $\chi$ is the bond dimension of the \gls{MPS}.
In most cases, it is therefore necessary to truncate the bond dimension to the maximal bond dimension $m_{\rm max}$ immediately after the enrichment step.
After the truncation, the change done to the state becomes manifest and in most cases, the subspace enlargement and subsequent truncation increases the energy.
To monitor the effect of the subspace expansion on the state, one compares the energy after the truncation on the last bond $E_{\rm last}$ with the energy after the \gls{DMRG} optimization on the current site $E_{\rm opt}$
and the energy after the truncation on the current bond $E_{\rm trunc}$.
This provides the energy differences $\Delta_{\rm opt} = E_{\rm last} - E_{\rm opt}$ and $\Delta_{\rm trunc} = E_{\rm trunc} - E_{\rm opt}$.
On the one hand, it is, of course, important to keep $\Delta_{\rm opt} > \Delta_{\rm trunc}$, as otherwise the energy actually grows and one does not approach the ground state.
On the other hand, one has to apply a strong enough mixing so that the global energy minimum can be found.
Comparing both energy differences provides a measure for the competing effects of the subspace expansion.
An optimal expansion strategy maximizes the energy gain when optimizing the current site tensor (measured by $\Delta_{\rm opt}$), while at the same time it minimizes the perturbation of the optimized site tensor when expanding the next bond (measured by $\Delta_{\rm trunc}$).
It also needs to account for the previously described situation, in which the site tensors are eigenstates of the effective Hamiltonian in which no expansion should occur, or, the mixing factor should be decreased, at least.
As suggested in \cite{Hubig2015}, we update the mixing factor with a multiplicative adaption factor $\eta$ such that $\alpha^\prime = \eta\alpha$.
This allows for a gradual change of the expansion rates that mimics the increase or decrease of correlations throughout the system.
The overall strategy is to increase the mixing factor if the ratio $\nicefrac{\Delta_{\rm trunc}}{\Delta_{\rm opt}}$ is very small (or even negative) but to decrease it if the ratio exceeds $0.3$. 
The exact procedure to choose the adaption factor $\eta$ is given in \cref{algo:choose_eta}.
\begin{algorithm}
	\caption
	{
		\label{algo:choose_eta}
		Procedure to choose the adaption factor $\eta$ for the update of the mixing factor $\alpha$ from the energy differences $\Delta_{\rm opt}$ and $\Delta_{\rm trunc}$.
	}
	\begin{algorithmic}
		\State $\delta_{\rm opt} = \frac{\Delta_{\rm opt}}{|E_{\rm opt}|}$
		\State $\delta_{\rm trunc} = \frac{\Delta_{\rm trunc}}{|E_{\rm opt}|}$
		\If{$\left(|\delta_{\rm opt}| < 10^{-14} \quad {\rm or} \quad |\delta_{\rm trunc}| < 10^{-14}\right)$}
			\State $\eta = 1$
		\ElsIf{$\delta_{\rm trunc} < 0$}
			\State $\eta = 2 (1+\frac{|\delta_{\rm trunc}|}{|\delta_{\rm opt}|})$
		\ElsIf{$\frac{|\delta_{\rm trunc}|}{|\delta_{\rm opt}|} < 0.05$}
			\State $\eta = 1.2 - \frac{|\delta_{\rm trunc}|}{|\delta_{\rm opt}|}$
		\ElsIf{$\frac{|\delta_{\rm trunc}|}{|\delta_{\rm opt}|} > 0.3$}
			\State $\eta = \frac{1}{1.01}$
		\Else
			\State $\eta = 1$
		\EndIf
	\end{algorithmic}
\end{algorithm}
Further, the adaption factor is restricted to $0.99 \leq \eta \leq 1.01$ and the mixing factor to $10^{-8} \leq \alpha \leq 100$.
If the mixing factor falls below $10^{-8}$ this is interpreted as a signal for the site tensor being an eigenstate of the effective single-site Hamiltonian and the subspace expansion is suspended entirely.
To achieve convergence of our \gls{DMRG3S+LBO} method, it turns out to be beneficial to postpone the adaptation of the mixing factor $\alpha$ in the first few sweeps to keep it at a relatively high value.
Furthermore, we noticed that the algorithm is still prone to getting stuck despite the use of the subspace expansion.
This behavior is especially common in the \gls{CDW} phase of the Holstein model when polarons become heavy and hard to move around in the lattice.
It is therefore beneficial to use initial states that are already close to the targeted ground state and possess key features of it.
We therefore pursue the strategy of first calculating the ground state at $\gamma = 0$ for a relatively small maximal bond dimension, in an initialization run.
In the following, we increase the coupling $\gamma$ from run to run while keeping the maximal bond dimension fixed, until the desired coupling strength is reached.
Then, the bond dimension is increased from run to run until the desired precision is reached.
\begin{figure}[!t]
	\centering
	\def\fileA{data/holstein/mps_with_lbo/L_51_N_25/t_1p0_w_1p0_g_2p0/convergence_example/holgs_if_DMRGOM2L51t1o1g2bdim20tr14cp31domax5sweeps500ndirty30.o}
	\def\fileB{data/holstein/mps_with_lbo/L_51_N_25/t_1p0_w_1p0_g_2p0/convergence_example/holgs_if_DMRGOM2L51t1o1g2bdim20tr14cp31domax5sweeps500ndirty1.o}
	\def\fileC{data/holstein/mps_with_lbo/L_51_N_25/t_1p0_w_1p0_g_2p0/convergence_example/holgs_if_DMRGOM2L51t1o1g2bdim20tr14cp31domax5sweeps501ndirty30.o}

	\tikzsetnextfilename{lbo_convergence_vs_sse}
	\begin{tikzpicture}
		\begin{groupplot}
		[
			group style = 
			{
				group size 		=	1 by 3,
				horizontal sep		=	1em,
				vertical sep		=	0em,
				x descriptions at	=	edge bottom,
				y descriptions at	=	edge left
			},
			width	= 0.95\textwidth-16.1pt,
			height	= 0.25\textheight,
			xlabel	= {sweeps},
			xmin	= 0,
			xmax	= 500,
			ymax	= 5,
			ticklabel style	= {font=\scriptsize},
			ylabel style	= {font=\scriptsize},
			xlabel style	= {font=\scriptsize},
		]
		
		\nextgroupplot
		[
			xmin	= 0,
			xmax	= 500,
			ylabel	= {$\alpha$},
			ymode = log,
			title		= {\subfloat[\label{fig:lbo_mixingfactor:t-1p0:w-1p0:g-2p0}]{}},
			title style	= {yshift=-0.195\textheight, xshift=-0.40\textwidth},
		]
			\addplot
			[
				color	= colorA,
				thick,
			] 
				table
				[
					x expr = \thisrowno{2},
					y expr = \thisrowno{9},
				]
				{\fileA.alphasmin};
			\addplot
			[
				color	= colorA,
				each nth point=15,
				only marks,
				mark = o,
				thick
			] 
				table
				[
					x expr = \thisrowno{2},
					y expr = \thisrowno{9},
				]
				{\fileA.alphasmin};

			\addplot
			[
				color	= colorB,
				thick,
			] 
				table
				[
					x expr = \thisrowno{2},
					y expr = \thisrowno{9},
				]
				{\fileB.alphasmin};
			\addplot
			[
				color	= colorB,
				each nth point=15,
				only marks,
				mark = triangle,
				thick
			] 
				table
				[
					x expr = \thisrowno{2},
					y expr = \thisrowno{9},
				]
				{\fileB.alphasmin};

			\addplot
			[
				color	= colorC,
				thick,
			] 
				table
				[
					x expr = \thisrowno{2},
					y expr = \thisrowno{9},
				]
				{\fileC.alphasmin};
			\addplot
			[
				color	= colorC,
				each nth point=15,
				only marks,
				mark = diamond,
				thick
			] 
				table
				[
					x expr = \thisrowno{2},
					y expr = \thisrowno{9},
				]
				{\fileC.alphasmin};

			\node[anchor=north east,inner sep = 0pt] (FirstInset) at (500,8) {};
		\nextgroupplot
		[
			ylabel	= {$\delta_{\text{self}} E$},
			ymode = log,
			title		= {\subfloat[\label{fig:lbo_mixingfactor_deltaSelfE:t-1p0:w-1p0:g-2p0}]{}},
			title style	= {yshift=-0.195\textheight, xshift=-0.40\textwidth},
		]
			\addplot
			[
				color	= colorA,
				thick,
			]
				table
				[
					x expr	= \thisrowno{2},
					y expr	= \thisrowno{0},
				]
					{\fileA.alphasmin};

			\addplot
			[
				color	= colorA,
				each nth point=15,
				only marks,
				mark = o,
				thick
			] 
				table
				[
					x expr = \thisrowno{2},
					y expr = \thisrowno{0},
				]
					{\fileA.alphasmin};

			\addplot
			[
				color	= colorB,
				thick,
			]
				table
				[
					x expr	= \thisrowno{2},
					y expr	= \thisrowno{0},
				]
					{\fileB.alphasmin};

			\addplot
			[
				color	= colorB,
				each nth point=15,
				only marks,
				mark = triangle,
				thick
			]
				table
				[
					x expr	= \thisrowno{2},
					y expr	= \thisrowno{0},
				]
					{\fileB.alphasmin};

			\addplot
			[
				color	= colorC,
				thick,
			]
				table
				[
					x expr	= \thisrowno{2},
					y expr	= \thisrowno{0},
				]
					{\fileC.alphasmin};
			\addplot
			[
				color	= colorC,
				each nth point=15,
				only marks,
				mark = diamond,
				thick
			]
				table
				[
					x expr	= \thisrowno{2},
					y expr	= \thisrowno{0},
				]
					{\fileC.alphasmin};
			\node[anchor=north east,inner sep = 0pt] (SecondInset) at (500,8) {};

		\nextgroupplot
		[
			ylabel	= {$\delta E$},
			ymode = log,
			title		= {\subfloat[\label{fig:lbo_mixingfactor_deltaE:t-1p0:w-1p0:g-2p0}]{}},
			title style	= {yshift=-0.195\textheight, xshift=-0.40\textwidth},
			legend style	= {font=\scriptsize,at={(0.97,0.97)}, anchor=north east},
		]
			\addplot
			[
				color	= colorA,
				thick,
				forget plot,
			]
				table
				[
					x expr	= \thisrowno{2},
					y expr	= \thisrowno{1},
				]
					{\fileA.alphasmin};

			\addplot
			[
				color	= colorA,
				each nth point=15,
				only marks,
				mark = o,
				thick
			] 
				table
				[
					x expr = \thisrowno{2},
					y expr = \thisrowno{1},
				]
					{\fileA.alphasmin};
			\addlegendentry{$\mathcal{R}_1: \ket{\psi_{\text{in}, 1}}, s_{\alpha=1}=30$};

			\addplot
			[
				color	= colorB,
				thick,
				forget plot,
			]
				table
				[
					x expr	= \thisrowno{2},
					y expr	= \thisrowno{1},
				]
					{\fileB.alphasmin};

			\addplot
			[
				color	= colorB,
				each nth point=15,
				only marks,
				mark = triangle,
				thick
			]
				table
				[
					x expr	= \thisrowno{2},
					y expr	= \thisrowno{1},
				]
					{\fileB.alphasmin};
			\addlegendentry{$\mathcal{R}_2: \ket{\psi_{\text{in}, 1}}, s_{\alpha=1}=\phantom{0}1$};

			\addplot
			[
				color	= colorC,
				thick,
				forget plot,
			]
				table
				[
					x expr	= \thisrowno{2},
					y expr	= \thisrowno{1},
				]
					{\fileC.alphasmin};
			\addplot
			[
				color	= colorC,
				each nth point=15,
				only marks,
				mark = diamond,
				thick
			]
				table
				[
					x expr	= \thisrowno{2},
					y expr	= \thisrowno{1},
				]
					{\fileC.alphasmin};
			\addlegendentry{$\mathcal{R}_3: \ket{\psi_{\text{in}, 2}}, s_{\alpha=1}=30$};
		\end{groupplot}
		\begin{axis}
		[
			at = (FirstInset),
			anchor = north east,
			width = 0.5\textwidth,
			height = 0.1625\textheight,
			xshift = -1em,
			yshift = -1em,
			xlabel = {site $j$},
			ylabel = {$\braket{\hat n_j}_{\text{init}}$},
			xmin = 0,
			xmax = 50,
			ticklabel style	= {font=\scriptsize},
			ylabel style	= {font=\scriptsize},
			xlabel style	= {font=\scriptsize},
		]
			\addplot
			[
				color = colorA,
				thick,
			]
				table
				[
					x expr	= \thisrowno{0},
					y expr	= \thisrowno{1},
				]
					{\fileA.densities};
			\addplot
			[
				color = colorA,
				thick,
				only marks,
				mark = o,
			]
				table
				[
					x expr	= \thisrowno{0},
					y expr	= \thisrowno{1},
				]
					{\fileA.densities};
			\addplot
			[
				color = colorB,
				thick,
			]
				table
				[
					x expr	= \thisrowno{0},
					y expr	= \thisrowno{1},
				]
					{\fileB.densities};
			\addplot
			[
				color = colorB,
				thick,
				only marks,
				mark = triangle,
			]
				table
				[
					x expr	= \thisrowno{0},
					y expr	= \thisrowno{1},
				]
					{\fileB.densities};
			\addplot
			[
				color = colorC,
				thick,
			]
				table
				[
					x expr	= \thisrowno{0},
					y expr	= \thisrowno{1},
				]
					{\fileC.densities};
			\addplot
			[
				color = colorC,
				thick,
				only marks,
				mark = diamond,
			]
				table
				[
					x expr	= \thisrowno{0},
					y expr	= \thisrowno{1},
				]
					{\fileC.densities};
		\end{axis}
		
		\begin{axis}
		[
			at = (SecondInset),
			anchor = north east,
			width = 0.5\textwidth,
			height = 0.125\textheight,
			xshift = -1em,
			yshift = -1em,
			xlabel = {site $j$},
			ylabel = {$\braket{\hat n_j}_{\text{final}}$},
			xmin = 0,
			xmax = 50,
			ticklabel style	= {font=\scriptsize},
			ylabel style	= {font=\scriptsize},
			xlabel style	= {font=\scriptsize},
		]
			\addplot
			[
				color = colorA,
				thick,
			]
				table
				[
					x expr	= \thisrowno{0},
					y expr	= \thisrowno{2},
				]
					{\fileA.densities};
			\addplot
			[
				color = colorA,
				thick,
				only marks,
				mark = o,
			]
				table
				[
					x expr	= \thisrowno{0},
					y expr	= \thisrowno{2},
				]
					{\fileA.densities};
			\addplot
			[
				color = colorB,
				thick,
			]
				table
				[
					x expr	= \thisrowno{0},
					y expr	= \thisrowno{2},
				]
					{\fileB.densities};
			\addplot
			[
				color = colorB,
				thick,
				only marks,
				mark = triangle,
			]
				table
				[
					x expr	= \thisrowno{0},
					y expr	= \thisrowno{2},
				]
					{\fileB.densities};
			\addplot
			[
				color = colorC,
				thick,
			]
				table
				[
					x expr	= \thisrowno{0},
					y expr	= \thisrowno{2},
				]
					{\fileC.densities};
			\addplot
			[
				color = colorC,
				thick,
				only marks,
				mark = diamond,
			]
				table
				[
					x expr	= \thisrowno{0},
					y expr	= \thisrowno{2},
				]
					{\fileC.densities};
		\end{axis}
	\end{tikzpicture}
	\caption
	{
		\label{fig:lbo_mixingfactor}
		\protect\subref{fig:lbo_mixingfactor:t-1p0:w-1p0:g-2p0} Evolution of the subspace-expansion mixing factor during the sweeping. 
		\protect\subref{fig:lbo_mixingfactor_deltaSelfE:t-1p0:w-1p0:g-2p0} Evolution of the relative energy distance during the sweeping to the best energy estimate in the runs.
		\protect\subref{fig:lbo_mixingfactor_deltaE:t-1p0:w-1p0:g-2p0} Evolution of the relative energy distance during the sweeping to the best energy estimate in the run with the initial state $\ket{\psi_{\text{in},1}}$ and $s_{\alpha=1} = 30$. 
		Inset of \protect\subref{fig:lbo_mixingfactor:t-1p0:w-1p0:g-2p0}: Electron density in the initial states. 
		Inset of \protect\subref{fig:lbo_mixingfactor_deltaSelfE:t-1p0:w-1p0:g-2p0}: Electron density in the final state. 
		The system parameters are $L = 51$, $N = 25$, $\nicefrac{\omega_0}{\hopping}=1$, and $\nicefrac{\gamma}{\hopping}=2$.
		In all runs, $\chi = 20$, $M_{\text{ph}} = 31$, and $d_0 = 5$.
	}
\end{figure}
We illustrate potential pitfalls of our method in \cref{fig:lbo_mixingfactor}.
In this example, three runs with $L = 51$, $\nicefrac{\omega_0}{\hopping} = 1$, and $\nicefrac{\gamma}{\hopping} = 2$, which corresponds to the \gls{CDW} phase, are compared.
The maximal bond dimension is set to $m_{\rm max} = 20$, the maximum phonon number per site to $n_{\rm ph} = 31$, and $d_o = 5$ optimal modes per symmetry block are considered.
In the first run $\mathcal{R}_1$, the initial state is the ground state at $\nicefrac{\omega_0}{\hopping} = 1$, $\nicefrac{\gamma}{\hopping} = 1.5$, $m_{\rm max} = 20$, $n_{\rm ph} = 31$, and $d_o = 5$ ($\ket{\Psi_{\rm in, 1}}$).
Furthermore, the mixing factor $\alpha$ is fixed to $1$ in the first $30$ sweeps ($s_{\alpha = 1} = 30$).
In the second run $\mathcal{R}_2$, the initial state is the same as in the first run ($\ket{\Psi_{\rm in, 1}}$), but the fixation of $\alpha$ is only done in the first sweep ($s_{\alpha = 1} = 1$).
For the third run $\mathcal{R}_3$, the initial state is chosen to be the ground state at $\gamma = 0$ and $m_{\rm max} = 15$ ($\ket{\Psi_{\rm in, 2}}$), while $\alpha$ is again fixed for the first $30$ sweeps ($s_{\alpha = 1} = 30$).
The inset of \cref{fig:lbo_mixingfactor}\subref{fig:lbo_mixingfactor:t-1p0:w-1p0:g-2p0} shows the electron density on the sites in the different initial states.
As expected, the density profile is flatter in the ground state at $\gamma = 0$ ($\ket{\Psi_{\rm in, 2}}$) and has more structure in the ground state at $\nicefrac{\gamma} {\hopping} = 1.5$ ($\ket{\Psi_{\rm in, 1}}$).
In the main panel of \cref{fig:lbo_mixingfactor}\subref{fig:lbo_mixingfactor:t-1p0:w-1p0:g-2p0}, the mixing factor $\alpha$ is plotted against the sweeps.
After the first $30$ sweeps where $\alpha$ is fixed, the mixing factor falls off quickly to $\approx 10^{-7}$ in $\mathcal{R}_1$.
In $\mathcal{R}_2$, the mixing factor falls off immediately with a small peak at around the 40\textsuperscript{th} sweep before $\alpha$ also settles at $\approx 10^{-7}$.
In $\mathcal{R}_3$, the decay of $\alpha$ is more gradual but after the 160\textsuperscript{th} sweep, it also has a value of $\approx 10^{-7}$ with small fluctuations.
In \cref{fig:lbo_mixingfactor}\subref{fig:lbo_mixingfactor_deltaSelfE:t-1p0:w-1p0:g-2p0}, the relative energy difference between the lowest energy found during the run and the current energy estimate in the middle of a specific sweep is presented:
\begin{align}
	\delta_{\rm self}E = \frac{E_{\rm sweep} - E_{\rm min, run}}{|E_{\rm min, run}|}\;.
\end{align}
The evolution of $\delta_{\rm self}E$ can be extracted from a single run.
In contrast, \cref{fig:lbo_mixingfactor}\subref{fig:lbo_mixingfactor_deltaE:t-1p0:w-1p0:g-2p0} displays the energy difference between the current energy estimate in the middle of a sweep and the lowest energy between all three runs:
\begin{align}
	\delta E = \frac{E_{\rm sweep} - E_{\rm min}}{|E_{\rm min}|}\;.
\end{align}
It can, of course, only be extracted from a comparison between different runs.
The runs $\mathcal{R}_1$ and $\mathcal{R}_2$ show a plateau in $\delta_{\rm self}E$ in \cref{fig:lbo_mixingfactor}\subref{fig:lbo_mixingfactor_deltaSelfE:t-1p0:w-1p0:g-2p0} until approximately the \nth{40} sweep.
Then, the energy quickly drops over the course of about 20 sweeps and $\delta_{\rm self}E$ stays below $10^{-10}$ after the \nth{60} sweep.
In contrast, the energy drops slower in $\mathcal{R}_3$, but after the \nth{300} sweep, $\delta_{\rm self}E$ also stays below $10^{-10}$.
While \cref{fig:lbo_mixingfactor}\subref{fig:lbo_mixingfactor_deltaSelfE:t-1p0:w-1p0:g-2p0} suggests that all three runs are converged, the data for $\delta E$ in \cref{fig:lbo_mixingfactor}\subref{fig:lbo_mixingfactor_deltaE:t-1p0:w-1p0:g-2p0} gives evidence that this is not the case.
The data for $\mathcal{R}_1$ is the same in \cref{fig:lbo_mixingfactor}\subref{fig:lbo_mixingfactor_deltaSelfE:t-1p0:w-1p0:g-2p0} and \cref{fig:lbo_mixingfactor}\subref{fig:lbo_mixingfactor_deltaE:t-1p0:w-1p0:g-2p0} since in $\mathcal{R}_1$ the lowest energy was found.
$\mathcal{R}_2$ finds an energy that is about $\delta E \approx 10^{-6}$ above the one found by $\mathcal{R}_1$.
The energy found in $\mathcal{R}_3$ is $\delta E \approx 10^{-2}$ above the one of $\mathcal{R}_1$.
The inset of \cref{fig:lbo_mixingfactor}\subref{fig:lbo_mixingfactor_deltaSelfE:t-1p0:w-1p0:g-2p0} shows the real-space electron-density profile in the final states of the runs.
The final states of $\mathcal{R}_1$ and $\mathcal{R}_2$ have the expected profile of a charge-density wave.
The final state of $\mathcal{R}_3$ does not have such a structure in the electron density at all. 
This example illustrates the impact of initial states and the choice of the mixing factor on the convergence of the DMRG3S+LBO.
Choosing an initial state that already possesses structural features of the ground state one is trying to find can prevent the algorithm from getting stuck in local minima.
This can be achieved by gradually changing the parameters in the Hamiltonian (here $\gamma$) by performing several runs of complete ground-state searches.
In this procedure, the initial state of the subsequent run is chosen to be the optimized ground state of the previous parameter set.
Finally, the mixing factor should be fixed at a large value for the first few runs to explore a large enough portion of the Hilbert space before it can be lowered to achieve convergence.
\subsubsection{Impact of $n_{\rm ph}$ and $d_{\rm o}$ on the precision of the ground state in the DMRG3S+LBO method}
\begin{figure}[t!]
	\def\fileA{data/holstein/mps_with_lbo/L_51_N_25/t_1p0_w_1p0_g_1p5/energies.nph-15.domax-5}%
	\def\fileB{data/holstein/mps_with_lbo/L_51_N_25/t_1p0_w_1p0_g_1p5/energies.nph-31.domax-5}%
	\def\fileC{data/holstein/mps_with_lbo/L_51_N_25/t_1p0_w_1p0_g_1p5/energies.nph-31.domax-10}%
	\def\fileD{data/holstein/mps_with_lbo/L_51_N_25/t_1p0_w_1p0_g_1p5/energies.nph-63.domax-10}%
	\centering
	\tikzsetnextfilename{deltaE_vs_mmax__Var_vs_mmax__t-1p0_w-1p0x}
	\begin{tikzpicture}
		\begin{groupplot}
		[
			group style = 
			{
				group size 			=	2 by 1,
				horizontal sep		=	5em,
			},
			width	= 0.475\textwidth,
			height	= 0.25\textheight,
			ymode	= log,
			ticklabel style	= {font=\scriptsize},
			ylabel style	= {font=\scriptsize},
			xlabel style	= {font=\scriptsize},
		]
			\nextgroupplot
			[
				title			=	{\subfloat[\label{fig:deltaE_vs_mmax:t-1p0:w-1p5}]{}},
				title style	= {yshift=-0.5em, xshift=-0.275\textwidth},	
				xlabel	= {$m_{\mathrm{max}}$},
				ylabel	= {$\delta E$},
				legend style	= {font=\scriptsize,at={(0.97,0.87)}, anchor=north east},
				extra description/.code={\node[anchor=north,font=\scriptsize] at (0.5,0.99) {$L=51$, $\nicefrac{\omega_0}{\hopping}=1.0$, $\nicefrac{\gamma}{\hopping}=1.5$};},
			]	
				\addplot
				[
					color			= colorA,
					mark			= x,
					unbounded coords	= jump,
					thick,
				]
					table
					[
						x expr	= \thisrowno{4},
						y expr	= \thisrowno{0},
					]
						{\fileA.deltaEVar};
				\addlegendentry{$n_{\text{ph}} = 15$, $d_{\text{o}} = \phantom{1}5$}
				\addplot
				[
					color			= colorB,
					mark			= +,
					unbounded coords	= jump,
					thick,
				]
					table
					[
						x expr	= \thisrowno{4},
						y expr	= \thisrowno{0},
					]
						{\fileB.deltaEVar};
				\addlegendentry{$n_{\text{ph}} = 31$, $d_{\text{o}} = \phantom{1}5$}
				\addplot
				[
					color			= colorC,
					mark			= o,
					unbounded coords	= jump,
					thick,
				]
					table
					[
						x expr	= \thisrowno{4},
						y expr	= \thisrowno{0},
					]
						{\fileC.deltaEVar};
				\addlegendentry{$n_{\text{ph}} = 31$, $d_{\text{o}} = 10$}
				\addplot
				[
					color			= colorD,
					mark			= diamond,
					unbounded coords	= jump,
					thick,
				]
					table
					[
						x expr	= \thisrowno{4},
						y expr	= \thisrowno{0},
					]
						{\fileD.deltaEVar};
				\addlegendentry{$n_{\text{ph}} = 63$, $d_{\text{o}} = 10$}
			\nextgroupplot
			[
				ymode	= log,
				xlabel	= {$m_{\text{max}}$},
				ylabel	= {$\operatorname{Var}[\hat H] / E^2_{\text{min}}$},
				legend style	= {font=\scriptsize,at={(0.97,0.87)}, anchor=north east},
				extra description/.code={\node[anchor=north,font=\scriptsize] at (0.5,0.99) {$L=51$, $\nicefrac{\omega_0}{\hopping}=1.0$, $\nicefrac{\gamma}{\hopping}=1.5$};},
				title			=	{\subfloat[\label{fig:Var_vs_mmax:t-1p0:w-1p5}]{}},
				title style	= {yshift=-0.5em, xshift=-0.275\textwidth},	
			]	
				\addplot
				[
					color			= colorA,
					mark			= x,
					unbounded coords	= jump,
					thick,
				]
					table
					[
						x expr	= \thisrowno{4},
						y expr	= \thisrowno{1},
					]
						{\fileA.deltaEVar};
						\addlegendentry{$n_{\text{ph}} = 15$, $d_{\text{o}} = \phantom{1}5$}
				\addplot
				[
					color			= colorB,
					mark			= +,
					unbounded coords	= jump,
					thick,
				]
					table
					[
						x expr	= \thisrowno{4},
						y expr	= \thisrowno{1},
					]
						{\fileB.deltaEVar};
						\addlegendentry{$n_{\text{ph}} = 31$, $d_{\text{o}} = \phantom{1}5$}
				\addplot
				[
					color			= colorC,
					mark			= o,
					unbounded coords	= jump,
					thick,
				]
					table
					[
						x expr	= \thisrowno{4},
						y expr	= \thisrowno{1},
					]
						{\fileC.deltaEVar};
						\addlegendentry{$n_{\text{ph}} = 31$, $d_{\text{o}} = 10$}
				\addplot
				[
					color			= colorD,
					mark			= diamond,
					unbounded coords	= jump,
					thick,
				]
					table
					[
						x expr	= \thisrowno{4},
						y expr	= \thisrowno{1},
					]
						{\fileD.deltaEVar};
						\addlegendentry{$n_{\text{ph}} = 63$, $d_{\text{o}} = 10$}
		\end{groupplot}
	\end{tikzpicture}
	\caption
	{
		\label{fig:deltaE_vs_mmax__Var_vs_mmax}
		\protect\subref{fig:deltaE_vs_mmax:t-1p0:w-1p5} Relative deviation of the ground\hyp state energy obtained with \gls{DMRG3S+LBO} from the smallest energy found, plotted against the maximal bond dimension $m_{\rm max}$.
		\protect\subref{fig:Var_vs_mmax:t-1p0:w-1p5} Variance of the energy plotted against the maximal bond dimension $m_{\rm max}$.
		%
	}
\end{figure}
In the \gls{DMRG3S+LBO} method, the maximal bond dimension $m_{\rm max}$, the maximal number of phonons per site $n_{\rm ph}$, 
and the maximal number of optimal modes per block $d_{\rm o}$ determine the reachable precision of the ground states.
All of these parameters have a different influence.
The maximal number of phonons per site $n_{\rm ph}$ changes the dimension of the Hilbert space and, in principle, the nature of the whole system.
The ultimate goal is to get results that do not change upon further increase of $n_{\rm ph}$ and are therefore indistinguishable from the case $n_{\rm ph} = \infty$.
For smaller $n_{\rm ph}$, results can be converged to a certain precision with respect to $m_{\rm max}$ and $d_{\rm o}$, but an increase of $n_{\rm ph}$ will still improve the precision substantially with respect to the $n_{\rm ph} = \infty$ limit.
In \cref{fig:deltaE_vs_mmax__Var_vs_mmax}, we compare the precision of the ground\hyp state energy for different choices of $m_{\rm max}$, $n_{\rm ph}$, and $d_{\rm o}$.
The system parameters are $L = 51$, $N = 25$, $\nicefrac{\omega_0}{\hopping} = 1$, and $\nicefrac{\gamma}{\hopping} = 1.5$.
\Cref{fig:deltaE_vs_mmax__Var_vs_mmax}\subref{fig:deltaE_vs_mmax:t-1p0:w-1p5} shows the relative difference between the energies at the end of each run to the best energy found:
\begin{align}
\delta E = \frac{E_0(m_{\rm max},n_{\rm ph},d_{\rm o}) - E_{\rm min}}{|E_{\rm min}|}\;.
\end{align}
As a function of $m_{\rm max}$, the energy decreases as expected.
For a certain choice of $n_{\rm ph}$ and $d_{\rm o}$, the energy can, however, only reach a certain precision that does not substantially improve as the bond dimension is increased further.
With $n_{\rm ph} = 31$ and $d_{\rm o} = 10$, the maximal precision of our implementation of the \gls{DMRG3S+LBO} method can be reached, which turns out to be a relative energy difference of about $10^{-12}$.
Further increasing $n_{\rm ph}$ does not improve the energies.
\Cref{fig:deltaE_vs_mmax__Var_vs_mmax}\subref{fig:Var_vs_mmax:t-1p0:w-1p5} shows the variance of the energy
\begin{align}
	{\rm Var}[\hat H] = \bra{\psi(m_{\rm max},n_{\rm ph},d_{\rm o})} (H - E_0(m_{\rm max},n_{\rm ph},d_{\rm o}))^2 \ket{\psi(m_{\rm max},n_{\rm ph},d_{\rm o})}\;
\end{align}
versus the maximal bond dimension $m_{\rm max}$.
The variance can be taken as a measure of proximity of a certain state to an eigenstate of the Hamiltonian.
As expected, a small choice of $d_{\rm o}$ limits the reachable precision in the variance.
However, limiting $n_{\rm ph}$ does not limit the precision in such a way.
As mentioned earlier, changing $n_{\rm ph}$ alters the Hilbert space and the system altogether and, therefore, one can be close to the ground state in the case of a small $n_{\rm ph}$, 
but the state is still substantially different from the ground state in the limit $n_{\rm ph} = \infty$.
This difference, however, does not manifest itself in the variance.
\subsection{\label{sec:methods:pp-dmrg}Projected Purification}
\begin{figure}[!h]
	\subfloat[\label{fig:pp-dmrg:sketch}]
	{
		\centering
		\tikzsetnextfilename{ppdmrg_sketch}
		\begin{tikzpicture}
			\node
			[
				minimum height=0.33\textheight
			]
			{
				\includegraphics[width=0.45\textwidth]{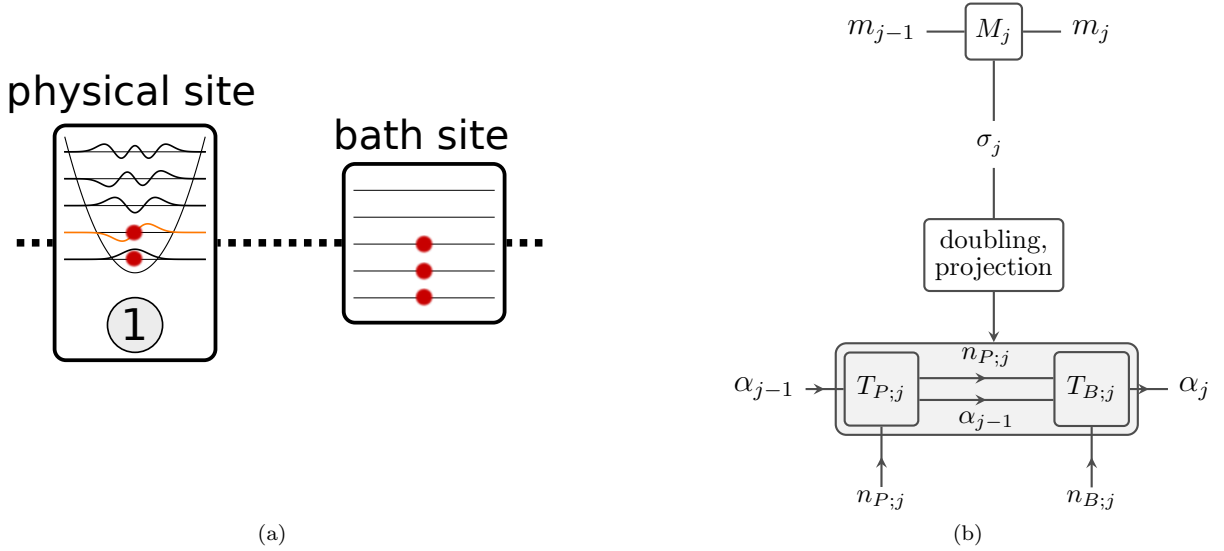}
			};
		\end{tikzpicture}
	}%
	\hfill%
	\subfloat[\label{fig:pp-mpstensor}]
	{
		\centering
		\tikzsetnextfilename{mpstensor2double}
		\begin{tikzpicture}[decoration=brace]
			\begin{scope}[node distance = 0.8 and 0.5]
				\node[site] (site1) {$M_j$};
				\node[ghost] (site2) [left=of site1] {$m_{j-1}$};
				\node[ghost] (site3) [right=of site1] {$m_{j}$};
				\node[ld] (sigma1) [below=of site1] {$\sigma^{\phantom{B}}_{j}$};
				
				\draw[-, draw=black!70, thick] (site1) -- (sigma1);
				\draw[-, draw=black!70, thick] (site1) -- (site2);
				\draw[-, draw=black!70, thick] (site1) -- (site3);
				
				\node[ghost] [below=7em of sigma1] (canchor) {};
				\node[site] at (site2|-canchor) (site21) {$T_{P;j}$};
				\node[site] at (site3|-canchor) (site22) {$T_{B;j}$};
				
				\node[ghost] (index1) [left=of site21] {$\ingoing{\biwqn}_{j-1}$};
				\node[ghost] (index2) [right=of site22] {$\outgoing{\biwqn}_j$};
				
				\node[ld] (sigma21) [below=of site21] {$\ingoing{n}_{P;j}$};
				\node[ld] (sigma22) [below=of site22] {$\ingoing{n}_{B;j}$};
			\end{scope}
			
			\begin{scope}[on background layer]
				\node[draw,thick,fill=black!5,draw=black!70,rounded corners,fit=(site21) (site22),inner sep=0.25em] (pbsite) {};
			\end{scope}
			
			\begin{scope}
				\coordinate (arrow_end) at (pbsite.north-|canchor);
				\draw[->, thick, draw=black!70] ($(sigma1.south)-(0,.25em)$) to node[midway, rounded corners=2pt, thick, draw=black!70, fill=white, align=center, font=\footnotesize\linespread{0.8}\selectfont] {doubling,\\ projection} (pbsite.north-|canchor);
				\draw[->-, thick, draw=black!70] (index1) to (site21);
				\draw[->-, thick, draw=black!70] (site22) to (index2);
				\draw[->-, thick, draw=black!70] (sigma21) to (site21);
				\draw[->-, thick, draw=black!70] (sigma22) to (site22);
				
				\draw[->-, thick, draw=black!70] ($(site21.east)+(0,.35em)$) to node[above, font=\footnotesize] {$\going n_{P;j}$} ($(site22.west)+(0,.35em)$);
				\draw[->-, thick, draw=black!70] ($(site21.east)-(0,.35em)$) to node[below, font=\footnotesize] {$\biwqn_{j-1}$} ($(site22.west)-(0,.35em)$);
			\end{scope}
		\end{tikzpicture}
	}
	\caption
	{
		\label{fig:pp-dmrg}
		\protect\subref{fig:pp-dmrg:sketch} A schematic representation of the local Hilbert space doubling via the introduction of a bath site for each local degree of freedom as used in \gls{PP-2DMRG} is shown.
		\protect\subref{fig:pp-mpstensor} The decomposition and projection of a general \gls{MPS} tensor (top) into the subspace $\mathcal P$ enforcing the local gauge conditions is depicted.
		The decomposition of the introduced auxiliary index into irreducible representations of the local conservation law generated by $\hat n_{P;j} + \hat n_{B;j}$ as described in \cite{10.21468/SciPostPhys.10.3.058} is sketched by the double bond $(n_{P;j}, \biwqn_{j-1})$.
	}
\end{figure}
In the following, we describe the projected\hyp purification ansatz \cite{10.21468/SciPostPhys.10.3.058}, which is based on a doubling of the Hilbert space and a subsequent projection into an invariant subspace.
This scheme allows us to formulate operators and states that transform under a restored global $U(1)$ symmetry and to efficiently represent and truncate the local phononic degrees of freedom.
A simple pictorial representation of the ansatz is sketched in \cref{fig:pp-dmrg}\subref{fig:pp-dmrg:sketch}.
Every physical site with electron and phonon degrees of freedom is accompanied by a phonon bath site that acts as a reservoir.
Creating or annihilating a phonon on the physical site means that it hops from or to the bath site.
This way, an artificial $U(1)$ symmetry is generated, namely the combined number of phonons on the physical and bath site is conserved.
\subsubsection{\label{sec:methods:pp-dmrg:operators}Projected Purified Operators}
Consider a Hilbert space $\mathcal H = \mathcal H^{\otimes L}_d$ with basis states labeled by the eigenvalues of the local density operators $\hat n_j = \operatorname{diag} (0, 1, \ldots, d-1)$.
We define a doubling $\mathcal H_{PB} = \mathcal H_P \otimes \mathcal H_B$, introducing two copies $\mathcal H_{P/B}$ of the original Hilbert space $\mathcal H$ as shown in \cref{fig:pp-dmrg}\subref{fig:pp-dmrg:sketch} and refer to $\mathcal H_P$ and $\mathcal H_B$ as the physical and bath Hilbert space, respectively.
Operators acting on (local) Hilbert spaces $\mathcal H_{P/B(;d)}$ will be equipped with an additional label, for instance, $\hat n_{P/B;j}$ denotes the density operators acting on the local physical or bath degrees of freedom at site $j$.
A projected purified operator $\hat O_{PP}: \mathcal P \rightarrow \mathcal P$ acts on a subspace $\mathcal P \subset \mathcal{H}_{PB}$ of the purified Hilbert space and satisfies the constraints
\begin{align}
	\left[\hat O_{PP}, \hat n_{P;j}+\hat n_{B;j} \right] = 0
\end{align}
for every $j$.
Defining global operators $\hat N_{P/B} = \sum_{j} \hat n_{P/B;j}$, any projected purified operator manifestly conserves the global $U(1)$ symmetry generated by $\hat N_P + \hat N_B$:
\begin{align}
	\left[\hat O_{PP}, \hat N_{P}+\hat N_{B} \right] = 0 \; .
\end{align}
An explicit construction scheme for the subspace $\mathcal P$ can be found by fixing gauge constraints between the physical and bath site to 
\begin{align}
	\hat n_{P;j} + \hat n_{B;j} \equiv (d-1) \; . \label{eq:PP:gauge-fixing}
\end{align}
In \cite{10.21468/SciPostPhys.10.3.058}, we showed that this choice implies $\operatorname{dim} \mathcal P = \operatorname{dim} \mathcal H$ and that there is a one\hyp to\hyp one mapping between states and operators in these Hilbert spaces.
It follows that for any operator $\hat O$ acting on the original Hilbert space, an analogous projected purified operator can be constructed.
This is achieved by introducing balancing operators $\hat \beta^{[\dagger]}_{B;j}$
\begin{align}
	\hat \beta^{\nodagger}_{B;j} = \sum_{n_{B;j}=1}^{d-1} \ket{n_{B;j}-1}\bra{n_{B;j}}, \quad \ket{n_{B;j}}\in\mathcal H_{B;d} \; .
\end{align}
Writing $\hat O$ in terms of ladder operators $\hat b^{[\dagger]}_{j}$, the mapping to the corresponding projected purified operator $\hat O_{PP}$ is achieved by pairing up ladder operators with conjugated balancing operators:
\begin{align}
	\hat b^{\dagger}_j \rightarrow \hat b^{\dagger}_{P;j} \hat \beta^{\nodagger}_{B;j}, \quad
	\hat b^{\nodagger}_j \rightarrow \hat b^{\nodagger}_{P;j} \hat \beta^{\dagger}_{B;j} \; .
\end{align}
The phononic part of the Holstein model breaking the global $U(1)$ symmetry in $\mathcal H$ is then represented by a projected purified operator $\hat H_{PP;f-ph}$ acting only on $\mathcal P$
\begin{align}
	\hat H_{PP;f-ph}
	=
	\gamma \sum_j \hat n^f_j \left(\hat b^\dagger_{P;j} \hat \beta^\nodagger_{B;j} + \hat b^\nodagger_{P;j} \hat \beta^\dagger_{B;j} \right) \; .
\end{align}
Note that $\hat H_{PP;f}$ and $\hat H_{PP;ph}$ remain unaltered except for a formal replacement of local operators acting on $\mathcal H_d$ with their counterparts acting on $\mathcal P$:
\begin{align}
	\hat c^{[\dagger]}_j \rightarrow \hat c^{[\dagger]}_{P;j} \hat{\mathbf 1}^{\nodagger}_{B;j}, \quad 
	\hat n^f_j \rightarrow \hat n^f_{P;j} \hat{\mathbf 1}^{\nodagger}_{B;j} \quad \text{and} \quad
	\hat b^{\dagger}_j \hat b^\nodagger_j \rightarrow \hat b^{\dagger}_{P;j} \hat b^\nodagger_{P;j} \hat{\mathbf 1}^{\nodagger}_{B;j} \; .
\end{align}
\subsubsection{\label{sec:methods:pp-dmrg:mps}Projected Purified \gls{MPS}}
The gauge\hyp fixing condition \cref{eq:PP:gauge-fixing} allows us to construct a projector to $\mathcal P$.
Using the action of this projector on \glspl{MPS} in the enlarged Hilbert space, we derived a condition on the combined physical and bath system's site tensors \cite{10.21468/SciPostPhys.10.3.058}.
In a practical ground\hyp state\hyp search calculation, it suffices to create an initial state whose physical indices reflect the gauge fixing.
For instance, let $T^{n_{P;j}}_{j;\alpha_{j-1},\gamma_{j}}$ and $T^{n_{B;j}}_{j;\gamma_{j},\alpha_{j}}$ be $U(1)$-invariant \gls{MPS} site tensors with physical and bath local degrees of freedom labeled by $n_{P;j}$ and $n_{B;j}$, respectively.
The condition for the \gls{MPS} to represent a projected purified state $\ket{\psi}_{PP}\in \mathcal P$ is then given by
\begin{align}
	\sum_{\gamma_j}T^{n_{P;j}}_{j;\alpha_{j-1},\gamma_{j}}T^{n_{B;j}}_{j;\gamma_{j},\alpha_{j}} \neq 0 \Leftrightarrow n_{P;j} + n_{B;j} \equiv (d-1) \; ,\label{eq:PP:gauge-fixing:MPS}
\end{align}
and shown graphically in \cref{fig:pp-dmrg}\subref{fig:pp-mpstensor}.
Since $\mathcal P$ is a subspace of the enlarged Hilbert space $\mathcal H_{PB}$, projected purified operators acting on such states conserve this condition.
This is an important point as it minimizes the implementational effort.
In fact, by doubling the Hilbert space and using projected purified operator representations only, \gls{DMRG} codes that are able to initialize states $\ket{\psi}_{PP}$ obeying \cref{eq:PP:gauge-fixing:MPS} can readily work in $\mathcal P$ and thereby exploit restored global $U(1)$ symmetries.
For the dimensions of the indices representing the local degrees of freedom, this implies $\operatorname{dim} n_{P/B;j} \equiv 1$.
\subsubsection{\label{sec:methods:pp-dmrg:1site-rdm}Truncation and Connection to \gls{1RDM}}
\begin{figure}
	\centering
	\tikzsetnextfilename{pp_truncation}
	\begin{tikzpicture}
		\node [align = center, inner sep = 0pt]
		(Contraction)
		{
			\(
				\begin{tikzpicture}[decoration=brace, baseline, anchor=base]
					\clip (-0.75,-2.25) rectangle (4,2.25);
					\begin{scope}[node distance = 0.75 and 0.575]
						\node[ghost]		(LeftBoundary)	{};
						\node[op]			(OpN)	[right=of LeftBoundary, minimum height=2.75em]	{$\going{\hat n}^{\noprime}_{P;j}$};
						\node[site]	(bra)	[above=of OpN]			{$T^{\nodagger}_{P;j}$};
						\node[site]	(ket)	[below=of OpN]			{$T^{\dagger}_{P;j}$};
						\node[ghost]		(RightBoundary)	[right=of OpN]	{};
						
						\draw[->-, out=180, in=180, looseness=1.25]	(ket) to node[midway,right,ld, inner sep = 2pt] {$\ingoing{\biwqn}_{j-1}$} (bra);
						
						\draw[->-, out=0, in=0, looseness=1.25]	(bra) to node[midway,left,ld, inner sep = 2pt] {$\ingoing{\biwqn}_{j-1}$} (ket);
						\draw[->-, out=0, in=0, looseness=1.5]	(bra) to node[midway,right,ld, inner sep = 2pt] {$\going n_{P;j}$} (ket);
						\draw[-<-,]					(bra) to node[midway,right,ld, inner sep = 2pt] {$\going n_{P;j}$} (OpN);
						\draw[-<-,]					(OpN) to node[midway,right,ld, inner sep = 2pt] {$\going n_{P;j}$} (ket);
					\end{scope}
				\end{tikzpicture}
				=
				\sum\limits_{n^\noprime_{P;j}} n^\noprime_{P;j} \rho_{n^\noprime_{P;j},n^\noprime_{P;j}}
				\Rightarrow
				T^\noprime_{P;j} =
			\)
		};
		
		\node
		[
			right		= -.5em of Contraction.east,
			align		= center,
		]
		(MPSRDM3)
		{
			\begin{tikzpicture}
			[
				baseline=(MPS.center),
				font	= \scriptsize,
			]
				\begin{scope}[node distance = 0]
					\matrix (MPS) 
					[
						matrix of math nodes,
						left delimiter=(,
						right delimiter=),
						nodes in empty cells,
						nodes = 
						{
							minimum height=1.5em,
							anchor=center,
							text width=1.5em,
							text depth=.25ex,
							align=center,
							inner sep=0pt
						}
					]
					{
						\phantom{1}		& \phantom{1} 	& \phantom{1}	& \phantom{1}	& \phantom{1}\\
						\phantom{1}		& \phantom{1}	& \phantom{1} 	& \phantom{1}	& \phantom{1}\\
						\phantom{1}		& \phantom{1}	& \phantom{1}	& \phantom{1}	& \phantom{1}\\
						\phantom{1}		& \phantom{1} 	& \phantom{1} 	& \phantom{1}	& \phantom{1}\\
						\phantom{1}		& \phantom{1} 	& \phantom{1} 	& \phantom{1}	& \phantom{1}\\
					};
					\node [left=of MPS-1-1] (LeftAnchor) {$\phantom{0}$};
					\node [left=of LeftAnchor, font=\tiny] (m00) {$0$};
					\node [left=of m00, font=\tiny, xshift=-.5em] (n00) {$0$};
					\node [font=\tiny, anchor=center] at (n00|-MPS-2-1.center) (n01) {$0$};
					\node [font=\tiny, anchor=center] at (m00|-MPS-2-1.center) (m01) {$1$};
					\node [font=\tiny, anchor=center] at (n01|-MPS-3-1.center) (n02) {$1$};
					\node [font=\tiny, anchor=center] at (m01|-MPS-3-1.center) (m02) {$0$};
					\node [font=\tiny, anchor=center] at (n02|-MPS-4-1.center) (n03) {$0$};
					\node [font=\tiny, anchor=center] at (m02|-MPS-4-1.center) (m03) {$2$};
					\node [font=\tiny, anchor=center] at (n03|-MPS-5-1.center) (n04) {$\vdots$};
					\node [font=\tiny, anchor=center] at (m03|-MPS-5-1.center) (m04) {$\vdots$};
					
					\node [font=\tiny, above=of n00] (nheader) {$\going n_{P;j}$};
					\node [font=\tiny, above=of m00] (mheader) {$\biwqn_{j-1}$};
					
					\node [above=of MPS-1-1, yshift=1.em, font=\tiny] (m10) {$0$};
					\node [font=\tiny, anchor=center] at (m10-|MPS-1-2.east) (m11) {$1$};
					\node [font=\tiny, anchor=center] at (m11-|MPS-1-4.center) (m12) {$2$};
					\node [font=\tiny, anchor=center] at (m12-|MPS-1-5.center) (m13) {$\cdots$};
					
					\node [font=\tiny, left=of m10] (mmheader) {$\biwqn_{j}$};
				\end{scope}
				\begin{scope}[on background layer, node distance = 0]
					\node [draw=blue!30, fill=blue!10, inner sep=-1pt, fit=(MPS-1-1)] (N1) {};
					\node [draw=blue!30, fill=blue!10, inner sep=-1pt, fit=(MPS-2-2)(MPS-2-3)] (N12) {};
					\node [draw=blue!30, fill=blue!10, inner sep=-1pt, fit=(MPS-3-2)(MPS-3-3)] (N22) {};
					\node [ghost, inner sep=-1pt, fit=(MPS-2-2)(MPS-3-3)] (N2) {};
					\node [draw=blue!30, fill=blue!10, inner sep=-1pt, fit=(MPS-4-4)(MPS-4-5)] (N13) {};
					\node [ghost, inner sep=-1pt, fit=(MPS-4-4)(MPS-5-5)] (N3) {};
				\end{scope}
				\begin{scope}[node distance = 0]
					\node [font=\tiny, anchor = center] at (N1.center) (N0Label) {$0$};
					\node [font=\tiny, anchor = center] at (N12.center) (N1Label) {$0$};
					\node [font=\tiny, anchor = center] at (N22.center) (N1Label) {$1$};
					\node [font=\tiny, anchor = center] at (N13.center) (N1Label) {$0$};
					\node [font=\tiny, anchor = south east, xshift=1.5em, yshift=-1.5em] at (N3.center) {$\ddots$};
					\draw [-, draw=black] ($(N1.south west)+(-2.5pt,.05pt)$) -- ($(N1.north west)+(-2.5pt,.25pt)$);
					\draw [-, draw=black] ($(N1.south east)+(+2.5pt,.05pt)$) -- ($(N1.north east)+(+2.5pt,+.25pt)$);
					\draw [-, draw=black] ($(N12.south west)+(-2.5pt,.05pt)$) -- ($(N12.north west)+(-2.5pt,.25pt)$);
					\draw [-, draw=black] ($(N12.south east)+(+2.5pt,.05pt)$) -- ($(N12.north east)+(+2.5pt,+.25pt)$);
					\draw [-, draw=black] ($(N22.south west)+(-2.5pt,.05pt)$) -- ($(N22.north west)+(-2.5pt,.25pt)$);
					\draw [-, draw=black] ($(N22.south east)+(+2.5pt,.05pt)$) -- ($(N22.north east)+(+2.5pt,+.25pt)$);
					\node [font=\tiny] at ($(N1.north east)+(+2.pt,+1.pt)$) {$2$};
					\node [font=\tiny] at ($(N12.north east)+(+2.pt,+1.pt)$) {$2$};
					\node [font=\tiny] at ($(N22.north east)+(+2.pt,+1.pt)$) {$2$};
					\node [right=of N1, font=\tiny] {$\leq \rho^\noprime_{0,0}$};
					\node [right=of N12, font=\tiny] {$ \leq \rho^\noprime_{0,0}$};
					\node [right=of N22, font=\tiny] {$ \leq \rho^\noprime_{1,1}$};
				\end{scope}
			\end{tikzpicture}
		};
	\end{tikzpicture}
	\caption
	{
		\label{fig:pp-dmrg:truncation}
		Schematic representation of the connection between the irreducible representations of tensor blocks $T^{n^{\protect\noprime}_{P;j}}$ and the diagonal elements of the \gls{1RDM} $\rho_{n^{\protect\noprime}_j, n^{\protect\noprime}_j}$.
		On the right\hyp hand side, the blue boxes indicate irreducible representations of the tensor blocks $T^{n^{\protect\noprime}_{P;j}}$ with $n^{\protect\noprime}_{P;j} = 0,1,2,\ldots$ constituting the site tensor $T^{\protect\noprime}_{P;j}$.
		Note that if $\rho_{n^{\protect\noprime}_j, n^{\protect\noprime}_j}$ is smaller than the given truncation threshold, complete tensor blocks $T^{n^{\protect\noprime}_{P;j}}$ can be discarded.
	}
\end{figure}
Projected purified states $\ket{\psi}_{PP}\in\mathcal P$ exhibit an intimate relation to the \gls{1RDM} $\hat \rho_j = \operatorname{Tr}_{k\neq j} \ket{\psi}\bra{\psi}$ of the corresponding state in the original Hilbert space $\mathcal H$, which is displayed graphically in \cref{fig:pp-dmrg:truncation}.
The crucial observation is that in a mixed-canonical \gls{MPS} representation, when tracing out the auxiliary indices of the physical Hilbert space $\mathcal H_P$ at the orthogonality center $j$, one obtains the diagonal elements
\begin{align}
	\rho_{n_{P;j},n_{P;j}} = \sum_{\alpha_{j-1},\gamma_j} T^{n_{P;j}}_{j; \alpha_{j-1},\gamma_j} \left[ T^{n_{P;j}}_{j; \alpha_{j-1},\gamma_j} \right]^\dagger \; , \label{eq:1rdm-schmidt-values}
\end{align}
of $\rho_{n^\noprime_j,n^\prime_j}$.
A truncation scheme across the auxiliary bond $\gamma_j$:
\begin{align}
	T^{n_{P;j}}_{j; \alpha_{j-1},\gamma_j} = \sum_{\sigma} U^{n_{P;j}}_{j;\alpha_{j-1},\sigma} \Lambda^\noprime_{\sigma} V^\noprime_{\sigma, \gamma_j} \;,
\end{align}
discarding only the smallest singular values $\Lambda_{\tilde \sigma}$ so that the truncated weight fulfills $\sum_{\tilde \sigma} \Lambda^2_{\tilde \sigma} < \delta$, approximates the single-site reduced density-matrix in an optimal way with respect to the $1$-norm
\begin{align}
	\Tr (\hat \rho_j - \hat{\tilde \rho}_j) < \delta \; .
\end{align}
Such a truncation can discard complete tensor blocks $T^{n_{P/B;j}}$ if $\rho_{n_{P;j},n_{P;j}} < \delta$ as follows immediately from \cref{eq:1rdm-schmidt-values} and thus reduces the number of local degrees of freedom $n_{P/B;j}$ in the state representation.
Finally, the full \gls{1RDM} can be obtained from contracting physical and bath site tensors in a mixed\hyp canonical representation
\begin{align}
	M^{n^\noprime_j}_{j; \alpha_{j-1},\alpha_j} 
	&= 
	\sum_{\gamma_j}T^{n_{P;j}}_{j;\alpha_{j-1},\gamma_{j}}T^{n_{B;j}}_{j;\gamma_{j},\alpha_{j}} \delta_{n^\noprime_j, n_{P;j}} \\
	\Rightarrow \rho_{n^\noprime_j, n^\prime_j}
	&=
	\sum_{\alpha_{j-1}, \alpha_{j-1}} M^{n^\noprime_j}_{j; \alpha_{j-1},\alpha_j}\left[ M^{n^\prime_j}_{j; \alpha_{j-1},\alpha_j} \right]^\dagger \; .
\end{align}
\begin{figure}[t!]
	\centering
	\tikzsetnextfilename{PPDMRG-optimal_modes_occ-L_51-N_25-t_1p0-w_1p0-g_2p0-site_25}
	\begin{tikzpicture}
		\begin{groupplot}
		[
			group style = 
			{
				group size 			=	2 by 1,
				x descriptions at	=	edge bottom,
				y descriptions at	=	edge left
			},
			height			= 	0.35\textheight,
			width 			= 	0.475\textwidth-7pt,
			ymode	= log,
			xmin 			= 	0, 
			ymin			=	1e-18,
			ymax			=	0,
			xlabel	= {optimal mode $d_{\rm o}$},
			ylabel	= {weight $w_{\rm o}$},
			ticklabel style	= {font=\scriptsize},
			xlabel style	= {font=\scriptsize,yshift=0pt},
			ylabel style	= {font=\scriptsize,yshift=0pt},
		]
			\nextgroupplot
			[
				axis on top,
				xlabel style	=	{yshift=0.3em},
				title		=	{\subfloat[\label{fig:ppdmrg:optimal-modes-occ:L-51_N-25:t-1p0_w-1p0_g-1p5:site_25}]{}},
				title style	= {yshift=-0.5em, xshift=-0.225\textwidth},	
			]
				\foreach [evaluate=\i as \clfrac using (\i*100/19)] \i in {0,1,...,19} {
					\edef\cmd{
						\noexpand\addplot
						[
							color			= red!\clfrac!blue,
							mark			= x,
							unbounded coords	= jump,
							thick
						]
						table
						[
							x expr	= \noexpand\coordindex,
							y expr	= \noexpand\thisrowno{\i},
						]
						{data/holstein/L_51_N_25_t_1p0_w_1p0_g_1p5/optimal-modes.25.gnu};
					}
					\cmd
				}
				\coordinate (inset_position_A) at (axis cs:10,0.5);
			
			\nextgroupplot
			[
				colorbar,
				point meta min	= 100,
				point meta max	= 2000,
				colormap	= {bluered}{rgb255=(0,0,255) rgb255=(255,0,0)},
				every colorbar/.append style =
				{
					width	= {10pt},
					xshift	= -0.5em,
					ytick	= {100,1000,2000},
					yticklabel style= {font=\tiny},
					ylabel style	= {font=\scriptsize},
					ylabel	= {max. bond dimension},
				},
				title			=	{\subfloat[\label{fig:ppdmrg:optimal-modes-occ:L-51_N-25:t-1p0_w-1p0_g-1p5:site_24}]{}},
				title style	= {yshift=-0.5em, xshift=-0.225\textwidth},	
			]
				\foreach [evaluate=\i as \clfrac using (\i*100/19)] \i in {0,1,...,19} {
					\edef\cmd{
						\noexpand\addplot
						[
							color			= red!\clfrac!blue,
							mark			= x,
							unbounded coords	= jump,
							thick
						]
							table
							[
								x expr	= \noexpand\coordindex,
								y expr	= \noexpand\thisrowno{\i},
							]
							{data/holstein/L_51_N_25_t_1p0_w_1p0_g_1p5/optimal-modes.24.gnu};
					}
					\cmd
				}
				\coordinate (inset_position_B) at (axis cs:10,0.5);
		\end{groupplot}

		\begin{axis}
		[
			at	= {(inset_position_A)},
			anchor	= north west,
			width	= 0.25\textwidth,
			ymode	= log,
			ymin	= 1e-16,
			title	= {diagonal elements $\rho_{n^\noprime_{25},n^\noprime_{25}}$},
			xlabel	= {$n_{25}$},
			title style	= {font=\scriptsize,yshift=-5pt},
			ticklabel style	= {font=\scriptsize},
			xlabel style	= {font=\scriptsize,yshift=5pt},
			ylabel style	= {font=\scriptsize,yshift=0pt},
			legend image post style	= {scale=0.5},
			legend style	= {font=\tiny,at={(0.03,0.03)}, anchor=south west},
		]
			\foreach [evaluate=\i as \clfrac using (\i*100/19)] \i in {19,18,...,0} {
				\edef\cmd{
					\noexpand\addplot
					[
						color			= red!\clfrac!blue,
						mark			= none,
						forget plot,
					]
					table
					[
						x expr	= \noexpand\coordindex,
						y expr	= \noexpand\thisrowno{\i}+1e-20,
					]
					{data/holstein/L_51_N_25_t_1p0_w_1p0_g_1p5/ph-occ.25.gnu};
				}
				\cmd
			}
			
			\addplot
			[
				color	= colorD,
				mark	= x,
			]
			table
			[
				x expr	= \coordindex,
				y expr	= \thisrowno{0},
			]
			{data/holstein/L_51_N_25_t_1p0_w_1p0_g_1p5/ph-occ.decoupled};
			\addlegendentry{$\hopping=0$}
		\end{axis}
		\begin{axis}
		[
			at	= {(inset_position_B)},
			anchor	= north west,
			width	= 0.25\textwidth,
			ymode	= log,
			ymin	= 1e-16,
			title	= {diagonal elements $\rho_{n^\noprime_{24},n^\noprime_{24}}$},
			xlabel	= {$n_{24}$},
			title style	= {font=\scriptsize,yshift=-5pt},
			ticklabel style	= {font=\scriptsize},
			xlabel style	= {font=\scriptsize,yshift=5pt},
			ylabel style	= {font=\scriptsize,yshift=0pt},
		]
			\foreach [evaluate=\i as \clfrac using (\i*100/19)] \i in {19,18,...,0} 
			{
				\edef\cmd
				{
					\noexpand\addplot
					[
						color			= red!\clfrac!blue,
						mark			= none,
					]
						table
						[
							x expr	= \noexpand\coordindex,
							y expr	= \noexpand\thisrowno{\i}+1e-20,
						]
						{data/holstein/L_51_N_25_t_1p0_w_1p0_g_1p5/ph-occ.24.gnu};
				}
				\cmd
			}
		\end{axis}
	\end{tikzpicture}
	\caption
	{
		\label{fig:ppdmrg:optimal-modes-occ:L-51_N-25:t-1p0_w-1p0_g-1p5}
		Weight $w_{\rm o}$ of optimal modes $d_{\rm o}$ as a function of the maximal bond dimension at the auxiliary bonds $\gamma_{25}$ \protect\subref{fig:ppdmrg:optimal-modes-occ:L-51_N-25:t-1p0_w-1p0_g-1p5:site_25} and \protect\subref{fig:ppdmrg:optimal-modes-occ:L-51_N-25:t-1p0_w-1p0_g-1p5:site_24} using the projected purification.
		Data is extracted from the single\hyp site reduced density matrix $\rho_{n^\protect\noprime_{25/24},n^\prime_{25/24}}$ at the center site ($j=25/24$) in the calculated ground state of the Holstein model with $L=51$ sites and $N=25$ fermions, $\nicefrac{\omega}{\hopping}=1.0, \nicefrac{\gamma}{\hopping}=1.5$.
		The inset shows the diagonal elements $\rho_{n^\protect\noprime_{25},n^\protect\noprime_{25}}$ indicating the immediate effect of truncations.
		For comparison, the phonon excitation probabilities obtained for $\hopping=0$ at occupied sites (left) are overlayed, indicated by yellow crosses.
	}
\end{figure}
Taking operators and states into their projected purified representation implies certain numerical advantages in \gls{MPS} calculations when the local dimension $d$ of the problem is large and the system does not conserve global $U(1)$ symmetries, initially.
When re-establishing a global $U(1)$-symmetry in the subspace $\mathcal P$, local degrees of freedom $\hat n_{P/B;j}$ decompose into one-dimensional representations.
As a consequence, contractions scaling as $d^l$ for some exponent $l\in\mathbb N$ can be parallelized over the local degrees of freedom.
Additionally, the connection to the \gls{1RDM} reveals that if for some $n_0\in \left\{0,\ldots,d-1 \right\}$ the diagonal elements $\rho_{n\pm n_0, n\pm n_0}$ are decaying fast enough, the number of relevant tensor blocks can be reduced drastically upon truncation.
An important conclusion is that having restored the global $U(1)$ symmetry w.r.t.~phononic degrees of freedom allows us to employ a local solver based on two-site tensors, which is less prone of getting stuck in local minima (\gls{PP-2DMRG}).
Furthermore, the truncation argument implies that the dimension of the auxiliary bond $\gamma_j$ between physical and bath site tensors controls the quality of the approximation of $\rho_{n^\noprime_j, n^\prime_j}$.
As a consequence, the same scaling analysis in terms of auxiliary bond dimensions, which is usually employed in \gls{DMRG} algorithms to extrapolate observables towards their exact value, can be used for the projected purification.
This is demonstrated in \cref{fig:ppdmrg:optimal-modes-occ:L-51_N-25:t-1p0_w-1p0_g-1p5}\subref{fig:ppdmrg:optimal-modes-occ:L-51_N-25:t-1p0_w-1p0_g-1p5:site_25}, where we show the weight $w_{\rm o}$ of the optimal modes $d_{\rm o}$ of the single-site reduced density-matrix as a function of the auxiliary bond dimension $\gamma_j$ used to approximate the ground state of the Holstein model near the phase boundary~\eqref{eq:comp-pp-with-lbo:2}.
Upon increasing the bond dimension we find a well-converging behavior for the weights $w_{\rm o}(d_{\rm o})$.
Note that in these calculations, we allowed for a maximum number of $n_{\rm ph} = 63$ phonons per lattice site.
However, as shown in the inset, the maximum occupation number for the phonons kept after truncation is $n_{\rm ph} \leq 20$, i.e., a significant amount of tensor blocks is discarded due to the vanishing contribution of high-occupation phonon modes.
Therein, we also indicate the diagonal elements calculated for the atomic limit $\hopping=0$ using the Lang-Firsov transformation of the Holstein Hamiltonian~\cite{lang1963kinetic}.
It can be seen that the phonon excitation probability (i.e., $\rho_{n^\noprime_j,n^\noprime_j}$), obtained from projected purification, is already very close to the limit of decoupled fermions, which is the asymptotic distribution when sending $\hopping\rightarrow 0$.
\subsubsection{\label{sec:methods:pp-dmrg:truncation-effects}Effects of Truncation}
\def\t{1p0}
\def\w{1p0}
\def\gs{0p5, 1p5, 2p0}
\def\L{51}
\def\N{25}
\def\blas{gnu}
\def\chis{	100, 120, 140, 160, 180,
			200, 220, 240, 260, 280,
			300, 320, 340, 360, 380,
			400, 420, 440, 460, 480,
			500, 600, 700, 800, 900,
			1000, 1100, 1200, 1300, 1400, 1500, 1600, 1700, 1800, 1900, 2000}
\def\d{1em14}
\def\nph{63}
\def\dataroot{data/holstein/symmps_with_bath_sites/}%
\edef\datadir{\dataroot/t_\noexpand\t_w_\noexpand\w_g_\noexpand\g/L_\noexpand\L_N_\noexpand\N/\noexpand\blas/m_\noexpand\m_d_\noexpand\d_nph_\noexpand\nph}%
\def\m{100}
\def\g{0p5}
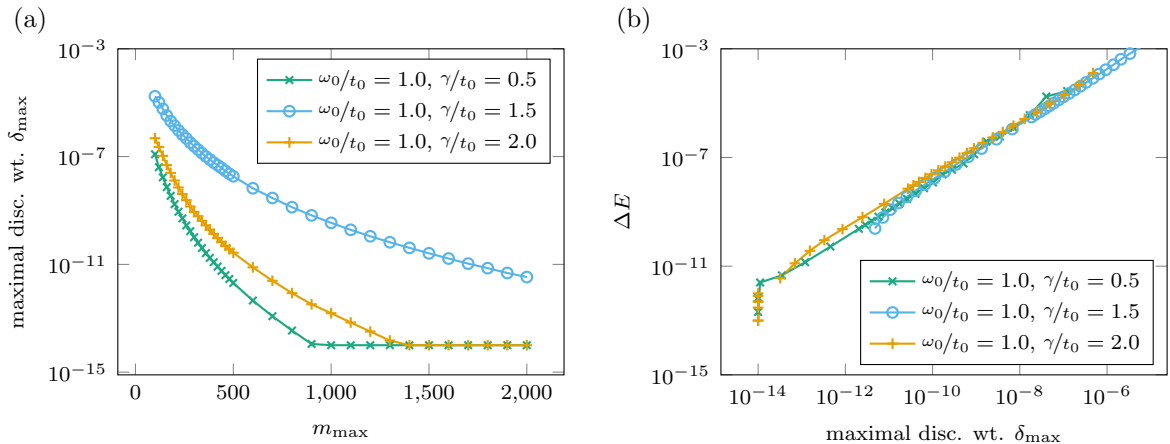
\begin{figure}[t!]
	\centering
	\tikzsetnextfilename{disc-wt-vs-chi-max_t-1p0_w-1p0}
	\begin{tikzpicture}
		\begin{groupplot}
		[
			group style = 
			{
				group size 			=	2 by 1,
				horizontal sep		=	5em,
			},
			width	= 0.475\textwidth,
			height	= 0.25\textheight,
			ymax	= 1e-03,
			ymode	= log,
			ticklabel style	= {font=\scriptsize},
			ylabel style	= {font=\scriptsize,yshift=0pt},
		]
			\nextgroupplot
			[
				title			=	{\subfloat[\label{fig:disc-wt-vs-chi-max:t-1p0:w-1p0}]{}},
				title style	= {yshift=-0.5em, xshift=-0.275\textwidth},	
				ymode	= log,
				xlabel	= {$m_{\mathrm{max}}$},
				ylabel	= {maximal disc. wt. $\delta_{\text{max}}$},
				ticklabel style	= {font=\scriptsize},
				xlabel style	= {font=\scriptsize,yshift=0pt},
				ylabel style	= {font=\scriptsize,yshift=0pt},
				legend style	= {font=\scriptsize,at={(0.97,0.97)}, anchor=north east},
			]	
				\def\g{0p5}				
				\addplot
				[
					color			= colorB,
					mark			= x,
					unbounded coords	= jump,
					thick,
				]
				table
				[
					x expr	= \thisrowno{0},
					y expr	= \thisrowno{3},
				]
				{\datadir/../gs_energies.dat};
				\addlegendentry{$\nicefrac{\omega_0}{\hopping}=1.0$, $\nicefrac{\gamma}{\hopping}=0.5$}
				\def\g{1p5}
				\addplot
				[
					color			= colorC,
					mark			= o,
					unbounded coords	= jump,
					thick,
				]
				table
				[
					x expr	= \thisrowno{0},
					y expr	= \thisrowno{3},
				]
				{\datadir/../gs_energies.dat};
				\addlegendentry{$\nicefrac{\omega_0}{\hopping}=1.0$, $\nicefrac{\gamma}{\hopping}=1.5$}
				\def\g{2p0}
				\addplot
				[
					color			= colorD,
					mark			= +,
					unbounded coords	= jump,
					thick,
				]
				table
				[
					x expr	= \thisrowno{0},
					y expr	= \thisrowno{3},
				]
				{\datadir/../gs_energies.dat};
				\addlegendentry{$\nicefrac{\omega_0}{\hopping}=1.0$, $\nicefrac{\gamma}{\hopping}=2.0$}
			\nextgroupplot
			[
				xmode	= log,
				ymode	= log,
				ymin	= 1e-15,
				xlabel	= {maximal disc. wt. $\delta_{\text{max}}$},
				ylabel	= {$\Delta E$},
				ticklabel style	= {font=\scriptsize},
				xlabel style	= {font=\scriptsize,yshift=0pt},
				ylabel style	= {font=\scriptsize,yshift=0pt},
				legend style	= {font=\scriptsize,at={(0.97,0.03)}, anchor=south east},
				title			=	{\subfloat[\label{fig:energy-dist-vs-disc-wt:t-1p0:w-1p0}]{}},
				title style	= {yshift=-0.5em, xshift=-0.275\textwidth},	
			]	
				\def\g{0p5}
				\addplot
				[
					color			= colorB,
					mark			= x,
					unbounded coords	= jump,
					thick,
				]
				table
				[
					x expr	= \thisrowno{4},
					y expr	= \thisrowno{0},
				]
				{\datadir/../gs_energies_diff.dat};
				\addlegendentry{$\nicefrac{\omega_0}{\hopping}=1.0$, $\nicefrac{\gamma}{\hopping}=0.5$}
				\def\g{1p5}
				\addplot
				[
					color			= colorC,
					mark			= o,
					unbounded coords	= jump,
					thick,
				]
				table
				[
					x expr	= \thisrowno{4},
					y expr	= \thisrowno{0},
				]
				{\datadir/../gs_energies_diff.dat};
				\addlegendentry{$\nicefrac{\omega_0}{\hopping}=1.0$, $\nicefrac{\gamma}{\hopping}=1.5$}
				\def\g{2p0}
				\addplot
				[
					color			= colorD,
					mark			= +,
					unbounded coords	= jump,
					thick,
				]
				table
				[
					x expr	= \thisrowno{4},
					y expr	= \thisrowno{0},
				]
				{\datadir/../gs_energies_diff.dat};
				\addlegendentry{$\nicefrac{\omega_0}{\hopping}=1.0$, $\nicefrac{\gamma}{\hopping}=2.0$}
			\end{groupplot}
		\end{tikzpicture}
	\caption
	{
		\label{fig:truncation:t-1p0:w-1p0}
		Truncation effects in the \gls{PP-2DMRG} method at three different points in parameter space, i.e., in the \gls{LL} phase \protect\eqref{eq:comp-pp-with-lbo:3} (green), near the phase boundary \protect\eqref{eq:comp-pp-with-lbo:2} (blue), and in the \gls{CDW} phase \protect\eqref{eq:comp-pp-with-lbo:3} (yellow).
		In \protect\subref{fig:disc-wt-vs-chi-max:t-1p0:w-1p0}, the maximal discarded weight $\delta_{\text{max}}$ as a function of the maximally allowed bond dimension $m_{\protect\rm max}$ is shown demonstrating the impact of strong fluctuations near the phase boundary on the required bond dimensions to achieve a particular accuracy.
		\protect\subref{fig:energy-dist-vs-disc-wt:t-1p0:w-1p0} shows the scaling of the relative distance $\Delta E$ to the minimally found ground\hyp state energies with the maximal discarded weight revealing that near the phase boundary, the maximally allowed number of states $m_{\protect\rm max}=2000$ is not sufficient to achieve the same numerical precision as in the \gls{LL} and \gls{CDW} phase.
	}
\end{figure}
Performing ground\hyp state searches using \gls{PP-2DMRG} allows to monitor the actual discarded weight $\delta_{\overbar j}$ on a bond $\overbar j = (j,j+1)$ after each¸ local\hyp optimization step as a function of the maximally allowed bond dimension $m_{\rm max}$.
As a consequence, an estimate for the quality of the obtained ground\hyp state energy can be constructed from the maximally found discarded weight per bond
\begin{align}
	\Delta = 2L \max_{\overbar j}\delta_{\overbar j} \equiv 2L\delta_{\rm max} \; .
\end{align}
This provides an alternative convergence test for the approximation of the ground state as we expect the difference of the approximated energy $E$ to the exact ground\hyp state energy $E_{\rm ex}$ to scale as $\frac{E-E_{\rm ex}}{E_{\rm ex}} \sim \Delta$.
For \gls{PP-2DMRG}, this is important, since the phonon correlations are encoded into the auxiliary bond between the physical\hyp{} and bath\hyp site tensors.
If the number of phonons is large and their excitation probability is broadly distributed, this implies the requirement of significantly larger auxiliary bond dimensions to keep the discarded weight below a certain threshold \cite{10.21468/SciPostPhys.10.3.058}.
The necessity to monitor the discarded weight can be illustrated by studying the system near the phase boundary where strong fluctuations in both the fermionic and phononic subsystem are present, which have to be captured by the auxiliary bonds. 
The bond dimensions between the original lattice sites are multiplied by the additional number of states required to capture the phononic degrees of freedom on the auxiliary bonds $\gamma_j$.
Thus, near the phase boundary, where we have strong fluctuations on large length scales also in the fermionic systems, the value of $m_{\rm max}$ on the auxiliary bonds between physical and bath sites has to be chosen comparably large in order to achieve a certain quality $\Delta$ of the approximated ground states.
We demonstrate this observation in \cref{fig:truncation:t-1p0:w-1p0}\subref{fig:disc-wt-vs-chi-max:t-1p0:w-1p0} (blue curve) where the chosen cutoff $m_{\rm max} = 2000$ is not sufficient to reach the desired discarded weight per bond $\delta_{\overbar j}=10^{-14}$.
However, it should be pointed out that using the restored global $U(1)$ symmetries, $m_{\rm max} = 2000$ is not fixed because of numerical limitations, but only to allow consistent comparisons.
\subsubsection{\label{sec:methods:pp-dmrg:benchmark-setup}Benchmark Setup}
For the benchmark simulations, the maximum allowed bond dimensions are varied from $m_{\rm max} = 100$ up to $m_{\rm max} = 2000$ and the number of phonons is bound by $n_{\rm ph} = 63$.
The initial state for the ground\hyp state searches are constructed as a Fermi sea of non-interacting, spinless fermions and no phonons in the physical system ($n_{P;j} \equiv 0$).
In order to achieve convergence, the maximally allowed number of sweeps for a single ground\hyp state calculation was set to $200$ (which is never reached) and the calculations were terminated once the relative fluctuations in the approximated ground\hyp state energy as a function of the number of sweeps fell below a certain threshold.
For the high-precision calculations, this threshold was set to $10^{-14}$ while in case of the finite-size extrapolation the condition to terminate the ground\hyp state search was relaxed to a threshold of $10^{-8}$.
\section{\label{sec:comparison}Comparing the Methods}
The Holstein model provides an excellent testing platform to compare the numerical behavior of the presented \gls{DMRG} methods with respect to two important problem settings: numerical high\hyp precision calculations at intermediate system sizes and finite-size extrapolations, thereby relaxing conditions on the numerical precision as a tradeoff for reaching larger system sizes.
For the high-precision calculations, we fix the system size to $L=51$ lattice sites at nearly half filling with $N=25$ fermions.
The finite-size extrapolation is performed for systems with $L=51, 101, 151, 201$ lattice sites where the fermion number is set to $N=\frac{L-1}{2}$.
Note that we always use an odd number of lattice sites.
One reason for this is the observation that in the \gls{CDW} phase, the sublattice symmetry between even and odd sites is broken and the unit cell is enlarged to consist of two lattice sites.
In a system with an even $L$, this yields two different possibilities of arranging the unit cells.
These two choices will hybridize to form the ground state.
For \gls{DMRG}, this is a delicate situation as the site tensors are optimized in a way to minimize entanglement entropy while sweeping through the system.
In such situations, an often encountered consequence is that algorithms get stuck in low-lying excited states, which minimize the entanglement entropy but not the global energy.
To circumvent this problem, we use odd system sizes, thereby fixing the arrangement of the unit cells in the \gls{CDW} phase.
This has the pleasant side effect that the \gls{CDW} order parameter actually acquires a finite value in the \gls{CDW} phase, which is not the case in finite systems with an even number of sites.
\subsection{\label{sec:comparison:gs-scaling}Convergence Analysis of Ground\hyp State Energies}
In order to compare the methods' capabilities of approximating the ground state, we perform a scaling analysis in the maximum bond dimension $m_{\rm max}$.
We calculate the relative distance of the approximated ground\hyp state energies $E_{0}(m_{\rm max}) = \braket{\psi(m_{\rm max}) | \hat H | \psi(m_{\rm max})}$ to the lowest energy found from all methods, $E_{\rm min}$,
\begin{align}
	\Delta E(m_{\rm max}) = \frac{E_{\rm min} - E_0(m_{\rm max})}{E_{\rm min}} \; .
\end{align}
For the \gls{DMRG3S+LBO} and \gls{PP-2DMRG} methods, we also systematically evaluated the variance for the found ground\hyp state approximations $\ket{\psi(m_{\rm max})}$
\begin{align}
	\operatorname{Var}[\hat H] = \braket{\psi(m_{\rm max}) | (\hat H - E_0(m_{\rm max}))^2 | \psi(m_{\rm max})} \; .
\end{align}
In practice, a systematic dependency of the variance on the ground\hyp state energy allows us to extrapolate toward the limit of infinite bond dimension $E_0(m_{\rm max}\rightarrow\infty)$ \cite{Hubig2018}.
During the numerical calculations, the maximally allowed discarded weight per bond is set to $\delta=10^{-14}$ for the \gls{DMRG3S+LBO} and \gls{PP-2DMRG} methods.
The \gls{PS-DMRG} calculations were not limited with respect to $\delta$.
Here, for the given parameter sets, the discarded weight is not a helpful quantity to decide whether a calculation converges or not and therefore, for \gls{PS-DMRG}, we focus on the bond dimension as the sole control parameter.
Having discussed each method's peculiarities in \cref{sec:methods:ps-dmrg,sec:methods:lbo-dmrg,sec:methods:pp-dmrg}, in the following, we only show one dataset per method and parameter set.
For the comparison, we pick those datasets yielding the lowest ground\hyp state energies for the largest bond dimension.
\paragraph*{\gls{LL} phase}
%
\pgfplotstableread[header=false,comment chars=!]{data/holstein/L_51_N_25_t_1p0_w_1p0_g_0p5/emin.dat}{\dat}

\pgfplotstablegetelem{0}{3}\of\dat 
\edef\Emin{\pgfplotsretval}
\pgfplotstablegetelem{0}{6}\of\dat 
\edef\code{\pgfplotsretval}

\makeatletter
\ifnum\pdf@strcmp{\code}{symmps:}=0 %
	\expandafter\@firstoftwo
\else
	\expandafter\@secondoftwo
\fi
{
	\def\code{\symmps}
	\pgfplotstablegetelem{0}{7}\of\dat 
	\edef\details{\pgfplotsretval}
	
}
{
	\def\code{DMRG3S}
	\pgfplotstablegetelem{0}{9}\of\dat 
	\edef\nph{\pgfplotsretval}
	\pgfplotstablegetelem{0}{12}\of\dat 
	\edef\dO{\pgfplotsretval}
	\def\details{$n_{\mathrm{ph}}=\nph, d_{\mathrm{o}}=\dO$}
}%
\makeatother

\edef\EminWIpOGOpV{\Emin}
\def \CodeWIpOGOpV {\code}
\def \DetailsWIpOGOpV {\details}
\begin{figure}[t!]
	\centering
	\tikzsetnextfilename{gs_energy_distance_and_variance_t-1p0_w-1p0_g-0p5}
	\begin{tikzpicture}
		\begin{groupplot}
		[
			group style = 
			{
				group size 		=	2 by 1,
				x descriptions at	=	edge bottom,
				y descriptions at	=	edge left
			},
			height			= 	0.25\textheight,
			width 			= 	0.45\textwidth-27pt,
			ymode	= log,
			ylabel	= {$\Delta E$},
			ticklabel style	= {font=\scriptsize},
			ylabel style	= {font=\scriptsize,yshift=0pt},
		]
			\nextgroupplot
			[
				axis on top,
				xlabel style	=	{yshift=0.3em},
				xlabel	= {$m_{\mathrm{max}}$},
				extra description/.code={\node[anchor=north,font=\scriptsize] at (0.5,0.99) {$\nicefrac{\omega_0}{\hopping}=1.0$, $\nicefrac{\gamma}{\hopping}=0.5$};},
				title			=	{\subfloat[\label{fig:gs-energy-distance:t-1p0:w-1p0:g-0p5}]{}},
				title style	= {yshift=-0.5em, xshift=-0.25\textwidth},	
				ticklabel style	= {font=\scriptsize},
				xlabel style	= {font=\scriptsize,yshift=0pt},
				ylabel style	= {font=\scriptsize,yshift=0pt},
				xmax		= 350,
				ymin		= 5e-14,
				ymax		= 5e-05,
			]
				\addplot
				[
					color			= colorB,
					mark			= x,
					unbounded coords	= jump,
					thick,
				]
				table
				[
					x expr	= \thisrowno{0}/3,
					y expr	= \thisrowno{1},
				]
				{data/holstein/L_51_N_25_t_1p0_w_1p0_g_0p5/ediff.symmps.gnu}; \label{pl:m-de:symmps:w-1p0:g-0p5}
				
				\addplot
				[
					color			= colorC,
					mark			= triangle,
					unbounded coords	= jump,
					thick,
				]
				table
				[
					x expr	= \thisrowno{0},
					y expr	= \thisrowno{1},
				]
				{data/holstein/L_51_N_25_t_1p0_w_1p0_g_0p5/ediff.dmrg3s.nph.23.do.10}; \label{pl:m-de:lbo:w-1p0:g-0p5:n-23:d-10}
				
				\addplot
				[
					color			= colorD,
					mark			= square,
					unbounded coords	= jump,
					thick,
				]
				table
				[
					x expr	= \thisrowno{0}/3,
					y expr	= \thisrowno{1},
				]
				{data/holstein/L_51_N_25_t_1p0_w_1p0_g_0p5/ediff.pseudo_site.nph.15}; \label{pl:m-de:ps:w-1p0:g-0p5:n-15}
			\nextgroupplot
			[
				xmode	= log,
				ymode	= log,
				xlabel	= {$\operatorname{Var}[\hat H] / E^2_0(m_{\rm max})$},
				title	= {},
				yticklabels	= {,,},
				extra description/.code={\node[anchor=north,font=\scriptsize] at (0.5,0.99) {$\nicefrac{\omega_0}{\hopping}=1.0$, $\nicefrac{\gamma}{\hopping}=0.5$};},
				title			=	{\subfloat[\label{fig:gs-var-energy:t-1p0:w-1p0:g-0p5}]{}},
				title style	= {yshift=-0.5em, xshift=-0.2\textwidth},						
				ticklabel style	= {font=\scriptsize},
				xlabel style	= {font=\scriptsize,yshift=0pt},
				ylabel style	= {font=\scriptsize,yshift=0pt},
				xmax		= 1e-06,
				ymin		= 5e-14,
				ymax		= 5e-05,
			]	
				\addplot
				[
					color			= colorB,
					mark			= x,
					unbounded coords	= jump,
					thick,
					restrict expr to domain={y}{3e-13:1e0}
				]
				table
				[
					x expr	= \thisrowno{2},
					y expr	= \thisrowno{1},
				]			{data/holstein/L_51_N_25_t_1p0_w_1p0_g_0p5/ediff.symmps.gnu}; \label{pl:var-de:symmps:w-1p0:g-0p5}
				
				\addplot
				[
					color			= colorC,
					mark			= triangle,
					unbounded coords	= jump,
					thick,
				]
				table
				[
					x expr	= \thisrowno{2},
					y expr	= \thisrowno{1},
				]
				{data/holstein/L_51_N_25_t_1p0_w_1p0_g_0p5/ediff.dmrg3s.nph.23.do.10}; \label{pl:var-de:lbo:w-1p0:g-0p5:n-23:d-10}
		\end{groupplot}
		\node[draw,fill=white,inner sep=1pt, right=1.5em, anchor=west] at (group c2r1.east)
		{
			\scriptsize
			\begin{tabular}{cc}
				\multicolumn{2}{c}{\gls{PP-2DMRG}} \\
				\multicolumn{2}{c}{\ref{pl:m-de:symmps:w-1p0:g-0p5} $m_{\rm max}/3$} \\
				\multicolumn{2}{c}{$n_{\rm ph}=63$} \\
				& \\[-1em]
				\multicolumn{2}{c}{\gls{PS-DMRG}} \\
				\multicolumn{2}{c}{\ref{pl:m-de:ps:w-1p0:g-0p5:n-15} $m_{\rm max}/3$} \\
				\multicolumn{2}{c}{$n_{\rm ph}=15$} \\
				&\\[-1em]
				\multicolumn{2}{c}{\gls{DMRG3S+LBO}}\\
				\multicolumn{2}{c}{\ref{pl:m-de:lbo:w-1p0:g-0p5:n-23:d-10} $m_{\rm max}$} \\
				$n_{\rm ph}=23$, & $d_{\rm o}=10$
			\end{tabular}
		};
	\end{tikzpicture}
	\caption
	{
		\label{fig:t-1p0:w-1p0:g-0p5}
		Relative deviations from the minimal ground\hyp state energy found \protect\subref{fig:gs-energy-distance:t-1p0:w-1p0:g-0p5} and their scaling w.r.t.~the relative variance \protect\subref{fig:gs-var-energy:t-1p0:w-1p0:g-0p5} using \gls{PS-DMRG} ($\Delta E$ only), \gls{PP-2DMRG} and \gls{DMRG3S+LBO} for the parameter set \eqref{eq:comp-pp-with-lbo:1}.
		We vary the maximal bond dimensions and for the case of \gls{PS-DMRG} and \gls{DMRG3S+LBO}, we only show data points belonging to those datasets that yield the lowest ground\hyp state energies.
		The maximal bond dimensions $m_{\rm max}$ in \protect\subref{fig:gs-energy-distance:t-1p0:w-1p0:g-0p5} are rescaled by a factor of $3$ in the case of the \gls{PS-DMRG} and \gls{PP-2DMRG} calculations.
	}
\end{figure}
\Cref{fig:t-1p0:w-1p0:g-0p5} displays the results of the scaling analysis in the \gls{LL} phase \cref{eq:comp-pp-with-lbo:1}.
The maximum bond dimension $m_{\rm max}$ used during the \gls{PS-DMRG} and \gls{PP-2DMRG} calculations are rescaled by a factor of $3$.
For the latter, this yields a dependency of the ground\hyp state energies on $m_{\rm max}$ comparable to the data obtained using \gls{DMRG3S+LBO}, as shown in \cref{fig:t-1p0:w-1p0:g-0p5}\subref{fig:gs-energy-distance:t-1p0:w-1p0:g-0p5}.
The smallest ground\hyp state energy reached is $E_{\rm min}/\hopping= \pgfmathprintnumber[std, precision=14,zerofill]{\EminWIpOGOpV}$ obtained by \gls{PP-2DMRG}
with a relative precision $\sim 10^{-13}$, which is consistent with the chosen discarded weight $\delta = 10^{-14}$.
The relative variance $\frac{\operatorname{Var}[\hat H]}{E^2_0(m_{\rm max})}$ as a function of the relative distance $\Delta E$ is shown in \cref{fig:t-1p0:w-1p0:g-0p5}\subref{fig:gs-var-energy:t-1p0:w-1p0:g-0p5}.
For both the \gls{DMRG3S+LBO} and \gls{PP-2DMRG} methods, we find a similar systematic dependence up to a very high precision.
Note that the saturation of $\frac{\operatorname{Var}[\hat H]}{E^2_0(m_{\rm max})}$ at $\Delta E(m_{\rm max}) < 10^{-12}$ is numerically consistent with the overall truncation error $\propto L \cdot \delta$ introduced by the chosen discarded weight per bond $\delta = 10^{-14}$.
\paragraph*{Phase boundary}
%
\pgfplotstableread[header=false,comment chars=!]{data/holstein/L_51_N_25_t_1p0_w_1p0_g_1p5/emin.dat}{\dat}

\pgfplotstablegetelem{0}{3}\of\dat 
\edef\Emin{\pgfplotsretval}
\pgfplotstablegetelem{0}{6}\of\dat 
\edef\code{\pgfplotsretval}

\makeatletter
\ifnum\pdf@strcmp{\code}{symmps:}=0 %
	\expandafter\@firstoftwo
\else
	\expandafter\@secondoftwo
\fi
{
	\def\code{\symmps}
	\pgfplotstablegetelem{0}{7}\of\dat 
	\edef\details{\pgfplotsretval}
	
}
{
	\def\code{DMRG3S}
	\pgfplotstablegetelem{0}{9}\of\dat 
	\edef\nph{\pgfplotsretval}
	\pgfplotstablegetelem{0}{12}\of\dat 
	\edef\dO{\pgfplotsretval}
	\def\details{$n_{\mathrm{ph}}=\nph, d_{\mathrm{o}}=\dO$}
}%
\makeatother

\edef\EminWIpOGIpV{\Emin}
\def \CodeWIpOGIpV {\code}
\def \DetailsWIpOGIpV {\details}
\begin{figure}[t!]
	\centering
	\tikzsetnextfilename{gs_energy_distance_and_variance_t-1p0_w-1p0_g-1p5}
	\begin{tikzpicture}
		\begin{groupplot}
		[
			group style = 
			{
				group size 		=	2 by 1,
				x descriptions at	=	edge bottom,
				y descriptions at	=	edge left
			},
			height			= 	0.25\textheight,
			width 			= 	0.45\textwidth-27pt,
			ymode	= log,
			ylabel	= {$\Delta E$},
			ticklabel style	= {font=\scriptsize},
			ylabel style	= {font=\scriptsize,yshift=0pt},
		]
			\nextgroupplot
			[
				axis on top,
				xlabel style	=	{yshift=0.3em},
				xlabel	= {$m_{\mathrm{max}}$},
				extra description/.code={\node[anchor=north,font=\scriptsize] at (0.5,0.99) {$\nicefrac{\omega_0}{\hopping}=1.0$, $\nicefrac{\gamma}{\hopping}=1.5$};},
				title			=	{\subfloat[\label{fig:gs-energy-distance:t-1p0:w-1p0:g-1p5}]{}},
				title style	= {yshift=-0.5em, xshift=-0.25\textwidth},	
				ticklabel style	= {font=\scriptsize},
				xlabel style	= {font=\scriptsize,yshift=0pt},
				ylabel style	= {font=\scriptsize,yshift=0pt},
				xmax	= 700,
				xmin	= 10,
				ymin	= 5e-13,
			]
				\addplot
				[
					color			= colorB,
					mark			= x,
					unbounded coords	= jump,
					thick,
				]
				table
				[
					x expr	= \thisrowno{0}/5,
					y expr	= \thisrowno{1},
				]
				{data/holstein/L_51_N_25_t_1p0_w_1p0_g_1p5/ediff.symmps.mkl}; \label{pl:m-de:symmps:w-1p0:g-1p5}
				
				\addplot
				[
					color			= colorC,
					mark			= triangle,
					unbounded coords	= jump,
					thick,
				]
				table
				[
					x expr	= \thisrowno{0},
					y expr	= \thisrowno{1},
				]
				{data/holstein/L_51_N_25_t_1p0_w_1p0_g_1p5/ediff.dmrg3s.nph.63.do.10}; \label{pl:m-de:lbo:w-1p0:g-1p5:n-63:d-10}
				
				\addplot
				[
					color			= colorD,
					mark			= square,
					unbounded coords	= jump,
					thick,
				]
				table
				[
					x expr	= \thisrowno{0}/1.5,
					y expr	= \thisrowno{1},
				]
				{data/holstein/L_51_N_25_t_1p0_w_1p0_g_1p5/ediff.pseudo_site.nph.15}; \label{pl:m-de:ps:w-1p0:g-1p5:n-15}
			\nextgroupplot
			[
				xlabel	= {$\operatorname{Var}[\hat H] / E^2_0(m_{\rm max})$},
				extra description/.code={\node[anchor=north,font=\scriptsize] at (0.5,0.99) {$\nicefrac{\omega_0}{\hopping}=1.0$, $\nicefrac{\gamma}{\hopping}=1.5$};},
				title			=	{\subfloat[\label{fig:gs-var-energy:t-1p0:w-1p0:g-1p5}]{}},
				title style	= {yshift=-0.5em, xshift=-0.2\textwidth},	
				ticklabel style	= {font=\scriptsize},
				xlabel style	= {font=\scriptsize,yshift=0pt},
				ylabel style	= {font=\scriptsize,yshift=0pt},
				xmax	= 1e-05,
				ymin	= 5e-13,
				xmode	= log
			]	
				\addplot
				[
					color			= colorB,
					mark			= x,
					unbounded coords	= jump,
					thick,
				]
				table
				[
					x expr	= \thisrowno{2},
					y expr	= \thisrowno{1},
				]
				{data/holstein/L_51_N_25_t_1p0_w_1p0_g_1p5/ediff.symmps.mkl}; \label{pl:var-de:symmps:w-1p0:g-1p5}
				
				\addplot
				[
					color			= colorC,
					mark			= triangle,
					unbounded coords	= jump,
					thick,
				]
				table
				[
					x expr	= \thisrowno{2},
					y expr	= \thisrowno{1},
				]
				{data/holstein/L_51_N_25_t_1p0_w_1p0_g_1p5/ediff.dmrg3s.nph.63.do.10}; \label{pl:var-de:lbo:w-1p0:g-1p5:n-63:d-10}
		\end{groupplot}
		\node[draw,fill=white,inner sep=1pt,right=1.5em, anchor=west] at (group c2r1.east)
		{
				\scriptsize
				\begin{tabular}{cc}
					\multicolumn{2}{c}{\gls{PP-2DMRG}} \\
					\multicolumn{2}{c}{\ref{pl:m-de:symmps:w-1p0:g-1p5} $m_{\rm max}/5$} \\
					\multicolumn{2}{c}{$n_{\rm ph}=63$} \\
					& \\[-1em]
					\multicolumn{2}{c}{\gls{PS-DMRG}} \\
					\multicolumn{2}{c}{\ref{pl:m-de:ps:w-1p0:g-1p5:n-15} $m_{\rm max}/1.5$} \\
					\multicolumn{2}{c}{$n_{\rm ph}=15$} \\
					&\\[-1em]
					\multicolumn{2}{c}{\gls{DMRG3S+LBO}}\\
					\multicolumn{2}{c}{\ref{pl:m-de:lbo:w-1p0:g-1p5:n-63:d-10} $m_{\rm max}$} \\
					$n_{\rm ph}=63$, & $d_{\rm o}=10$
				\end{tabular}
		};
	\end{tikzpicture}
	\caption
	{
		\label{fig:t-1p0:w-1p0:g-1p5}
		Relative deviations from the minimal ground\hyp state energy found \protect\subref{fig:gs-energy-distance:t-1p0:w-1p0:g-1p5} and their scaling w.r.t.~the relative variance \protect\subref{fig:gs-var-energy:t-1p0:w-1p0:g-1p5} using \gls{PS-DMRG} ($\Delta E$ only), \gls{PP-2DMRG} and \gls{DMRG3S+LBO} for the parameter set \eqref{eq:comp-pp-with-lbo:2}.
		We vary the maximal bond dimensions and for the case of \gls{PS-DMRG} and \gls{DMRG3S+LBO}, we only show data points belonging to those datasets that yield the lowest ground\hyp state energies.
		The maximal bond dimensions $m_{\rm max}$ in \protect\subref{fig:gs-energy-distance:t-1p0:w-1p0:g-1p5} are rescaled by a factor of $1.5$ in the case of the \gls{PS-DMRG} calculations and a factor of $5$ in the case of the \gls{PP-2DMRG} calculations.
	}
\end{figure}
\Cref{fig:t-1p0:w-1p0:g-1p5} shows the results near the phase boundary between the \gls{LL} and \gls{CDW} phase.
The scaling of $\Delta E(m_{\rm max})$ displayed in \cref{fig:t-1p0:w-1p0:g-1p5}\subref{fig:gs-energy-distance:t-1p0:w-1p0:g-1p5} reveals that the appearance of heavy polarons and strong fluctuations in the fermion system require a much larger bond dimension to approximate the ground state with a high precision.
This is reflected by a rescaling of $m_{\rm max}$ with a factor of $1.5$ for the \gls{PS-DMRG} method and a factor of $5$ for \gls{PP-2DMRG} in order to achieve a convergence behavior similar to the one observed in \gls{DMRG3S+LBO}.
Here, the ground\hyp state approximations obtained using \gls{DMRG3S+LBO} yield the smallest energy, which is given by $E_{\rm min}/\hopping= \pgfmathprintnumber[std, precision=12,zerofill]{\EminWIpOGIpV}$
with a relative precision of $\lesssim 10^{-11}$ as can be seen by the variance displayed in \cref{fig:t-1p0:w-1p0:g-1p5}\subref{fig:gs-var-energy:t-1p0:w-1p0:g-1p5}.
\paragraph*{\gls{CDW} phase}
%
\pgfplotstableread[header=false,comment chars=!]{data/holstein/L_51_N_25_t_1p0_w_1p0_g_2p0/emin.dat}{\dat}

\pgfplotstablegetelem{0}{3}\of\dat 
\edef\Emin{\pgfplotsretval}
\pgfplotstablegetelem{0}{6}\of\dat 
\edef\code{\pgfplotsretval}

\makeatletter
\ifnum\pdf@strcmp{\code}{symmps:}=0 %
	\expandafter\@firstoftwo
\else
	\expandafter\@secondoftwo
\fi
{
	\def\code{\symmps}
	\pgfplotstablegetelem{0}{7}\of\dat 
	\edef\details{\pgfplotsretval}
	
}
{
	\def\code{DMRG3S}
	\pgfplotstablegetelem{0}{9}\of\dat 
	\edef\nph{\pgfplotsretval}
	\pgfplotstablegetelem{0}{12}\of\dat 
	\edef\dO{\pgfplotsretval}
	\def\details{$n_{\mathrm{ph}}=\nph, d_{\mathrm{o}}=\dO$}
}%
\makeatother

\edef\EminWIpOGIIpO{\Emin}
\def \CodeWIpOGIIpO {\code}
\def \DetailsWIpOGIIpO {\details}
\begin{figure}[t!]
	\tikzsetnextfilename{gs_energy_distance_and_variance_t-1p0_w-1p0_g-2p0}
	\begin{tikzpicture}
		\begin{groupplot}
		[
			group style = 
			{
				group size 		=	2 by 1,
				x descriptions at	=	edge bottom,
				y descriptions at	=	edge left
			},
			height			= 	0.25\textheight,
			width 			= 	0.45\textwidth-27pt,
			ymode	= log,
			ylabel	= {$\Delta E$},
			ticklabel style	= {font=\scriptsize},
			ylabel style	= {font=\scriptsize,yshift=0pt},
		]
			\nextgroupplot
			[
				axis on top,
				xlabel style	=	{yshift=0.3em},
				xlabel	= {$m_{\mathrm{max}}$},
				extra description/.code={\node[anchor=north,font=\scriptsize] at (0.5,0.99) {$\nicefrac{\omega_0}{\hopping}=1.0$, $\nicefrac{\gamma}{\hopping}=2.0$};},
				title			=	{\subfloat[\label{fig:gs-energy-distance:t-1p0:w-1p0:g-2p0}]{}},
				title style	= {yshift=-0.5em, xshift=-0.25\textwidth},	
				ticklabel style	= {font=\scriptsize},
				xlabel style	= {font=\scriptsize,yshift=0pt},
				ylabel style	= {font=\scriptsize,yshift=0pt},
				xmax	= 175,
				xmin	= 10,
				ymin	= 5e-14,
			]
				\addplot
				[
					color			= colorB,
					mark			= x,
					unbounded coords	= jump,
					thick,
				]
				table
				[
					x expr	= \thisrowno{0}/7.5,
					y expr	= \thisrowno{1},
				]
				{data/holstein/L_51_N_25_t_1p0_w_1p0_g_2p0/ediff.symmps.gnu}; \label{pl:m-de:symmps:w-1p0:g-2p0}
				
				\addplot
				[
					color			= colorC,
					mark			= triangle,
					unbounded coords	= jump,
					thick,
				]
				table
				[
					x expr	= \thisrowno{0},
					y expr	= \thisrowno{1},
				]
				{data/holstein/L_51_N_25_t_1p0_w_1p0_g_2p0/ediff.dmrg3s.nph.63.do.10}; \label{pl:m-de:lbo:w-1p0:g-2p0:n-63:d-10}
				
				\addplot
				[
					color			= colorD,
					mark			= square,
					unbounded coords	= jump,
					thick,
				]
				table
				[
					x expr	= \thisrowno{0}/4.5,
					y expr	= \thisrowno{1},
				]
				{data/holstein/L_51_N_25_t_1p0_w_1p0_g_2p0/ediff.pseudo_site.nph.63}; \label{pl:m-de:ps:w-1p0:g-2p0:n-63}
			\nextgroupplot
			[
				xlabel	= {$\operatorname{Var}[\hat H] / E^2_0(m_{\rm max})$},
				extra description/.code={\node[anchor=north,font=\scriptsize] at (0.5,0.99) {$\nicefrac{\omega_0}{\hopping}=1.0$, $\nicefrac{\gamma}{\hopping}=2.0$};},
				title			=	{\subfloat[\label{fig:gs-var-energy:t-1p0:w-1p0:g-2p0}]{}},
				title style	= {yshift=-0.5em, xshift=-0.2\textwidth},	
				ticklabel style	= {font=\scriptsize},
				xlabel style	= {font=\scriptsize,yshift=0pt},
				ylabel style	= {font=\scriptsize,yshift=0pt},
				xmax	= 1e-06,
				ymin	= 5e-14,
				xmode	= log
			]
				\addplot
				[
					color			= colorB,
					mark			= x,
					unbounded coords	= jump,
					thick,
				]
				table
				[
					x expr	= \thisrowno{2},
					y expr	= \thisrowno{1},
				]
				{data/holstein/L_51_N_25_t_1p0_w_1p0_g_2p0/ediff.symmps.gnu}; \label{pl:var-de:symmps:w-1p0:g-2p0}
				
				\addplot
				[
					color			= colorC,
					mark			= triangle,
					unbounded coords	= jump,
					thick,
				]
				table
				[
					x expr	= \thisrowno{2},
					y expr	= \thisrowno{1},
				]
				{data/holstein/L_51_N_25_t_1p0_w_1p0_g_2p0/ediff.dmrg3s.nph.63.do.10}; \label{pl:var-de:lbo:w-1p0:g-2p0:n-63:d-10}
		\end{groupplot}
		\node[draw,fill=white,inner sep=1pt,right=1.5em, anchor=west] at (group c2r1.east)
		{
			\scriptsize
			\begin{tabular}{cc}
				\multicolumn{2}{c}{\gls{PP-2DMRG}} \\
				\multicolumn{2}{c}{\ref{pl:m-de:symmps:w-1p0:g-2p0} $m_{\rm max}/7.5$} \\
				\multicolumn{2}{c}{$n_{\rm ph}=63$} \\
				& \\[-1em]
				\multicolumn{2}{c}{\gls{PS-DMRG}} \\
				\multicolumn{2}{c}{\ref{pl:m-de:ps:w-1p0:g-2p0:n-63} $m_{\rm max}/4$} \\
				\multicolumn{2}{c}{$n_{\rm ph}=63$} \\
				&\\[-1em]
				\multicolumn{2}{c}{\gls{DMRG3S+LBO}}\\
				\multicolumn{2}{c}{\ref{pl:m-de:lbo:w-1p0:g-2p0:n-63:d-10} $m_{\rm max}$,} \\
				$n_{\rm ph}=63$ & $d_{\rm o}=10$
			\end{tabular}
		};
	\end{tikzpicture}
	\caption
	{
		\label{fig:t-1p0:w-1p0:g-2p0}
		Relative deviations from the minimal ground\hyp state energy found \protect\subref{fig:gs-energy-distance:t-1p0:w-1p0:g-2p0} and their scaling w.r.t.~the relative variance \protect\subref{fig:gs-var-energy:t-1p0:w-1p0:g-2p0} using \gls{PS-DMRG} ($\Delta E$ only), \gls{PP-2DMRG} and \gls{DMRG3S+LBO} for the parameter set \eqref{eq:comp-pp-with-lbo:3}.
		We vary the maximal bond dimensions and for the case of \gls{PS-DMRG} and \gls{DMRG3S+LBO}, we only show data points belonging to those datasets that yield the lowest ground\hyp state energies.
		The maximal bond dimensions $m_{\rm max}$ in \protect\subref{fig:gs-energy-distance:t-1p0:w-1p0:g-2p0} are rescaled by a factor of $4$ in the case of the \gls{PS-DMRG} calculations and a factor of $7.5$ in the case of the \gls{PP-2DMRG} calculations.
	}
\end{figure}
In \cref{fig:t-1p0:w-1p0:g-2p0}, the scaling analysis is shown for the parameter set \eqref{eq:comp-pp-with-lbo:3} for which the system is in the \gls{CDW} phase.
All tested methods are capable of faithfully representing the ground state with a very high precision and comparably small bond dimensions with the lowest energy given by $E_{\rm min}/\hopping= \pgfmathprintnumber[std, precision=14,zerofill]{\EminWIpOGIIpO}$ with a relative precision of $10^{-14}$ using \gls{PP-2DMRG}.
For the \gls{PS-DMRG} method, we rescaled the maximally allowed bond dimension by a factor of $4$ while a factor of $7.5$ was required for the \gls{PP-2DMRG} method in order to achieve a similar scaling behavior of $\Delta E$ for the three methods.
This large rescaling factor is based on the broadly distributed excitation probabilities of the phonons requiring a larger local Hilbert\hyp space dimension.
Here, for all three methods tested, we use $N_{\rm ph}=63$ to achieve the best results.
\subsection{\label{sec:comparison:loc-observables}Local Observables - Oscillator Displacement}
%
\begin{figure}[t!]
	\centering
	\tikzsetnextfilename{gs_oscillator_displacement_t-1p0_w-1p0_g-1p5}
	\begin{tikzpicture}
		\begin{groupplot}
		[
			group style = 
			{
				group size 		=	2 by 1,
				horizontal sep		=	3em,
			},
			width	= 0.41\textwidth,
			height	= 0.25\textheight,
			ticklabel style	= {font=\scriptsize},
			ylabel style	= {font=\scriptsize,yshift=0pt},
			ylabel	= {$x_{\rm ph}(j)$},
		]
			\nextgroupplot
			[
				xlabel	= {even site $j$},
				extra description/.code={\node[anchor=north,font=\scriptsize] at (0.5,0.99) {$\nicefrac{\omega_0}{\hopping}=1.0$, $\nicefrac{\gamma}{\hopping}=1.5$};},
				title		= {\subfloat[\label{fig:t-1p0:w-1p0:g-1p5:ph-elong:even}]{}},
				title style	= {yshift=-0.5em, xshift=-0.225\textwidth},	
				ticklabel style	= {font=\scriptsize},
				xlabel style	= {font=\scriptsize,yshift=0pt},
				ylabel style	= {font=\scriptsize,yshift=0pt},
				ymin		= 1.05,
				ymax		= 1.48,
			]	
				\addplot
				[
					color			= colorB,
					mark			= x,
					unbounded coords	= jump,
					each nth point		= 2,
					opacity			= 0.5,
					only marks,
					thick,
				]
				table
				[
					x expr	= \coordindex,
					y expr	= -\thisrowno{35},
				]
				{data/holstein/L_51_N_25_t_1p0_w_1p0_g_1p5/ph-elong.real.symmps.mkl}; \label{pl:ph-elong:symmps:w-1p0:g-1p5:m-2000}

				\addplot
				[
					color			= colorB,
					mark			= none,
					unbounded coords	= jump,
					each nth point		= 2,
					opacity			= 0.5,
					thick,
				]
				table
				[
					x expr	= \coordindex,
					y expr	= \thisrowno{2},
				]
				{data/holstein/mps_with_lbo/L_51_N_25/t_1p0_w_1p0_g_1p5/densities.bdim-440.nph-63.domax-10}; \label{pl:ph-elong:lbo:w-1p0:g-1p5:m-440}
				
				\addplot
				[
					color			= colorB,
					mark			= o,
					unbounded coords	= jump,
					each nth point		= 2,
					only marks,
				]
				table
				[
					x expr	= \coordindex,
					y expr	= \thisrowno{5},
				]
				{data/holstein/pseudo_site_dmrg/phononqw1gamma15b32l51m200.dat}; \label{pl:ph-elong:ps:w-1p0:g-1p5:m-200}
				
				\coordinate (inset_position_A) at (axis cs:30,1.1);
			\nextgroupplot
			[
				xlabel	= {odd site $j$},
				extra description/.code={\node[anchor=north,font=\scriptsize] at (0.5,0.99) {$\nicefrac{\omega_0}{\hopping}=1.0$, $\nicefrac{\gamma}{\hopping}=1.5$};},
				title		= {\subfloat[\label{fig:t-1p0:w-1p0:g-1p5:ph-elong:odd}]{}},
				title style	= {yshift=-0.5em, xshift=-0.225\textwidth},	
				ticklabel style	= {font=\scriptsize},
				xlabel style	= {font=\scriptsize,yshift=0pt},
				ylabel style	= {font=\scriptsize,yshift=0pt},
				ymin		= 1.25,
				ymax		= 1.70,
			]					
				\addplot
				[
					color			= colorB,
					mark			= x,
					unbounded coords	= jump,
					each nth point**	= {2}{1},
					opacity			= 0.5,
					only marks,
					thick,
				]
				table
				[
					x expr	= \coordindex,
					y expr	= -\thisrowno{35},
				]
				{data/holstein/L_51_N_25_t_1p0_w_1p0_g_1p5/ph-elong.real.symmps.mkl};

				\addplot
				[
					color			= colorB,
					mark			= none,
					unbounded coords	= jump,
					each nth point**	= {2}{1},
					opacity			= 0.5,
					thick,
				]
				table
				[
					x expr	= \coordindex,
					y expr	= \thisrowno{2},
				]
				{data/holstein/mps_with_lbo/L_51_N_25/t_1p0_w_1p0_g_1p5/densities.bdim-440.nph-63.domax-10};

				\addplot
				[
					color			= colorB,
					mark			= o,
					unbounded coords	= jump,
					each nth point**	= {2}{1},
					only marks,
				]
				table
				[
					x expr	= \coordindex,
					y expr	= \thisrowno{5},
				]
				{data/holstein/pseudo_site_dmrg/phononqw1gamma15b32l51m200.dat};
				
				\coordinate (inset_position_B) at (axis cs:30,1.31);
			\end{groupplot}
			
			\begin{axis}
			[
				at	= {(inset_position_A)},
				name	= inset_even,
				anchor	= south,
				width	= 0.3\textwidth,
				height	= 0.15\textheight,
				title style	= {font=\tiny},
				ticklabel style	= {font=\tiny},
				scaled y ticks	= false,
				xlabel style	= {font=\tiny,yshift=0pt},
				ylabel style	= {font=\tiny,yshift=-25pt},
				ylabel	= {$ \Delta x_{\rm ph} $},
			]
				\addplot
				[
					color		= colorF,
					each nth point	= 2,
					mark		= square,
					mark repeat	= 5,
				]
				table
				[
					x expr	= \coordindex,
					y expr	= \thisrowno{0}+\thisrowno{1},
				]
				{data/holstein/L_51_N_25_t_1p0_w_1p0_g_1p5/ph-elong.max-bdims.all}; \label{pl:ph-elong:diff:w-1p0:g-1p5:lbo-pp:even}
				\node [draw=none, fill=none] at ({axis cs: 40,-1e-6}) {\tiny LBO-PP};
				\addplot
				[
					color		= colorF,
					each nth point	= 2,
					mark		= +,
					mark repeat	= 5,
				]
				table
				[
					x expr	= \coordindex,
					y expr	= \thisrowno{0}-\thisrowno{2},
				]
				{data/holstein/L_51_N_25_t_1p0_w_1p0_g_1p5/ph-elong.max-bdims.all}; \label{pl:ph-elong:diff:w-1p0:g-1p5:lbo-ps:even}
				\node [draw=none, fill=none] at ({axis cs: 45,4.e-6}) {\tiny LBO-PS};
				\addplot
				[
					color		= colorF,
					each nth point	= 2,
					mark		= triangle,
					mark repeat	= 5,
				]
				table
				[
					x expr	= \coordindex,
					y expr	= \thisrowno{1}+\thisrowno{2},
				]
				{data/holstein/L_51_N_25_t_1p0_w_1p0_g_1p5/ph-elong.max-bdims.all}; \label{pl:ph-elong:diff:w-1p0:g-1p5:pp-ps:even}
				\node [draw=none, fill=none] at ({axis cs: 35,-5.5e-6}) {\tiny PP-PS};
			\end{axis}
			\begin{axis}
			[
				at	= {(inset_position_B)},
				name	= inset_odd,
				anchor	= south,
				width	= 0.3\textwidth,
				height	= 0.15\textheight,
				title style	= {font=\tiny},
				ticklabel style	= {font=\tiny},
				scaled y ticks	= false,
				xlabel style	= {font=\tiny,yshift=0pt},
				ylabel style	= {font=\tiny,yshift=-25pt},
				ylabel	= {$ \Delta x_{\rm ph} $},
				ytick	= {-4e-6, 0, 4e-6},
			]
				\addplot
				[
					color		= colorF,
					each nth point**= {2}{1},
					mark		= square,
					mark repeat	= 5,
				]
				table
				[
					x expr	= \coordindex,
					y expr	= \thisrowno{0}+\thisrowno{1},
				]
				{data/holstein/L_51_N_25_t_1p0_w_1p0_g_1p5/ph-elong.max-bdims.all}; \label{pl:ph-elong:diff:w-1p0:g-1p5:lbo-pp:odd}
				\node [draw=none, fill=none] at ({axis cs: 25,0e-6}) {\tiny LBO-PP};
				\addplot
				[
					color		= colorF,
					each nth point**= {2}{1},
					mark		= +,
					mark repeat	= 5,
				]
				table
				[
					x expr	= \coordindex,
					y expr	= \thisrowno{0}-\thisrowno{2},
				]
				{data/holstein/L_51_N_25_t_1p0_w_1p0_g_1p5/ph-elong.max-bdims.all}; \label{pl:ph-elong:diff:w-1p0:g-1p5:lbo-ps:odd}
				\node [draw=none, fill=none] at ({axis cs: 8,-4e-6}) {\tiny LBO-PS};
				\addplot
				[
					color		= colorF,
					each nth point**= {2}{1},
					mark		= triangle,
					mark repeat	= 5,
				]
				table
				[
					x expr	= \coordindex,
					y expr	= \thisrowno{1}+\thisrowno{2},
				]
				{data/holstein/L_51_N_25_t_1p0_w_1p0_g_1p5/ph-elong.max-bdims.all}; \label{pl:ph-elong:diff:w-1p0:g-1p5:pp-ps:odd})
				\node [draw=none, fill=none] at ({axis cs: 13,3.5e-6}) {\tiny PP-PS};
			\end{axis}
			\node[draw,fill=white,inner sep=1pt,right=1.5em, anchor=west] at (group c2r1.east)
			{
				\scriptsize
				\begin{tabular}{c}
				\gls{PP-2DMRG} \\
				\ref{pl:ph-elong:symmps:w-1p0:g-1p5:m-2000} $m_{\rm max}=2000$ \\
				\\[-1em]
				\gls{DMRG3S+LBO} \\
				\ref{pl:ph-elong:lbo:w-1p0:g-1p5:m-440} $m_{\rm max}=440$ \\
				\\[-1em]
				\gls{PS-DMRG} \\
				\ref{pl:ph-elong:ps:w-1p0:g-1p5:m-200} $m_{\rm max}=200$
				\end{tabular}
			};
		\end{tikzpicture}
	\caption
	{
		\label{fig:t-1p0:w-1p0:g-1p5:ph-elong}
		Oscillator displacement \protect{$x_{\rm ph}(j)=\braket{\hat b^\dagger_j + \hat b^{\protect\nodagger}_j}$} in the ground state near the phase boundary.
		The data shown here is taken from the calculations with the largest bond dimension for each method.
		For convenience, even\hyp \protect\subref{fig:t-1p0:w-1p0:g-1p5:ph-elong:even} and odd\hyp site \protect\subref{fig:t-1p0:w-1p0:g-1p5:ph-elong:odd} expectation values are plotted separately.
		Insets display differences of the calculated oscillator displacements between the different methods.
	}
\end{figure}
Having studied the convergence of the ground\hyp state approximation for the different methods, we turn to physical observables next.
In the Holstein model, the \gls{CDW} phase is characterized by the formation of lattice distortions with a period of twice the lattice constant, i.e., the translation symmetry of the lattice is broken into two translational invariant sublattices $\mathcal A/ \mathcal B$ with even/odd lattice sites.
We evaluate the oscillator displacement of the $j$th lattice site for different maximum bond dimensions $m_{\rm max}$ near the phase boundary \cref{eq:comp-pp-with-lbo:2}
\begin{align}
	x_{\rm ph}(j) = \braket{\hat b^\dagger_j + \hat b^\nodagger_j} \; ,
\end{align}
breaking translational symmetry.
In \cref{fig:t-1p0:w-1p0:g-1p5:ph-elong}, the results are shown for $\nicefrac{\omega_0}{\hopping}=1.0$ and $\nicefrac{\gamma}{\hopping}=1.5$.
We separately plot even\hyp (\cref{fig:t-1p0:w-1p0:g-1p5:ph-elong}\subref{fig:t-1p0:w-1p0:g-1p5:ph-elong:even}) and odd\hyp site (\cref{fig:t-1p0:w-1p0:g-1p5:ph-elong}\subref{fig:t-1p0:w-1p0:g-1p5:ph-elong:odd}) expectation values to illustrate the behavior of $x_{\rm ph}(j)$ in the two sublattices.
All methods yield the same behavior, i.e., we find developing \gls{CDW} modulations in the two sublattices caused by finite displacements with larger values $x_{\rm ph}(j)$ for $j\in \mathcal B$ compared to those $x_{\rm ph}(j)$ for $j\in \mathcal A$.
In order to visualize the numerical deviations, in the insets, we plot the difference $\Delta x_{\rm ph}(j)$ between the displacements obtained from the datasets with the largest $m_{\rm max}$ for each method, as they exhibit the highest precision ($m_{\rm max}=2000$ for \gls{PP-2DMRG}, $m_{\rm max}=440$ for \gls{DMRG3S+LBO} and $m_{\rm max}=200$ for \gls{PS-DMRG}).
In the analyzed data, the deviations $\Delta x_{\rm ph}(j)$ rapidly decrease and, for the largest bond dimensions, are located around $10^{-6}$ as shown in \cref{fig:t-1p0:w-1p0:g-1p5:ph-elong}.
\subsection{\label{sec:comparison:finite-site-extrapolation}Finite-Size Extrapolation}
Practically, one is often interested in a finite-size extrapolation of intensive quantities such as the energy density $\frac{E_0}{L}$ or the order parameter
\begin{align}
	\mathcal{O}_{\rm disp}(L) = \frac{1}{L} \sum_{j} (-1)^j x_{\rm ph}(j) \; ,
\end{align}
where the extrapolation towards $L \rightarrow \infty$ allows us to approach the thermodynamic limit.
Here, we perform a scaling analysis of the ground\hyp state energy density $\epsilon(L) = \frac{E_0}{L}$  in units of $\hopping \equiv 1$ and $\mathcal{O}_{\rm disp}(L)$.
The extrapolations are done using fitting functions
\begin{align}
	\epsilon(L) &= \frac{A_\epsilon}{L} + \epsilon_{\infty} \\
	\mathcal{O}_{\rm disp}(L) &= \frac{A_\mathcal{O}}{L} + \frac{B_\mathcal{O}}{L^ 2} + \mathcal{O}_{\rm disp,\infty} \; .
\end{align}
For the order parameter, we add a contribution $\propto \frac{1}{L^2}$ to account for boundary effects, which are important for the parameter sets \eqref{eq:comp-pp-with-lbo:1} and \eqref{eq:comp-pp-with-lbo:2}.
\paragraph*{Energy density}
\def\Ls{51, 101, 151, 201}
\def\Ls{51, 101, 151, 201, 251, 301, 401, 501, 601, 701, 801, 901, 1001}
\def\dws{1em04, 1em06, 1em08, 1em10}
\def\dw{1em10}
\def\t{1p0}
\def\w{1p0}
\def\g{1p5}
\def\blas{mkl}
\def\m{2000}
\def\nph{63}
\def\nphlbo{31}
\def\xmininset{1e-10}%
\def\xmaxinset{2e-4}%
\def\xminMain{9e-4}%
\def\xmin{1001}%
\def\xmax{51}%
\def\ocgname{pp}%
\def\legendPosition{(0.03,0.03)}%
\def\dataroot{data/holstein/symmps_with_bath_sites/}
\edef\datadir{\dataroot/t_\noexpand\t_w_\noexpand\w_g_\noexpand\g/finite_size_scaling/L_\noexpand\L_N_\noexpand\N/\noexpand\blas/m_\noexpand\m_d_\noexpand\d_nph_\noexpand\nph}
\edef\dstdir{\dataroot/t_\noexpand\t_w_\noexpand\w_g_\noexpand\g/finite_size_scaling}
\setboolean{rebuildData}{false}%
\def\captionContent{%
	\label{fig:ppdmrg:finite-size-scaling:t-\t_w-\w_g-\g}
	\protect\subref{fig:ppdmrg:finite-size-scaling:t-\t_w-\w_g-\g:energy} Finite-size scaling of the ground\hyp state energy of the Holstein model.
	\protect\subref{fig:ppdmrg:finite-size-scaling:t-\t_w-\w_g-\g:order-parameter} Finite-size scaling for the \gls{CDW} order parameter of the Holstein model.
	The model is evaluated for parameters $\nicefrac{\omega_0}{\hopping}=\protect\ReplaceStr{\w}$, $\nicefrac{\gamma}{\hopping}=\protect\ReplaceStr{\g}$ at nearly half filling where we set $N=\frac{L-1}{2}$.
}%
\setboolean{removeData}{false}%
\def\dataroot{data/holstein/symmps_with_bath_sites/}%
\edef\datadir{\dataroot/t_\noexpand\t_w_\noexpand\w_g_\noexpand\g/finite_size_scaling/L_\noexpand\L_N_\noexpand\N/\noexpand\blas/m_\noexpand\m_d_\noexpand\d_nph_\noexpand\nph}%
\edef\dstdir{\dataroot/t_\noexpand\t_w_\noexpand\w_g_\noexpand\g/finite_size_scaling}%
\ifthenelse{\boolean{rebuildData}}%
{%
}{}%
\setboolean{rebuildCurrentBlock}{false}%
\setboolean{buildCurrentBlockInline}{false}%
\ifthenelse{\boolean{removeData}}%
{%
	\immediate\write18{rm -f data/holstein/finite_size_scaling/t_\t_w_\w_g_\g*gs_energy_fit}%
	\immediate\write18{rm -f data/holstein/finite_size_scaling/t_\t_w_\w_g_\g*order_parameter_fit}%
	\immediate\write18{rm -f \dstdir/scaling_output.dat}%
	\setboolean{rebuildData}{true}%
}{}%
\setboolean{rebuildCurrentBlock}{false}%
\setboolean{buildCurrentBlockInline}{false}%
\begin{figure}[!t]
	\centering
	\ifthenelse{\boolean{rebuildCurrentBlock}}%
	{%
		\tikzset{external/remake next=true}%
		\pgfplotstableread[header=false]{\dstdir/scaling.dat}{\DataCPU}%
		\pgfplotstablegetrowsof{\DataCPU}%
		\pgfmathtruncatemacro{\lastrowoftable}{int(\pgfplotsretval-1)}%
		\pgfplotstablegetelem{\lastrowoftable}{0}\of\DataCPU%
		\xdef\Lmaxforfit{\pgfplotsretval}%
	}%
	{%
		\ifthenelse{\boolean{buildCurrentBlockInline}}%
		{%
			\tikzset{external/export next=false}%
			\pgfplotstableread[header=false]{\dstdir/scaling.dat}{\DataCPU}%
			\pgfplotstablegetrowsof{\DataCPU}%
			\pgfmathtruncatemacro{\lastrowoftable}{int(\pgfplotsretval-1)}%
			\pgfplotstablegetelem{\lastrowoftable}{0}\of\DataCPU%
			\xdef\Lmaxforfit{\pgfplotsretval}%
		}{}%
	}%
	\tikzsetnextfilename{PPDMRG-finite-size-scalings-t_\t-w_\w-g_\g}
	\pgfdeclareplotmark{xocgC}%
	{%
		\pgfmathtruncatemacro{\ocgnumber}{int(\pgfplotspointmetatransformed)*2+1}%
		\pgfsetstrokecolor{colorC}%
		\pgfpathmoveto{\pgfqpoint{-.70710678\pgfplotmarksize}{-.70710678\pgfplotmarksize}}%
		\pgfpathlineto{\pgfqpoint{.70710678\pgfplotmarksize}{.70710678\pgfplotmarksize}}%
		\pgfpathmoveto{\pgfqpoint{-.70710678\pgfplotmarksize}{.70710678\pgfplotmarksize}}%
		\pgfpathlineto{\pgfqpoint{.70710678\pgfplotmarksize}{-.70710678\pgfplotmarksize}}%
		\node[actions ocg={ocg\ocgname\ocgnumber}{}{ocg\ocgname51 ocg\ocgname101 ocg\ocgname151 ocg\ocgname201 ocg\ocgname251 ocg\ocgname301 ocg\ocgname351 ocg\ocgname401 ocg\ocgname451 ocg\ocgname501 ocg\ocgname601 ocg\ocgname701 ocg\ocgname801 ocg\ocgname901 ocg\ocgname1001}, draw opacity=0] {\pgfusepathqstroke};%
	}%
	\tikzset{external/export next=true}%
	\tikzsetnextfilename{finite_size_scalings_t-1p0_w-1p0_g-1p5}
	\begin{tikzpicture}
		\begin{groupplot}
		[
			group style = 
			{
				group size 		=	2 by 1,
				horizontal sep		=	4em,
			},
			width	= 0.48\textwidth,
			height	= 0.3\textheight,
			ticklabel style	= {font=\scriptsize},
			ylabel style	= {font=\scriptsize},
			xlabel style	= {font=\scriptsize},
		]
		
			\nextgroupplot
			[
				xmin	= 0,
				xmax	= 0.02,
				xlabel	= {$1/L$},
				ylabel	= {$E_0/L$},
				legend style	= {font=\scriptsize, at={(0.95,0.075)}, anchor=south east},
				legend columns	= 1,
				title		= {\subfloat[\label{fig:ppdmrg:finite-size-scaling:t-\t_w-\w_g-\g:energy}]{}},
				title style	= {yshift=-0.5em, xshift=-0.25\textwidth},	
				point meta min = 0, 
				point meta max = 2000, 
				scaled y ticks=false,
				scaled x ticks=false,
				xticklabel style={
					/pgf/number format/fixed,
					/pgf/number format/precision=3},
			]
				\ifthenelse{\boolean{includeFits}}
				{
					\addplot
					[
						color	= colorC,
						scatter,
						mark	= xocgC,
						only marks,
						thick,
						point meta={\thisrowno{0}},
						error bars/.cd, 
						y dir=both,
						y explicit,
					] 
						table
						[
							x expr = 1.0/\thisrowno{0}, 
							y expr = \thisrowno{2}/\thisrowno{0},
							y error expr = \thisrowno{3}/\thisrowno{0}, 
						]
						{\dstdir/scaling.dat};
					\addlegendentry{\gls{PP-2DMRG}};
				}
				{
					\addplot
					[
						color	= colorC,
						mark	= x,
						only marks,
						thick,
						error bars/.cd, 
						y dir=both,
						y explicit,
					] 
						table
						[
							x expr = 1.0/\thisrowno{0}, 
							y expr = \thisrowno{2}/\thisrowno{0},
							y error expr = \thisrowno{3}/\thisrowno{0}, 
						]
						{\dstdir/scaling.dat};
					\addlegendentry{\gls{PP-2DMRG}};
				}
				\ifthenelse{\boolean{rebuildData}}%
				{%
					\makeatletter
					\addplot
					[
						no markers, 
						color=blue,
						thick,
						forget plot,
					] %
						gnuplot 
						[
							raw gnuplot
						] 
						{
							set fit quiet;
							set fit logfile '/dev/null';
							set print "\dstdir/scaling_output.dat" append;
							set fit errorvariables;
							f(x) = a+b*x;
							fit f(x) '\dstdir/scaling.dat' u (1.0/$1):($3/$1):($4/$1) yerrors via a,b;
							print a,a_err,b,b_err;
						};
					\makeatother
				}{}%
				\pgfplotstableread[header=false]{\dstdir/scaling_output.dat}{\Data}%
				\pgfplotstablegetelem{0}{0}\of\Data%
				\xdef\ppEapp{\pgfplotsretval}%
				\pgfplotstablegetelem{0}{1}\of\Data%
				\xdef\ppEaerrpp{\pgfplotsretval}%
				\pgfplotstablegetelem{0}{2}\of\Data%
				\xdef\ppEbpp{\pgfplotsretval}%
				\pgfplotstablegetelem{0}{3}\of\Data%
				\xdef\ppEberrpp{\pgfplotsretval}%
				\addplot
				[
					mark	= none,
					color	= colorC,
					thick,
					domain	= \xminMain:0.02,
					samples	= 100,
					forget plot,
				]
					{linearFct(\ppEbpp, \ppEapp)};
				\addplot
				[
					color	= colorA,
					mark	= o,
					only marks,
					thick,
				]
					table
					[
						x expr	= 1/\thisrowno{0},
						y expr	= \thisrowno{1}/\thisrowno{0},
					]
					{data/holstein/finite_size_scaling/t_\t_w_\w_g_\g.ps};
				\addlegendentry{\gls{PS-DMRG}};
				\ifthenelse{\boolean{rebuildData}}%
				{%
					\makeatletter
					\addplot
					[
						no markers, 
						color=blue,
						thick,
						forget plot,
					] %
						gnuplot 
						[
							raw gnuplot
						] 
						{
							set fit quiet;
							set fit logfile '/dev/null';
							set print "data/holstein/finite_size_scaling/t_\t_w_\w_g_\g.ps_gs_energy_fit" append;
							set fit errorvariables;
							f(x) = a+b*x;
							fit f(x) 'data/holstein/finite_size_scaling/t_\t_w_\w_g_\g.ps' u (1.0/$1):($2/$1):($3) yerrors via a,b;
							print a,a_err,b,b_err;
						};
					\makeatother
				}{}%
				\pgfplotstableread[header=false]{data/holstein/finite_size_scaling/t_\t_w_\w_g_\g.ps_gs_energy_fit}{\Data}%
				\pgfplotstablegetelem{0}{0}\of\Data%
				\xdef\ppEaps{\pgfplotsretval}%
				\pgfplotstablegetelem{0}{1}\of\Data%
				\xdef\ppEaerrps{\pgfplotsretval}%
				\pgfplotstablegetelem{0}{2}\of\Data%
				\xdef\ppEbps{\pgfplotsretval}%
				\pgfplotstablegetelem{0}{3}\of\Data%
				\xdef\ppEberrps{\pgfplotsretval}%
				\addplot
				[
					forget plot,
					mark	= none,
					color	= colorA,
					thick,
					domain	= 0:0.02,
					samples	= 100,
				]
					{linearFct(\ppEbps,\ppEaps)};
				\addplot
				[
					color	= colorB,
					mark	= square,
					only marks,
					thick,
				]
					table
					[
						x expr	= 1/\thisrowno{0},
						y expr	= \thisrowno{1}/\thisrowno{0},
					]
					{data/holstein/finite_size_scaling/t_\t_w_\w_g_\g.dmrg3s};
				\addlegendentry{\gls{DMRG3S+LBO}};
				\ifthenelse{\boolean{rebuildData}}%
				{%
					\makeatletter
					\addplot
					[
						no markers, 
						color=blue,
						thick,
						forget plot,
					] %
						gnuplot 
						[
							raw gnuplot
						] 
						{
							set fit quiet;
							set fit logfile '/dev/null';
							set print "data/holstein/finite_size_scaling/t_\t_w_\w_g_\g.dmrg3s_gs_energy_fit" append;
							set fit errorvariables;
							f(x) = a+b*x;
							fit f(x) 'data/holstein/finite_size_scaling/t_\t_w_\w_g_\g.dmrg3s' u (1.0/$1):($2/$1):($3) yerrors via a,b;
							print a,a_err,b,b_err;
						};
					\makeatother
				}{}%
				\pgfplotstableread[header=false]{data/holstein/finite_size_scaling/t_\t_w_\w_g_\g.dmrg3s_gs_energy_fit}{\Data}%
				\pgfplotstablegetelem{0}{0}\of\Data%
				\xdef\ppEalbo{\pgfplotsretval}%
				\pgfplotstablegetelem{0}{1}\of\Data%
				\xdef\ppEaerrlbo{\pgfplotsretval}%
				\pgfplotstablegetelem{0}{2}\of\Data%
				\xdef\ppEblbo{\pgfplotsretval}%
				\pgfplotstablegetelem{0}{3}\of\Data%
				\xdef\ppEberrlbo{\pgfplotsretval}%
				\addplot
				[
					forget plot,
					mark	= none,
					color	= colorB,
					thick,
					domain	= 0:0.02,
					samples	= 100,
				]
					{linearFct(\ppEblbo,\ppEalbo)};
				\coordinate (inset_position) at (axis description cs:0.975,0.125);
				\coordinate (inset_position_two) at (axis description cs:0.2,0.92);

			\nextgroupplot
			[
				xmin	= 0,
				xmax	= 0.02,
				xlabel	= {$1/L$},
				ylabel	= {$\mathcal{O}_{\rm disp}$},
				title		= {\subfloat[\label{fig:ppdmrg:finite-size-scaling:t-\t_w-\w_g-\g:order-parameter}]{}},
				title style	= {yshift=-0.5em, xshift=-0.25\textwidth},
				legend style	= 
				{
					font=\scriptsize, 
					at={(0.05,0.075)}, 
					anchor=south west
				},
				legend columns	= 1,
				scaled y ticks=false,
				scaled x ticks=false,
				yticklabel style={
					/pgf/number format/fixed,
					/pgf/number format/precision=2},
				xticklabel style={
					/pgf/number format/fixed,
					/pgf/number format/precision=3},
			]
				
				\addplot
				[
					color	= colorC,
					mark	= x,
					only marks,
					thick,
					restrict expr to domain	= {1/x}{\xmax:\xmin},
				]
					table
					[
						x expr	= 1/\thisrowno{0},
						y expr	= \thisrowno{6},
					]
					{data/holstein/finite_size_scaling/t_\t_w_\w_g_\g.symmps.\dw.\blas};
				\addlegendentry{\gls{PP-2DMRG}};
				\ifthenelse{\boolean{rebuildData}}%
				{%
					\makeatletter
						\addplot
						[
							no markers, 
							color=blue,
							thick,
							forget plot,
						] %
							gnuplot 
							[
								raw gnuplot
							] 
							{
								set fit quiet;
								set fit logfile '/dev/null';
								set print "data/holstein/finite_size_scaling/t_\t_w_\w_g_\g.symmps.\dw.\blas_order_parameter_fit" append;
								set fit errorvariables;
								f(x) = a+b*x+c*x*x;
								fit [1.0/\xmin:1.0/\xmax]f(x) 'data/holstein/finite_size_scaling/t_\t_w_\w_g_\g.symmps.\dw.\blas' u (1.0/$1):7 via a,b,c;
								print a,a_err,b,b_err,c,c_err;
							};
					\makeatother
				}{}%
				\pgfplotstableread[header=false]{data/holstein/finite_size_scaling/t_\t_w_\w_g_\g.symmps.\dw.\blas_order_parameter_fit}{\Data}
				\pgfplotstablegetelem{0}{0}\of\Data 
				\xdef\ppEapp{\pgfplotsretval}
				\pgfplotstablegetelem{0}{1}\of\Data 
				\xdef\ppEaerrpp{\pgfplotsretval}
				\pgfplotstablegetelem{0}{2}\of\Data 
				\xdef\ppEbpp{\pgfplotsretval}
				\pgfplotstablegetelem{0}{3}\of\Data 
				\xdef\ppEberrpp{\pgfplotsretval}
				\pgfplotstablegetelem{0}{4}\of\Data 
				\xdef\ppEcpp{\pgfplotsretval}
				\pgfplotstablegetelem{0}{5}\of\Data 
				\xdef\ppEcerrpp{\pgfplotsretval}
				\addplot
				[
					forget plot,
					mark	= none,
					color	= colorC,
					thick,
					domain	= 0:0.02,
					samples	= 100,
				]
				{quadFct(\ppEcpp,\ppEbpp,\ppEapp)};
				\addplot
				[
					color	= colorA,
					mark	= o,
					only marks,
					thick,
				]
				table
				[
					skip rows between index={0}{2},
					x expr	= 1/\thisrowno{0},
					y expr	= \thisrowno{3},
				]
				{data/holstein/finite_size_scaling/t_\t_w_\w_g_\g.ps};
				\addlegendentry{\gls{PS-DMRG}}
				\ifthenelse{\boolean{rebuildData}}%
				{%
					\makeatletter
					\addplot
					[
					no markers, 
					color=blue,
					thick,
					forget plot,
					] %
					gnuplot 
					[
					raw gnuplot
					] 
					{
						set fit quiet;
						set fit logfile '/dev/null';
						set print "data/holstein/finite_size_scaling/t_\t_w_\w_g_\g.ps_order_parameter_fit" append;
						set fit errorvariables;
						f(x) = a+b*x+c*x*x;
						fit [1.0/\xmin:1.0/\xmax]f(x) 'data/holstein/finite_size_scaling/t_\t_w_\w_g_\g.ps' u (1.0/$1):($4) via a,b,c;
						print a,a_err,b,b_err,c,c_err;
					};
					\makeatother
				}{}%
				\pgfplotstableread[header=false]{data/holstein/finite_size_scaling/t_\t_w_\w_g_\g.ps_order_parameter_fit}{\Data}
				\pgfplotstablegetelem{0}{0}\of\Data 
				\xdef\ppEaps{\pgfplotsretval}
				\pgfplotstablegetelem{0}{1}\of\Data 
				\xdef\ppEaerrps{\pgfplotsretval}
				\pgfplotstablegetelem{0}{2}\of\Data 
				\xdef\ppEbps{\pgfplotsretval}
				\pgfplotstablegetelem{0}{3}\of\Data 
				\xdef\ppEberrps{\pgfplotsretval}
				\pgfplotstablegetelem{0}{4}\of\Data 
				\xdef\ppEcps{\pgfplotsretval}
				\pgfplotstablegetelem{0}{5}\of\Data 
				\xdef\ppEcerrps{\pgfplotsretval}
				
				\addplot
				[
					forget plot,
					mark	= none,
					color	= colorA,
					thick,
					domain	= 0:0.02,
					samples	= 100,
				]
				{quadFct(\ppEcps,\ppEbps,\ppEaps)};
				
				\addplot
				[
					color	= colorB,
					mark	= square,
					only marks,
					thick,
				]
					table
					[
						x expr	= 1/\thisrowno{0},
						y expr	= \thisrowno{6},
					]
					{data/holstein/finite_size_scaling/t_\t_w_\w_g_\g.dmrg3s};
				\addlegendentry{\gls{DMRG3S+LBO}};

				\ifthenelse{\boolean{rebuildData}}%
				{%
					\makeatletter
						\addplot
						[
							no markers, 
							color=blue,
							thick,
							forget plot,
						] %
							gnuplot 
							[
								raw gnuplot
							] 
							{
								set fit quiet;
								set fit logfile '/dev/null';
								set print "data/holstein/finite_size_scaling/t_\t_w_\w_g_\g.dmrg3s_order_parameter_fit" append;
								set fit errorvariables;
								f(x) = a+b*x+c*x*x;
								fit [1.0/\xmin:1.0/\xmax]f(x) 'data/holstein/finite_size_scaling/t_\t_w_\w_g_\g.dmrg3s' u (1.0/$1):($7) via a,b,c;
								print a,a_err,b,b_err,c,c_err;
							};
					\makeatother
				}{}%
				\pgfplotstableread[header=false]{data/holstein/finite_size_scaling/t_\t_w_\w_g_\g.dmrg3s_order_parameter_fit}{\Data}
				\pgfplotstablegetelem{0}{0}\of\Data 
				\xdef\ppEalbo{\pgfplotsretval}
				\pgfplotstablegetelem{0}{1}\of\Data 
				\xdef\ppEaerrlbo{\pgfplotsretval}
				\pgfplotstablegetelem{0}{2}\of\Data 
				\xdef\ppEblbo{\pgfplotsretval}
				\pgfplotstablegetelem{0}{3}\of\Data 
				\xdef\ppEberrlbo{\pgfplotsretval}
				\pgfplotstablegetelem{0}{4}\of\Data 
				\xdef\ppEclbo{\pgfplotsretval}
				\pgfplotstablegetelem{0}{5}\of\Data 
				\xdef\ppEcerrlbo{\pgfplotsretval}
				
				\addplot
				[
					forget plot,
					mark	= none,
					color	= colorB,
					thick,
					domain	= 0:0.02,
					samples	= 100,
				]
				{quadFct(\ppEclbo,\ppEblbo,\ppEalbo)};
		\end{groupplot}
		\ifthenelse{\boolean{includeFits}}
		{
			\foreach \L in \Ls
			{
				\def\d{1em04}
				\pgfmathsetmacro{\N}{int(\L/2)}
				\IfFileExists{\datadir/../gs_energies.dat}%
				{
					\immediate\write18%
					{if [ \string$(cat \datadir/../gs_energies.dat | wc -l) = 1 ]; then touch  \datadir/../.NoValidResultsDetected.tmp; fi}%
					\IfFileExists{\datadir/../gs_fit.dat}%
					{
						\immediate\write18%
						{if [ \string$(cat \datadir/../gs_fit.dat | wc -l) = 0 ]; then touch  \datadir/../.NoValidResultsDetected.tmp; fi}%
						\immediate\write18%
						{if [ \string$(grep -v "NaN" \datadir/../gs_fit.dat | wc -l) = 0 ]; then touch  \datadir/../.NoValidResultsDetected.tmp; fi}%
						\immediate\write18%
						{if [ \string$(grep -v "inf" \datadir/../gs_fit.dat | wc -l) = 0 ]; then touch  \datadir/../.NoValidResultsDetected.tmp; fi}%
					}{}%
					
				}%
				{
					\immediate\write18{touch  \datadir/../.NoValidResultsDetected.tmp}%
				}%
				\IfFileExists{\datadir/../.NoValidResultsDetected.tmp}%
				{	
					\immediate\write18{rm -f  \datadir/../.NoValidResultsDetected.tmp}%
				}%
				{
					\pgfplotstableread[header=false]{\datadir/../gs_fit.dat}{\DataUZero}%
					\pgfplotstablegetelem{0}{2}\of\DataUZero%
					\pgfmathsetmacro{\Ea}{\pgfplotsretval}%
					\pgfplotstablegetelem{0}{3}\of\DataUZero%
					\edef\Eaerr{\pgfplotsretval}%
					\pgfplotstablegetelem{0}{4}\of\DataUZero%
					\pgfmathsetmacro{\Eb}{\pgfplotsretval}%
					\pgfmathparse{\eval{\Ea/\L}}%
					\edef\E{\pgfmathresult}%
					\pgfmathparse{\eval{\Eaerr}}%
					\edef\Eerr{\pgfmathresult}%
					\begin{axis}
					[
						y tick label style={
							/pgf/number format/.cd,
							fixed,
							fixed zerofill,
							precision=3,
							/tikz/.cd
						},
						name = ocgplot,
						ocg = {name=ocg\ocgname\L, ref=ocg\ocgname\L, status=invisible},
						inner sep=1pt,
						at	=	 {(inset_position_two)},
						anchor	= north west,
						width	=	0.39\textwidth,
						height	=	0.135\textheight,
						extra description/.code={\node[anchor=north,font=\tiny] at (0.5,0.95) {Scaling for \gls{PP-2DMRG}};},
						xmode	=	log,
						axis background/.style={fill=white, fill opacity=0.9},
						xlabel	=	{\tiny Discarded weight},
						ylabel	=	{\tiny $\nicefrac{E_0}{L}$},
						title	=	{\tiny $L=\L$, $E_0/L=\pgfmathprintnumber[std,precision=4]{\E} \pm \pgfmathprintnumber[sci,precision=1]{\Eerr}$},
						title style={yshift=-0.35em,xshift=-1em,fill=white, name=title},
						every tick label/.append style={font=\tiny,fill=white, fill opacity=0.9},
						xlabel style={name=xlabel,fill=white, fill opacity=0.9},
					]
						\addplot
						[
							color=blue,
							only marks,
							mark=x,
							forget plot,
							unbounded coords=jump,
							error bars/.cd, 
							y dir=both,
							y explicit,
						] 
							table
							[
								x expr = \thisrowno{3},
								y expr = \eval{(\thisrowno{1})/(\L)},
								y error expr = \eval{2.0*sqrt(\thisrowno{3}/\L)}, 
							]
							{\datadir/../gs_energies.dat};
					
						\addplot
						[
							domain	=	\xmininset:\xmaxinset,
							color=blue,
							dashed,
						]
							{linearFct(\Eb, \Ea)/\L};
					\end{axis}
				}%
			}
			\node[actions ocg={}{}{ocg\ocgname51 ocg\ocgname101 ocg\ocgname151 ocg\ocgname201 ocg\ocgname251 ocg\ocgname301 ocg\ocgname351 ocg\ocgname401 ocg\ocgname451 ocg\ocgname501 ocg\ocgname601 ocg\ocgname701 ocg\ocgname801 ocg\ocgname901 ocg\ocgname1001},fit = (title) (xlabel), inner sep = 0,yshift=-0.0575\textheight, inner xsep=2em, xshift=-0.0625\textwidth] {};
		}{}
	\end{tikzpicture}
	\caption
	{
		\captionContent
	}
\end{figure}%
\setboolean{removeData}{false}%
%
\def\t{1p0}%
\def\w{1p0}%
\def\g{0p5}%
\pgfplotstableread[header=false]{data/holstein/symmps_with_bath_sites/t_\t_w_\w_g_\g/finite_size_scaling/scaling_output.dat}{\Data}%
\pgfplotstablegetelem{0}{0}\of\Data%
\xdef\pEapp{\pgfplotsretval}%
\pgfplotstablegetelem{0}{1}\of\Data%
\xdef\pEaerrpp{\pgfplotsretval}%
\pgfplotstablegetelem{0}{2}\of\Data%
\xdef\pEbpp{\pgfplotsretval}%
\pgfplotstablegetelem{0}{3}\of\Data%
\xdef\pEberrpp{\pgfplotsretval}%
\pgfplotstableread[header=false]{data/holstein/finite_size_scaling/t_\t_w_\w_g_\g.ps_gs_energy_fit}{\Data}%
\pgfplotstablegetelem{0}{0}\of\Data%
\xdef\pEaps{\pgfplotsretval}%
\pgfplotstablegetelem{0}{1}\of\Data%
\xdef\pEaerrps{\pgfplotsretval}%
\pgfplotstablegetelem{0}{2}\of\Data%
\xdef\pEbps{\pgfplotsretval}%
\pgfplotstablegetelem{0}{3}\of\Data%
\xdef\pEberrps{\pgfplotsretval}%
\pgfplotstableread[header=false]{data/holstein/finite_size_scaling/t_\t_w_\w_g_\g.dmrg3s_gs_energy_fit}{\Data}%
\pgfplotstablegetelem{0}{0}\of\Data%
\xdef\pEalbo{\pgfplotsretval}%
\pgfplotstablegetelem{0}{1}\of\Data%
\xdef\pEaerrlbo{\pgfplotsretval}%
\pgfplotstablegetelem{0}{2}\of\Data%
\xdef\pEblbo{\pgfplotsretval}%
\pgfplotstablegetelem{0}{3}\of\Data%
\xdef\pEberrlbo{\pgfplotsretval}%
%
\def\t{1p0}%
\def\w{1p0}%
\def\g{1p5}%
\pgfplotstableread[header=false]{data/holstein/symmps_with_bath_sites/t_\t_w_\w_g_\g/finite_size_scaling/scaling_output.dat}{\Data}%
\pgfplotstablegetelem{0}{0}\of\Data%
\xdef\ppEapp{\pgfplotsretval}%
\pgfplotstablegetelem{0}{1}\of\Data%
\xdef\ppEaerrpp{\pgfplotsretval}%
\pgfplotstablegetelem{0}{2}\of\Data%
\xdef\ppEbpp{\pgfplotsretval}%
\pgfplotstablegetelem{0}{3}\of\Data%
\xdef\ppEberrpp{\pgfplotsretval}%
\pgfplotstableread[header=false]{data/holstein/finite_size_scaling/t_\t_w_\w_g_\g.ps_gs_energy_fit}{\Data}%
\pgfplotstablegetelem{0}{0}\of\Data%
\xdef\ppEaps{\pgfplotsretval}%
\pgfplotstablegetelem{0}{1}\of\Data%
\xdef\ppEaerrps{\pgfplotsretval}%
\pgfplotstablegetelem{0}{2}\of\Data%
\xdef\ppEbps{\pgfplotsretval}%
\pgfplotstablegetelem{0}{3}\of\Data%
\xdef\ppEberrps{\pgfplotsretval}%
\pgfplotstableread[header=false]{data/holstein/finite_size_scaling/t_\t_w_\w_g_\g.dmrg3s_gs_energy_fit}{\Data}%
\pgfplotstablegetelem{0}{0}\of\Data%
\xdef\ppEalbo{\pgfplotsretval}%
\pgfplotstablegetelem{0}{1}\of\Data%
\xdef\ppEaerrlbo{\pgfplotsretval}%
\pgfplotstablegetelem{0}{2}\of\Data%
\xdef\ppEblbo{\pgfplotsretval}%
\pgfplotstablegetelem{0}{3}\of\Data%
\xdef\ppEberrlbo{\pgfplotsretval}%
%
\def\t{1p0}%
\def\w{1p0}%
\def\g{2p0}%
\pgfplotstableread[header=false]{data/holstein/symmps_with_bath_sites/t_\t_w_\w_g_\g/finite_size_scaling/scaling_output.dat}{\Data}%
\pgfplotstablegetelem{0}{0}\of\Data%
\xdef\pppEapp{\pgfplotsretval}%
\pgfplotstablegetelem{0}{1}\of\Data%
\xdef\pppEaerrpp{\pgfplotsretval}%
\pgfplotstablegetelem{0}{2}\of\Data%
\xdef\pppEbpp{\pgfplotsretval}%
\pgfplotstablegetelem{0}{3}\of\Data%
\xdef\pppEberrpp{\pgfplotsretval}%
\pgfplotstableread[header=false]{data/holstein/finite_size_scaling/t_\t_w_\w_g_\g.ps_gs_energy_fit}{\Data}%
\pgfplotstablegetelem{0}{0}\of\Data%
\xdef\pppEaps{\pgfplotsretval}%
\pgfplotstablegetelem{0}{1}\of\Data%
\xdef\pppEaerrps{\pgfplotsretval}%
\pgfplotstablegetelem{0}{2}\of\Data%
\xdef\pppEbps{\pgfplotsretval}%
\pgfplotstablegetelem{0}{3}\of\Data%
\xdef\pppEberrps{\pgfplotsretval}%
\pgfplotstableread[header=false]{data/holstein/finite_size_scaling/t_\t_w_\w_g_\g.dmrg3s_gs_energy_fit}{\Data}%
\pgfplotstablegetelem{0}{0}\of\Data%
\xdef\pppEalbo{\pgfplotsretval}%
\pgfplotstablegetelem{0}{1}\of\Data%
\xdef\pppEaerrlbo{\pgfplotsretval}%
\pgfplotstablegetelem{0}{2}\of\Data%
\xdef\pppEblbo{\pgfplotsretval}%
\pgfplotstablegetelem{0}{3}\of\Data%
\xdef\pppEberrlbo{\pgfplotsretval}%
\begin{table}
	\centering
	\caption
	{
		\label{tab:e-densities}
		Energy densities $\varepsilon_{\infty} = \lim_{L\rightarrow \infty} \varepsilon (L)$ in units of $\hopping$ obtained from finite-size scaling of ground\hyp state energies.
	}
	\begin{tabular}{@{}llll}
		\toprule
		& \gls{PS-DMRG} & \gls{DMRG3S+LBO} & \gls{PP-2DMRG}\\
		\midrule
		\cref{eq:comp-pp-with-lbo:1} & 
			\printpgfnumberwitherror{\pEaps}{\pEaerrps}		&
			\printpgfnumberwitherror{\pEalbo}{\pEaerrlbo}	&
			\printpgfnumberwitherror{\pEapp}{\pEaerrpp}		\\
		\cref{eq:comp-pp-with-lbo:2} & 
			\printpgfnumberwitherror{\ppEaps}{\ppEaerrps}		&
			\printpgfnumberwitherror{\ppEalbo}{\ppEaerrlbo}	&
			\printpgfnumberwitherror{\ppEapp}{\ppEaerrpp}		\\
		\cref{eq:comp-pp-with-lbo:3} & 
			\printpgfnumberwitherror{\pppEaps}{\pppEaerrps}		&
			\printpgfnumberwitherror{\pppEalbo}{\pppEaerrlbo}	&
			\printpgfnumberwitherror{\pppEapp}{\pppEaerrpp}		\\
		\bottomrule
	\end{tabular}
\end{table}
The extrapolated energy density in the thermodynamic limit is shown in \cref{tab:e-densities}.
All methods agree within their confidence intervals for the investigated parameter sets \eqref{eq:comp-pp-with-lbo:1} and \eqref{eq:comp-pp-with-lbo:2}.
For illustrational purposes, in \cref{fig:ppdmrg:finite-size-scaling:t-1p0_w-1p0_g-1p5}\subref{fig:ppdmrg:finite-size-scaling:t-1p0_w-1p0_g-1p5:energy}, the finite\hyp size extrapolation is displayed for the parameter set \eqref{eq:comp-pp-with-lbo:2}.
In the \gls{CDW} phase \eqref{eq:comp-pp-with-lbo:3}, the finite-size extrapolations seem to underestimate the fitting error.
We attribute this to the fact that in the \gls{CDW} phase, and at weakened conditions on the precision, the formation of heavy polarons effectively suppresses the energy gains due to fermion delocalization.
As a consequence, domain walls in the fermionic system are harder to resolve and the calculations can get stuck in a local minimum.
In our calculations, it appears that all three methods are affected by this problem, though in a different way owing to their different approximations, and practically, convergence has to be checked very carefully.
\paragraph*{Order parameter}
%
\def\t{1p0}%
\def\w{1p0}%
\def\g{0p5}%
\pgfplotstableread[header=false]{data/holstein/finite_size_scaling/t_\t_w_\w_g_\g.symmps.\dw.\blas_order_parameter_fit}{\Data}%
\pgfplotstablegetelem{0}{0}\of\Data%
\xdef\pOPapp{\pgfplotsretval}%
\pgfplotstablegetelem{0}{1}\of\Data%
\xdef\pOPaerrpp{\pgfplotsretval}%
\pgfplotstablegetelem{0}{2}\of\Data%
\xdef\pOPbpp{\pgfplotsretval}%
\pgfplotstablegetelem{0}{3}\of\Data%
\xdef\pOPberrpp{\pgfplotsretval}%
\pgfplotstablegetelem{0}{4}\of\Data%
\xdef\pOPcpp{\pgfplotsretval}%
\pgfplotstablegetelem{0}{5}\of\Data%
\xdef\pOPcerrpp{\pgfplotsretval}%
\pgfplotstableread[header=false]{data/holstein/finite_size_scaling/t_\t_w_\w_g_\g.ps_order_parameter_fit}{\Data}%
\pgfplotstablegetelem{0}{0}\of\Data%
\xdef\pOPaps{\pgfplotsretval}%
\pgfplotstablegetelem{0}{1}\of\Data%
\xdef\pOPaerrps{\pgfplotsretval}%
\pgfplotstablegetelem{0}{2}\of\Data%
\xdef\pOPbps{\pgfplotsretval}%
\pgfplotstablegetelem{0}{3}\of\Data%
\xdef\pOPberrps{\pgfplotsretval}%
\pgfplotstablegetelem{0}{4}\of\Data%
\xdef\pOPcps{\pgfplotsretval}%
\pgfplotstablegetelem{0}{5}\of\Data%
\xdef\pOPcerrps{\pgfplotsretval}%
\pgfplotstableread[header=false]{data/holstein/finite_size_scaling/t_\t_w_\w_g_\g.dmrg3s_order_parameter_fit}{\Data}%
\pgfplotstablegetelem{0}{0}\of\Data%
\xdef\pOPalbo{\pgfplotsretval}%
\pgfplotstablegetelem{0}{1}\of\Data%
\xdef\pOPaerrlbo{\pgfplotsretval}%
\pgfplotstablegetelem{0}{2}\of\Data%
\xdef\pOPblbo{\pgfplotsretval}%
\pgfplotstablegetelem{0}{3}\of\Data%
\xdef\pOPberrlbo{\pgfplotsretval}%
\pgfplotstablegetelem{0}{4}\of\Data%
\xdef\pOcblbo{\pgfplotsretval}%
\pgfplotstablegetelem{0}{5}\of\Data%
\xdef\pOPcerrlbo{\pgfplotsretval}%
%
\def\t{1p0}%
\def\w{1p0}%
\def\g{1p5}%
\pgfplotstableread[header=false]{data/holstein/finite_size_scaling/t_\t_w_\w_g_\g.symmps.\dw.\blas_order_parameter_fit}{\Data}%
\pgfplotstablegetelem{0}{0}\of\Data%
\xdef\ppOPapp{\pgfplotsretval}%
\pgfplotstablegetelem{0}{1}\of\Data%
\xdef\ppOPaerrpp{\pgfplotsretval}%
\pgfplotstablegetelem{0}{2}\of\Data%
\xdef\ppOPbpp{\pgfplotsretval}%
\pgfplotstablegetelem{0}{3}\of\Data%
\xdef\ppOPberrpp{\pgfplotsretval}%
\pgfplotstablegetelem{0}{4}\of\Data%
\xdef\ppOPcpp{\pgfplotsretval}%
\pgfplotstablegetelem{0}{5}\of\Data%
\xdef\ppOPcerrpp{\pgfplotsretval}%
\pgfplotstableread[header=false]{data/holstein/finite_size_scaling/t_\t_w_\w_g_\g.ps_order_parameter_fit}{\Data}%
\pgfplotstablegetelem{0}{0}\of\Data%
\xdef\ppOPaps{\pgfplotsretval}%
\pgfplotstablegetelem{0}{1}\of\Data%
\xdef\ppOPaerrps{\pgfplotsretval}%
\pgfplotstablegetelem{0}{2}\of\Data%
\xdef\ppOPbps{\pgfplotsretval}%
\pgfplotstablegetelem{0}{3}\of\Data%
\xdef\ppOPberrps{\pgfplotsretval}%
\pgfplotstablegetelem{0}{4}\of\Data%
\xdef\ppOPcps{\pgfplotsretval}%
\pgfplotstablegetelem{0}{5}\of\Data%
\xdef\ppOPcerrps{\pgfplotsretval}%
\pgfplotstableread[header=false]{data/holstein/finite_size_scaling/t_\t_w_\w_g_\g.dmrg3s_order_parameter_fit}{\Data}%
\pgfplotstablegetelem{0}{0}\of\Data%
\xdef\ppOPalbo{\pgfplotsretval}%
\pgfplotstablegetelem{0}{1}\of\Data%
\xdef\ppOPaerrlbo{\pgfplotsretval}%
\pgfplotstablegetelem{0}{2}\of\Data%
\xdef\ppOPblbo{\pgfplotsretval}%
\pgfplotstablegetelem{0}{3}\of\Data%
\xdef\ppOPberrlbo{\pgfplotsretval}%
\pgfplotstablegetelem{0}{4}\of\Data%
\xdef\ppOPclbo{\pgfplotsretval}%
\pgfplotstablegetelem{0}{5}\of\Data%
\xdef\ppOPcerrlbo{\pgfplotsretval}%
%
\def\t{1p0}%
\def\w{1p0}%
\def\g{2p0}%
\pgfplotstableread[header=false]{data/holstein/finite_size_scaling/t_\t_w_\w_g_\g.symmps.\dw.\blas_order_parameter_fit}{\Data}%
\pgfplotstablegetelem{0}{0}\of\Data%
\xdef\pppOPapp{\pgfplotsretval}%
\pgfplotstablegetelem{0}{1}\of\Data%
\xdef\pppOPaerrpp{\pgfplotsretval}%
\pgfplotstablegetelem{0}{2}\of\Data%
\xdef\pppOPbpp{\pgfplotsretval}%
\pgfplotstablegetelem{0}{3}\of\Data%
\xdef\pppOPberrpp{\pgfplotsretval}%
\pgfplotstablegetelem{0}{4}\of\Data%
\xdef\pppOPcpp{\pgfplotsretval}%
\pgfplotstablegetelem{0}{5}\of\Data%
\xdef\pppOPcerrpp{\pgfplotsretval}%
\pgfplotstableread[header=false]{data/holstein/finite_size_scaling/t_\t_w_\w_g_\g.ps_order_parameter_fit}{\Data}%
\pgfplotstablegetelem{0}{0}\of\Data%
\xdef\pppOPaps{\pgfplotsretval}%
\pgfplotstablegetelem{0}{1}\of\Data%
\xdef\pppOPaerrps{\pgfplotsretval}%
\pgfplotstablegetelem{0}{2}\of\Data%
\xdef\pppOPbps{\pgfplotsretval}%
\pgfplotstablegetelem{0}{3}\of\Data%
\xdef\pppOPberrps{\pgfplotsretval}%
\pgfplotstablegetelem{0}{4}\of\Data%
\xdef\pppOPcps{\pgfplotsretval}%
\pgfplotstablegetelem{0}{5}\of\Data%
\xdef\pppOPcerrps{\pgfplotsretval}%
\pgfplotstableread[header=false]{data/holstein/finite_size_scaling/t_\t_w_\w_g_\g.dmrg3s_order_parameter_fit}{\Data}%
\pgfplotstablegetelem{0}{0}\of\Data%
\xdef\pppOPalbo{\pgfplotsretval}%
\pgfplotstablegetelem{0}{1}\of\Data%
\xdef\pppOPaerrlbo{\pgfplotsretval}%
\pgfplotstablegetelem{0}{2}\of\Data%
\xdef\pppOPblbo{\pgfplotsretval}%
\pgfplotstablegetelem{0}{3}\of\Data%
\xdef\pppOPberrlbo{\pgfplotsretval}%
\pgfplotstablegetelem{0}{4}\of\Data%
\xdef\pppOPclbo{\pgfplotsretval}%
\pgfplotstablegetelem{0}{5}\of\Data%
\xdef\pppOPcerrlbo{\pgfplotsretval}%
\begin{table}
	\centering
	\caption
	{
		\label{tab:order-parameter}
		Order parameters $\mathcal{O}_{\rm disp, \infty} = \lim_{L\rightarrow \infty} \mathcal{O}_{\rm disp}(L)$ obtained from the finite-size scaling of ground\hyp state values.
	}
	\begin{tabular}{@{}llll}
		\toprule
		& \gls{PS-DMRG} & \gls{DMRG3S+LBO} & \gls{PP-2DMRG}\\
		\midrule
		\cref{eq:comp-pp-with-lbo:1} & 
			\printpgfnumberwitherror{\pOPaps}{\pOPaerrps}		&
			\printpgfnumberwitherror{\pOPalbo}{\pOPaerrlbo}		&
			\printpgfnumberwitherror{\pOPapp}{\pOPaerrpp}		\\
		\cref{eq:comp-pp-with-lbo:2} & 
			\printpgfnumberwitherror{\ppOPaps}{\ppOPaerrps}		&
			\printpgfnumberwitherror{\ppOPalbo}{\ppOPaerrlbo}	&
			\printpgfnumberwitherror{\ppOPapp}{\ppOPaerrpp}		\\
		\cref{eq:comp-pp-with-lbo:3} & 
			\printpgfnumberwitherror{\pppOPaps}{\pppOPaerrps}	&
			\printpgfnumberwitherror{\pppOPalbo}{\pppOPaerrlbo}	&
			\printpgfnumberwitherror{\pppOPapp}{\pppOPaerrpp}	\\
		\bottomrule
	\end{tabular}
\end{table}
The extrapolated values for the \gls{CDW} order-parameter in the thermodynamic limit are shown in \cref{tab:order-parameter}; and the explicitly calculated system-size dependencies are plotted in \cref{fig:ppdmrg:finite-size-scaling:t-1p0_w-1p0_g-1p5}\subref{fig:ppdmrg:finite-size-scaling:t-1p0_w-1p0_g-1p5:order-parameter} for the parameter set \eqref{eq:comp-pp-with-lbo:2}.
In the \gls{LL} phase \cref{eq:comp-pp-with-lbo:1}, we expect $\mathcal{O}_{\rm disp,\infty} \equiv 0$.
With a relative precision $\sim \mathcal O(10^{-5})$, this is found for the \gls{PS-DMRG} method, whereas the \gls{PP-2DMRG} and \gls{DMRG3S+LBO} calculations are about an order of magnitude smaller $\sim \mathcal O(10^{-6})$.
Near the phase boundary \cref{eq:comp-pp-with-lbo:2}, all methods yield a small but finite value coinciding with a relative precision of $\sim \mathcal O(10^{-2})$, which is comparably small and most probably a signature of the strong fluctuations.
Note that it is not too surprising that near the phase boundary, the methods may produce different outcomes for the order parameter.
In the \gls{CDW} phase, the three methods coincide up to $4$ significant digits yielding a finite value of the order parameter, as expected.
In particular, the \gls{DMRG3S+LBO} and \gls{PP-2DMRG} methods coincide up to a relative precision of $10^{-5}$.
However, at \eqref{eq:comp-pp-with-lbo:3} all methods do not agree with respect to their error bounds obtained from the finite-size scaling, which was also found in the ground\hyp state energy extrapolation.
A more careful convergence analysis is essential to achieve consistent extrapolations, which, however, is not the goal of this comparison.
Instead, we want to point out that for the case of heavy polarons, ensuring the convergence with respect to all numerical control parameters is of particular importance.
\section{\label{sec:discussion}Discussion}
\begin{table}
	\centering
	\begin{tabular}{@{}>{\centering\arraybackslash}m{.225\textwidth}  >{\centering\arraybackslash}m{.225\textwidth}  >{\centering\arraybackslash}m{.225\textwidth}  >{\centering\arraybackslash}m{.225\textwidth}@{}}
		\toprule
							& 
		\gls{PS-DMRG} 		& 
		\gls{DMRG3S+LBO}	& 
		\gls{PP-2DMRG}		\\
		\midrule
		System enlargement 										& 
		$\times \log_2(n_{\rm ph}+1)$							&
		$\times 1$												&
		$\times 2$												\\ \addlinespace[0.25em]
		Local Hilbert space truncation 							& 
		none													&
		w.r.t. eigenvalues of \gls{1RDM} $\hat \rho_{j}$		&
		w.r.t. diagonal elements of \gls{1RDM} $\hat \rho_{j}$	\\ \addlinespace[0.25em]
		Convergence parameter								& 
		$n_{\rm ph}$, $\delta(m_{\rm max})$					&
		$n_{\rm ph}$, $\delta(m_{\rm max})$, $d_{\rm o}$	&
		$n_{\rm ph}$, $\delta(m_{\rm max})$					\\ \addlinespace[0.25em]
		Phonon symmetries									& 
		none												&
		none												&
		$U(1)$												\\ \addlinespace[0.25em]
		Additional code requirements						& 
		pseudo-site operators								&
		\gls{1RDM} computation, 
		diagonalization \& \gls{DMRG} optimization of \gls{1RDM} trafo, 
		 \gls{MPS} and \gls{MPO} transformation	&
		balancing operators									\\
		\bottomrule
	\end{tabular}
	\caption{
		\label{tab:methods-comparison}
		Comparison of method specific properties when applied to the Holstein model.
		Local Hilbert space truncation refers to the possibility of reducing the local Hilbert space dimension of the phononic degrees of freedom.
		Listed, additional code requirements assume that the given implementation is capable of dealing with long-ranged couplings.
		It is also assumed that the implementation is capable to truncate site tensors w.r.t. control parameters such as the maximal bond dimension $m_{\rm max}$ or the discarded weight $\delta$ per bond.
	}
\end{table}
The technical properties of the discussed methods are summarized and compared in~\cref{tab:methods-comparison}.
It should be noted that both \gls{PS-DMRG} and \gls{PP-2DMRG} require fewer modifications to an existing code compared to an implementation of \gls{DMRG3S+LBO}.
In particular, \gls{PS-DMRG} can be readily used if the implementation supports combined fermionic and hardcore-bosonic lattice degrees of freedom.
In order to use \gls{PP-2DMRG}, the most relevant required modification is the implementation of the balancing operators which, however, are only local operators, i.e., they can be realized in a straightforward manner.
In contrast, the implementation of \gls{DMRG3S+LBO} is more involved and requires the evaluation, optimization and truncation of the \gls{1RDM} during each site update as well as a transformation of the local Hilbert space representation for both, the \gls{MPS} and \gls{MPO} site tensors.
The technical prerequisites in~\cref{tab:methods-comparison} are put in the context of the numerical behavior of the different approaches, described in~\cref{sec:comparison}.
\paragraph*{\gls{PS-DMRG}}
The extension of the system size by introducing pseudo sites controls the numerical behavior.
There are long-ranged couplings between the fermionic sites and the pseudo sites, as well as long-ranged hoppings between the fermionic sites with a maximum range spanning $\log_2(n_{\rm ph}+1)+2$ lattice sites.
The situation can be analyzed by considering a bipartition of the system right in the middle of the chain of pseudo sites.
In such a bipartition, the cut bond has to account for both, fermionic correlations between the physical lattice sites and off-diagonal phonon correlations on the current, physical lattice site.
For the latter, strong phonon fluctuations contribute to the entanglement entropy with a scaling that can be estimated as $S_{\rm PS} \sim N_{\rm PS}/2$ where $N_{\rm PS} = \log_2(n_{\rm ph}+1)$ is the number of pseudo sites.
This contribution to the entanglement entropy superimposes the fermionic correlations.
The convergence behavior of the \gls{PS-DMRG} high-precision calculations in the \gls{LL} phase (c.f.~\cref{fig:t-1p0:w-1p0:g-0p5}) and near the phase boundary (c.f.~\cref{fig:t-1p0:w-1p0:g-1p5}) can be related to this picture.
In both regions, the fermionic degrees of freedom exhibit critical behavior that translates to very long-ranged correlations when introducing pseudo sites.
A local \gls{DMRG}-optimization step in the middle of the chain of pseudo sites, subject to a finite bond dimension, then preferably optimizes with respect to the local correlations.
If the maximally allowed bond dimension $m_{\rm max}$ is not sufficient, this results in a situation in which the optimization is mainly constrained to the phonon system.
Upon increasing $m_{\rm max}$, at some point the bond dimension is large enough to reproduce the long-ranged fermionic correlations with high fidelity.
A similar reasoning can be employed to interpret the behavior of the~\gls{PS-DMRG} when increasing the number of pseudo sites (c.f.~\cref{fig:ps-dmrg:n-pseudosites}).
Here, larger values of $S_{\rm PS}$ in the middle of the chain of pseudo sites yield a slower convergence with respect to the bond dimension.
Note that this behavior can also serve as a convergence criterion, i.e., if the maximum phonon number $n_{\rm ph}$ is large enough so that $S_{\rm PS}$ saturates, the convergence of the ground\hyp state energy when scaling the bond dimension becomes independent of $n_{\rm ph}$.
Finally, once the system is in the \gls{CDW} phase, the fermions localize, allowing for a very fast convergence of the \gls{PS-DMRG} as seen in~\cref{fig:t-1p0:w-1p0:g-2p0}.
\paragraph*{\gls{DMRG3S+LBO}}
In comparison to the other methods, the~\gls{DMRG3S+LBO} method achieves excellent convergence of the ground\hyp state energy with the smallest required bond dimensions.
This is achieved by representing on\hyp site phonon\hyp correlations in terms of the truncated transformation matrices $R_j$ acting on the physical legs.
A careful optimization of the $R_j$'s in order to find the optimal modes representing the phononic~\gls{1RDM} is thus a necessary condition for a well\hyp behaved simulation.
For that reason, a convergence analysis with respect to the number of optimal modes $d_{\rm o}$ is crucial and requires various ground\hyp state searches, since $d_{\rm o}$ cannot be increased in the current implementation of the groundstate search (an adaptive increase of $d_{\rm{o}}$ is possible, though, see~\cite{Brockt2015, Stolpp2020, Jansen2020}).
During a~\gls{DMRG} sweep one has to perform various optimizations per lattice site.
A single site optimization updates both, the site tensor and the transformation matrix.
In the current setup, two subsequent optimizations are sufficient to achieve converged results, but dynamic and more involved convergence criteria may become necessary for more complicated systems.
Using a single-site~\gls{DMRG}-solver~\cite{Hubig2015}, the convergence behavior during~\gls{DMRG}-sweeps does not only depend on the maximally allowed bond dimension and the chosen optimal modes but also the initial state, as well as the subspace expansion mixing factor.
A reasonable choice of the initial state is of particular importance (c.f.~\cref{fig:lbo_mixingfactor}).
Here, the separation of the optimization of the transformation matrix, updating the~\gls{1RDM} approximation of the phononic degrees of freedom, from the local site\hyp tensor optimization, appears to be an important contribution to the tendency of the ground\hyp state search to get stuck.
In the scope of our investigations, a Fermi sea with power\hyp law correlations has proven to be a reliable initial guess state in the three investigated regions of the phase diagram.
\paragraph*{\gls{PP-2DMRG}}
The projected purification combines the ideas of the other methods and, thus, inherits some of their properties.
The convergence behavior is dominated by the off-diagonal correlations that build up in the phononic system, which control the growth of the bond dimension between physical and bath sites.
Since these bonds carry both, fermionic and phononic correlations, the number of states required to approximate the ground\hyp state to a given precision is in general larger than in the case of the~\gls{DMRG3S+LBO} method.
As already discussed for the case of the~\gls{PS-DMRG} method, this may translate to convergence problems, if the available maximal number of states $m_{\rm max}$ between physical and bath sites is not sufficient so that mainly the phononic degrees of freedom are optimized.
However, since the coupling range between the fermions is only increased from nearest\hyp neighbor to next\hyp to\hyp nearest\hyp neighbor, convergence issues typically are not drastic and can be overcome by warm-up sweeps in which the fermion\hyp phonon coupling is increased gradually.
Notably, in our implementations the~\gls{PP-2DMRG} is the only method that allows for a faithful extrapolation in terms of the discarded weight $\delta$.
On the one hand, this is based on the fact that by restoring the global $U(1)$\hyp symmetry, the~\gls{1RDM} can be truncated according to its diagonal elements and thereby the approximation quality of the phonon system is only controlled by $\delta$.
On the other hand, a small bond dimension yields local optimizations to be biased towards the phononic degrees of freedom as observed in the~\gls{PS-DMRG} method.
However, in our calculations we found these effects to be less prominent, since the coupling range between the fermions is increased only moderately.
Since the approximation quality of the phononic~\gls{1RDM} crucially depends on the discarded weight, a scaling analysis with respect to this control parameter is nevertheless of particular importance.
\section{\label{sec:conclusion}Conclusion}
We discussed three state\hyp of\hyp the\hyp art matrix\hyp product\hyp state methods to simulate numerically challenging systems with large local Hilbert spaces and broken $U(1)$-symmetries that appear in various physical problem settings~\cite{Holstein1959,Bloch:2005p988,Heidrich-Meisner2010,Ejima2011,Dorfner2016,Purdy2010,Thompson2013,Ritsch2013,Brouwer2011,Cook2011,Hui2015,Keselman2019}.
The presented methods exploit different representations and optimization schemes to reduce the computational costs.
The pseudosite method (\gls{PS-DMRG}) unfolds the large local Hilbert space into additional sites using a compact binary encoding of the local degrees of freedom.
Efficiently representing large local Hilbert spaces by means of the local basis optimization (\gls{DMRG3S+LBO}) directly operates on the physical degrees of freedom.
These are rotated into an optimal basis in which the \gls{1RDM} is diagonal and can be truncated faithfully.
The recently developed projected purification (\gls{PP-2DMRG}) interpolates between both approaches in the sense that the system is extended by pairing up each lattice site with a bath site, and truncated according to the diagonal elements of the \gls{1RDM}.
Being conceptually very different, it is not immediately clear which method is best suited for a given problem setting.
For that reason, we applied these methods to the Holstein model at half filling, which is a prototypical system featuring Einstein phonons with large local Hilbert spaces and broken $U(1)$-symmetries.
We performed two different common numerical analyses: a high-precision scaling analysis of the model's ground-state energy as a function of the maximally allowed bond dimension at an intermediate system size ($L=51$ sites) as well as finite-size extrapolations for intensive quantities such as the \gls{CDW} order parameter in the ground state up to systems with $L=201$ sites.
Our comparisons demonstrate that, in general, all methods characterize the different phases with high numerical precision and allow for an extrapolation of observables towards the thermodynamic limit.
Analyzing the different methods in more detail, we also identify situations in which it can be beneficial to use a particular method.
For instance, in the case of broadly distributed phonon\hyp excitation probabilities, the \gls{DMRG3S+LBO} and \gls{PP-2DMRG} methods benefit from their capability of truncations on the phononic degrees of freedom.
Here, in particular, the \gls{DMRG3S+LBO} method achieves very compact representations with the smallest bond dimensions found in our calculations (\cref{fig:t-1p0:w-1p0:g-0p5,fig:t-1p0:w-1p0:g-1p5,fig:t-1p0:w-1p0:g-2p0}).
This comes at the cost of a larger amount of numerical control parameters such as the number of phonons per lattice site in combination with the amount of optimal modes kept, or the mixing factor used in the ground\hyp state search.
As discussed in \cref{sec:methods:convergence}, using improper configurations can produce a strong dependency on the initial state, which becomes particularly important in the \gls{CDW} phase, where heavy polarons slow down the overall convergence of the ground\hyp state search.
In these situations, the $2$-site solvers used in the \gls{PS-DMRG} and \gls{PP-2DMRG} methods together with the fact that the only relevant control parameter is the maximum bond dimension seem to be easier to control.
Finally, it should be pointed out that \gls{PP-2DMRG} is capable of exploiting restored global $U(1)$ symmetries, which reduces computational costs associated with the local degrees of freedom.
Therefore, even though larger maximal bond dimensions are required to achieve the same numerical precision as the \gls{PS-DMRG} and \gls{DMRG3S+LBO} methods, this is compensated by the more efficient representation of matrix-product states and operators as can be seen in \cref{fig:t-1p0:w-1p0:g-0p5,fig:t-1p0:w-1p0:g-2p0} where the most accurate ground\hyp state approximations are found by \gls{PP-2DMRG}.
However, since the maximally allowed bond dimension is limited to $m_{\rm max}\leq 2000$ in our calculations, close to the phase boundary, the more compact representation of \gls{DMRG3S+LBO} yields the best approximation (see \cref{fig:t-1p0:w-1p0:g-1p5}).
Another important aspect is the question of the applicability of the described methods to study out-of-equilibrium setups.
The \gls{DMRG-LBO} already proved its capability to simulate the dynamics of systems with small fermion densities coupled to lattice phonons \cite{Brockt2015,Brockt2017}, global quenches \cite{Hashimoto2017,Stolpp2020} and also finite-temperature simulations \cite{Jansen2020} by using a Trotter decomposition of the time\hyp evolution operator.
A very natural, further development would be to employ time-evolution schemes such as the \gls{TDVP} \cite{PhysRevLett.107.070601,PhysRevB.94.165116} or the $W^{\rm II}$-representation \cite{PhysRevB.91.165112} for time-evolution allowing the efficient treatment of long-ranged interactions and larger time-steps (\gls{TDVP}).
From our previous discussion, we expect the \gls{DMRG-LBO} as well as the projected purification to be well-suited for an adoption of \gls{TDVP} as the time-evolution scheme (see \cite{Schroeder2016} for developments to combine \gls{LBO} with \gls{TDVP}), while due to the continuously required basis transformations of the Hamiltonian, the $W^{\rm II}$ method seems to be more suitable for the projected purification.
\section{Acknowledgments}
We thank K. Harms and D. Jansen for insightful discussions. 
TK acknowledges financial support by the ERC Starting Grant from the European Union's Horizon 2020 research and innovation program under grant agreement No. 758935.
This work was funded by the Deutsche Forschungsgemeinschaft (DFG, German Research Foundation) – 207383564; 217133147, via FOR 1807 (projects P4 and P7) and CRC 1073 (projects B03 and B09), respectively.
SP acknowledges support by the Deutsche Forschungsgemeinschaft (DFG, German Research Foundation) under Germany’s Excellence Strategy-426 EXC-2111-390814868.
We thank the TU Clausthal for providing access to the Nuku computational cluster.
\def\newblock{\ }%
\bibliographystyle{iopart-num}
\bibliography{Literatur}
\appendix
%

\end{document}